\documentclass[%
reprint,
twocolumn,
superscriptaddress,
nofootinbib,
amsmath,amssymb,
aps,
]{revtex4-2}
\usepackage{natbib}
\usepackage{booktabs}
\usepackage{url}
\usepackage{orcidlink}
\usepackage{multirow}
\usepackage{float}
\usepackage{graphicx}
\usepackage{dcolumn}
\usepackage{bm}

\begin{document}
	
	\def\aap{AA}
	\def\prl{Phys. Rev. Lett.}
	\def\aapr{AA Rev}
	\def\apjl{ApJL}
	\def\na{NewA}
	\def\apss{APSS}
	\def\aaps{AAPS}
	\def\mnras{MNRAS}
	\def\apj{ApJ}
	\def\apjs{ApJS}
	\def\aj{AJ}
	\def\pasp{PASP}
	\def\pasj{PASJ}
	\def\nat{Nat}
	\def\memsai{MmSAI}
	\def\prd{PhysRevD}
	\def\jcap{JCAP}
	\def\physrep{Phys. Rept.}
	\def\araa{Annual Review of Astronomy and Astrophysics}
	\def\ssr{Space Science Reviews}
	
	\newcommand{{\qsub}}{\bf{q}}
	\newcommand{{\rsub}}{{\bf{r}}}
	\newcommand{{\xlight}}{{\bf{n}}_{\rm{light}}}
	\newcommand{{\xlens}}{{\bf{n}}_{\rm{mac}}}
	\newcommand{{\data}}{\bf{D}}
	\newcommand{{\datan}}{{\bf{d}}_{\rm{n}}}
	\newcommand{{\datanprime}}{{{\bf{d}}_{\rm{n}}^{\prime}}}
	\newcommand{{\dimg}}{{\bf{d}}_{\mathcal{I}}}
	\newcommand{{\dimgprime}}{{\bf{d}}_{\mathcal{I}}^{\prime}}
	\newcommand{{\dptsrc}}{{\bf{d}}_{\rm{ptsrc}}}
	\newcommand{{\dptsrcprime}}{{\bf{d}}_{\rm{ptsrc}}^{\prime}}
	\newcommand{{\dfr}}{{\bf{d}}_{\rm{fr}}}
	\newcommand{{\dfrprime}}{{\bf{d}}_{\rm{fr}}^{\prime}}
	\newcommand{{\zlens}}{{z_{\rm{d}}}}
	\newcommand{{\zsrc}}{{z_{\rm{s}}}}
	
	\newcommand{{\M}}{\boldsymbol{\mathcal{M}}}
	
	\newcommand{\vdag}{(v)^\dagger}
	\newcommand\aastex{AAS\TeX}
	\newcommand\latex{La\TeX}
	
	\newcommand{{\qm}}{{\boldsymbol{q}_{\mathcal{M}}}}
	\newcommand{{\qemp}}{{\bf{\rm{q}_{\rm{emp}}}}}	
	\newcommand{{\qphys}}{{\bf{\rm{q}_{\rm{phys}}}}}
	
	\preprint{APS/123-QED}
	
	\title{JWST lensed quasar dark matter survey V: Hints of self-interacting dark matter from 29 quadruply imaged quasars}
	
	\author{D.~Gilman\orcidlink{0000-0002-5116-7287}}
\thanks{Brinson Prize Fellow}
\email{gilmanda@uchicago.edu}
\affiliation{Department of Astronomy and Astrophysics, University of Chicago, Chicago, IL 60637, USA}
\affiliation{Kavli Institute for Cosmological Physics, University of Chicago, Chicago, IL 60637, USA}

\author{A.~M.~Nierenberg\orcidlink{0000-0001-6809-2536}}
\affiliation{University of California, Merced, 5200 N Lake Road, Merced, CA 95341, USA}

\author{J.~Gurian\orcidlink{0000-0002-8677-1038}}
\affiliation{Perimeter Institute for Theoretical Physics, Waterloo, Ontario, N2L 2Y5, Canada}
\affiliation{Carnegie Observatories, 813 Santa Barbara Street, Pasadena, CA 91101, USA}

\author{H.~Paugnat\orcidlink{0000-0002-2603-6031}}
\affiliation{Department of Physics and Astronomy, University of California, Los Angeles, CA, 90095, USA}

\author{C.~Gannon\orcidlink{0009-0009-0443-3181}}
\affiliation{University of California, Merced, 5200 N Lake Road, Merced, CA 95341, USA}

\author{M.~N.~Martinez\orcidlink{0000-0002-8397-8412}}
\affiliation{University of California, Merced, 5200 N Lake Road, Merced, CA 95341, USA}

\author{T.~Treu\orcidlink{0000-0002-8460-0390}}
\affiliation{Department of Physics and Astronomy, University of California, Los Angeles, CA, 90095, USA}

\author{K.~N.~Abazajian\orcidlink{0000-0001-9919-6362}}
\affiliation{Department of Physics and Astronomy, University of California, Irvine, CA 92697-4575, USA}

\author{T.~Anguita\orcidlink{0000-0003-0930-5815}}
\affiliation{Instituto de Astrofisica, Departamento de Fisica y Astronomia, Universidad Andres Bello, Santiago, Chile}

\author{V.~N.~Bennert\orcidlink{0000-0003-2064-0518}}
\affiliation{Physics Department, California Polytechnic State University, San Luis Obispo, CA 93407, USA }
    
\author{A.~J.~Benson\orcidlink{0000-0001-5501-6008}}
\affiliation{Carnegie Observatories, 813 Santa Barbara Street, Pasadena, CA 91101, USA}

\author{S.~Birrer\orcidlink{0000-0003-3195-5507}}
\affiliation{Department of Physics and Astronomy, Stony Brook University, Stony Brook, NY 11794, USA}

\author{S.~G.~Djorgovski\orcidlink{0000-0002-0603-3087}}
\affiliation{California Institute of Technology, Pasadena, CA 91125, USA}

\author{S.~F.~Hoenig\orcidlink{0000-0002-6353-1111}}
\affiliation{School of Physics and Astronomy, University of Southampton, Southampton SO17 1BJ, United Kingdom}

\author{R.~E.~Keeley\orcidlink{0000-0002-0862-8789}}
\affiliation{Department of Physics and Astronomy, University of California, Irvine, CA 92697-4575, USA}

\author{A.~Kusenko\orcidlink{0000-0002-8619-1260}}
\affiliation{Department of Physics and Astronomy, University of California, Los Angeles, CA,  90095, USA}
\affiliation{Kavli Institute for the Physics and Mathematics of the Universe (WPI), UTIAS, The University of Tokyo, Kashiwa, Chiba 277-8583, Japan}

\author{M.~Millon\orcidlink{0000-0001-7051-497X}}
\affiliation{D\'epartement de Physique Th\'eorique, Universit\'e de Gen\`eve, 24 quai Ernest-Ansermet, CH-1211 Gen\`eve 4, Switzerland}

\author{T.~Morishita\orcidlink{0000-0002-8512-1404}}
\affiliation{Astronomical Institute, Graduate School of Science, Tohoku University, 6--3 Aoba, Sendai 980-8578, Japan}

\author{L.~A.~Moustakas\orcidlink{0000-0003-3030-2360}}
\affiliation{Jet Propulsion Laboratory, California Institute of Technology, 4800 Oak Grove Dr, Pasadena, CA 91109, USA}

\author{P.~Mozumdar\orcidlink{0000-0002-8593-7243}}
\affiliation{Department of Physics and Astronomy, University of California, Los Angeles, CA,  90095, USA}
\affiliation{Department of Physics and Astronomy, University of California, Davis, 1 Shields Ave., Davis, CA 95616, USA}

\author{D.~Paris\orcidlink{0000-0002-7409-8114}}
\affiliation{INAF -- Osservatorio Astronomico di Roma, via di Frascati 33, 00078 Monte Porzio Catone (RM), Italy}

\author{W.~Sheu\orcidlink{0000-0003-1889-0227}}
\affiliation{Department of Physics and Astronomy, University of California, Los Angeles, CA,  90095, USA}

\author{D.~Sluse\orcidlink{0000-0001-6116-2095}}
\affiliation{STAR Institute, University of Li\`ege, Quartier Agora, All\'ee du six Ao\^ut 19c, 4000 Li\`ege, Belgium}

\author{K.~C.~Wong\orcidlink{0000-0002-8459-7793}}
\affiliation{Research Center for the Early Universe, Graduate School of Science, The University of Tokyo, 7-3-1 Hongo, Bunkyo-ku, Tokyo 113-0033, Japan}
	
	\date{\today}

	\begin{abstract}
		Theories of self-interacting dark matter (SIDM) predict the eventual core collapse of dark matter halos, a process that transforms these structures into extremely efficient gravitational lenses. We present a population-level inference on the abundance of core collapsed halos and subhalos in the mass range $10^6 - 10^{10.7} M_{\odot}$ using 29 quadruply imaged quasars, drawing on the collective lensing signal of many low-mass perturbers. Our structure formation model predicts the abundance of collapsed (sub)halos as a function of halo mass and redshift. We quantify how our results are affected by systematic uncertainties, including the abundance of globular clusters (GCs) and subhalos, and the internal structure of deeply collapsed halos. Our data exhibits a strong preference for core collapsed halos in the lens model, with Bayes factors disfavoring CDM relative to SIDM ranging from 8:1 to 24:1, depending on the assumed SIDM subhalo abundance and the internal structure of collapsed halos. Explaining our data without core collapse requires both a GC abundance well above expectations and more subhalos than predicted by $N$-body simulations. These results pose a challenge to the CDM paradigm. We present a particle physics interpretation of these results in a companion paper.   
	\end{abstract}
	\maketitle
	
	\section{Introduction}
	The cold dark matter (CDM) model predicts an approximately scale-free halo mass function and a universal form for most halo density profiles \citep{NFW97, Springel++08a, Wang++20}. New physics beyond CDM manifests itself in small-scale structure as deviations from universality in halo density profiles and the halo mass function. For example, in warm dark matter models, free streaming suppresses density fluctuations below a characteristic length scale, precluding the formation of low-mass halos \citep{Bode++01, Schneider++12, Lovell++14}. Cosmic probes of small-scale structure have begun to measure halo properties on sub-galactic scales, with the potential to test key predictions of the CDM paradigm and potentially discover new physics \citep[see][and references therein]{Nadler++26}. 
	
	In self-interacting dark matter (SIDM), the class of dark matter theories considered in this work, a new timescale governs the evolution of dark matter halos \citep[see, e.g.,][and references therein]{Adhikari++25}. This timescale depends on the structural properties of a halo and on the elastic (or in some models, inelastic) scattering cross section between dark matter particles. The evolution of the halo profile begins as heat is conducted inwards, and the halo establishes an isothermal core. The direction of heat flow later reverses, triggering core collapse, a runaway contraction of the isothermal core that increases the central density by many orders of magnitude. \citet{LyndenBellWood68} originally studied this phenomenon in the context of dense star clusters. \citet{Balberg++02} later pointed out that the same process should occur in SIDM halos on a much shorter timescale than in CDM halos, because interactions between dark matter particles conduct heat more efficiently than purely gravitational encounters. As self-interactions occur more frequently in high density regions, such as halos, SIDM predicts an evolution of halo density profiles while leaving the large scale structure of the Universe unchanged \citep{SS2000,Vogelsberger++12,Rocha++13}. In many well-motivated SIDM frameworks \citep{Feng++09,Tulin++13,Petraki++14} the cross section has a velocity dependence, in which case the halo mass sets the characteristic velocity scale for the self-interactions \citep{Kaplinghat++16}. A velocity-dependent cross section can evade stringent upper limits $\sigma < 0.1-1 \ \rm{cm^2} \ \rm{g^{-1}}$ from galaxy cluster scales \citep{Randall++2008,Robertson++2017,Sagunski++21,Andrade++22} while exceeding $100 \ \rm{cm^2} \ \rm{g^{-1}}$ on sub-galactic scales. 
	
	Core formation, rather than core collapse, motivated the early investigations of SIDM and small-scale structure, because many dwarf galaxies appeared less concentrated than expected in CDM \citep{SS2000}. As the sample size of low-mass galaxies with kinematic measurements grows and understanding of various astrophysical systematics improves, it has become clear that low-mass galaxies actually exhibit a diversity of internal structure, rather than a systematic underdensity or overdensity \citep[e.g.,][]{Oman++2015,Santos-Santos++20,Sales++22,Raman++26}. Most current investigations of SIDM invoke both core expansion and collapse as possible explanations for the inferred diversity of low-mass galaxies \citep{Kaplinghat++16,Correa21,Slone++23,Nadler++23,Roberts++25}. Studies of low-mass galaxies will continue to provide insights into SIDM, and upcoming surveys will further increase the sample size and control over systematics \citep{Nadler++24}. However, inferences derived from dwarf galaxy properties will necessarily be tied to assumptions in the kinematic modeling and baryonic physics \citep{PontzenGovernato14,Pineda++17,Roper++23}, and the degree of tension with respect to CDM depends on the simulations and modeling assumptions \citep[e.g.][]{ManceraPina++26}. Further, the requirement that halos contain enough luminous tracers of their density profile imposes a minimum halo mass scale at which one can test SIDM predictions. This minimum halo mass scale translates to a minimum velocity scale where one can probe the SIDM cross section. 
	
	In this work, we use strong gravitational lensing to infer dark matter halo density profiles and subhalo abundance. Strong lensing occurs when a foreground deflector, typically a massive elliptical galaxy, produces multiple images of a background source. Dark matter substructure imparts milli-arcsecond scale distortions on lensed images, enabling the direct gravitational detection of dark halos across cosmological distance, even if the halos are too small to retain significant stellar mass \citep[for a review, see][]{Vegetti++24}. More concentrated halos act as more efficient gravitational lenses \citep{Minor++21,Gilman++22}, meaning SIDM models in which halo density profiles evolve with time should produce distinct strong lensing signatures from CDM \citep{Gilman++21, Gilman++23}. As a purely gravitational probe, lensing can push below the threshold of galaxy formation, and test predictions on scales below $10^7 M_{\odot}$, where halos are not expected to host galaxies \citep{Nadler++20}. 
	
	Recent investigations with galaxy-galaxy strong lens systems have resulted in detections of dark objects through their lensing perturbations on extended lensed arcs \citep{Vegetti++10,Hezaveh++16,Nightingale++24}. In many cases, the inferred density of these objects exceeds the central density of halos in $N$-body simulations of CDM structure formation by several orders of magnitude \citep{Minor++21,Ballard++24,Minor++25,Enzi++25,Lange++25,Amvrosiadis++26}, with the particularly notable example of the $\sim 10^6 M_{\odot}$ object detected in a lensed arc imaged through Very Long Baseline Interferometry \citep{Powell++25,Vegetti++26}. These studies have motivated investigations of whether core collapse in SIDM provides a better explanation for these detections than CDM \citep{Nadler++23,Tajalli++25,Li++25,Kong++26}. 
	
	The interpretation of these detections in terms of fundamental physics remains a challenging open question. First, connecting the inferred properties of a few individual structures to the population-level characteristics requires significant extrapolation and additional modeling assumptions. For example, analyses that include a single halo in the lens model typically assume that the collective effect of the surrounding halo population does not affect the inferred properties of the perturber. Single-halo detections also come with many open questions regarding the relative likelihood of various competing explanations of the data, for example, that the detected object is an unusual globular cluster, an ultra-compact dwarf galaxy, or a core collapsed SIDM halo  \citep{He++18,Vegetti++26}, all of which may be present in a lens system and affect the data simultaneously. Putting aside these various complications, if we accept an SIDM explanation for the properties of dense strong lensing perturbers, we should expect to discover a population of core collapsed halos around every galaxy, including all strong lens systems and the Milky Way. 
	
	In our analysis, we use a particular kind of strong lens system, quadruply imaged quasars (quads), to perform a population-level analysis of dark substructure in SIDM. The constraining power of quad lens systems stems from the relative magnifications (flux ratios) among lensed images, which depend on second derivatives of the gravitational potential projected onto the plane of the lens. These data are sensitive to the small fluctuations in the projected mass near an image caused by dark matter subhalos and field halos along the line of sight \citep{MaoSchneider,DalalKochanek02,Nierenberg++14,Gilman++19}. Quad lens systems can constrain any dark matter model that predicts a change in the internal structure or overall abundance of low-mass halos \citep{Hsueh++20,Gilman++20,Laroche++22,Gilman++26}, including SIDM \citep[e.g.,][]{Gilman++21,Gilman++23}.  
	
	The analysis methods applied to quad lens systems have improved considerably over the last decade. Current methods now perform end-to-end forward modeling of subhalo and line-of-sight halo populations, including detailed prescriptions for subhalo tidal evolution \citep{Du++24,Du++25}. Recent advances enable joint modeling of image positions, flux ratios, and extended lensed arcs \citep{Gilman++24,Gilman++25a,Gilman++26}, which aid in constraining the mass profile of the main deflector and isolating the small-scale perturbations by halos from other sources of lens modeling uncertainty. Current methods also include lens model complexity in the form of elliptical multipole expansions of the main deflector mass profile \citep{Paugnat++25b}, and include globular clusters alongside halos as sources of small-scale perturbation \citep{Gilman++25a}. 
	
	In parallel to the modeling, the observations of quad lenses have also advanced considerably, with JWST recently surveying a sample of 31 lensed quasars and measuring image magnifications from compact warm dust emission around the background quasar. The size of the warm dust region renders these data immune to contamination from stellar microlensing, while being compact enough to experience perturbation by halos on mass scales of $\sim 10^6 M_{\odot}$ \citep{Nierenberg++24}. As shown by \citet{Gilman++21}, the compact warm dust region observed by JWST is well suited to studying core collapse in SIDM, because the compact source is more sensitive to the collapsed region of a halo than the more extended nuclear narrow-line region around a background quasar. 
	
	In this work, we present population-level constraints on SIDM by inferring the core collapse timescale of dark matter halos as a function of halo mass. We implement flexible models for subhalo tidal evolution and the density profiles of deeply collapsed objects, two sources of systematic uncertainty in the model of dark substructure, and present inferences on the core collapse timescale under different assumptions for subhalo abundance, tidal evolution, and the effects of tidal stripping on the collapse time of subhalos. Unlike most other lensing investigations, we do not aim to detect the masses, positions, and density profiles of individual perturbers. Instead, we measure the collective effect of the population, while accounting for the relative likelihood that a given measurement arises from an SIDM halo, a CDM halo, or other non-halo sources of small-scale perturbation, such as globular clusters or angular complexity in the lens model. A companion paper \citep{Gilman++26toappear} presents an interpretation of these results in the context of an SIDM particle physics model, with a long-range interaction that causes late-time differences in halo properties relative to CDM.
	
	This paper is organized as follows: Section \ref{sec:inference} describes our Bayesian inference methodology connecting data to substructure properties.  Section \ref{sec:structureformationmodel} details our structure formation model, including how we model core collapse in subhalos and field halos, and how we model the density profiles of SIDM halos. Section \ref{sec:structureinference} presents the results of our analysis, which include inferences on the core collapse timescale under different assumptions for subhalo abundance and tidal evolution in SIDM. Section~\ref{sec:conclusions} summarizes our main results and gives concluding remarks. Throughout this work we assume cosmological parameters for flat $\Lambda$CDM given by \citep{Planck18}. 
	
	\section{Bayesian inference methodology}
	\label{sec:inference}
	In this section, we review the Bayesian inference methodology used in this work. We give a high-level overview in the following paragraphs, and refer to \citet{Gilman++25a} for a more comprehensive review. The material unique to this paper, which analyzes the same dataset as \citet{Gilman++25a} but in the context of an SIDM structure formation model, begins in Section~\ref{sec:structureformationmodel}.  
	
	\subsection{Data}
	The JWST lensed quasar survey measured flux ratios from the compact warm dust region surrounding the quasar in 31 quadruple-image lens systems. These data are highly sensitive to low-mass perturbers, with the sensitivity determined by the angular size of the source relative to the deflection angle produced by a perturber \citep{Dobler++06}. 
	
	In this work, we use 27 lenses observed through the program. We exclude 4 of the 31 systems in the JWST survey on the basis of showing evidence for morphological complexity in the form of a stellar disk or a velocity dispersion that implies they are not early-type galaxies (two systems), or because they appear in MIRI imaging to have two or three images instead of four (two systems), which our analysis pipeline does not currently accommodate. The flux ratio measurements for the remaining 27 systems are presented by \citet{Nierenberg++24,Keeley++24,Keeley++25}. The 2 additional lenses, which were not included in the JWST survey, have flux ratios measured from the nuclear narrow-line region \citep{Nierenberg++14,Nierenberg++20}. We use archival HST imaging, NIRCam observations, and MIRI imaging for the extended lensed arcs, as summarized by \citep{Gilman++25a}. 
	
	The dataset we analyze in this work differs with respect to \citet{Gilman++25a} in three ways. First, we incorporate new NIRCam and HST imaging of extended lensed arcs for systems that previously lacked high-resolution space-based imaging. Appendix \ref{sec:appA} discusses the lens modeling and observations for these systems. Second, we include one additional lens, B2045+265, which was observed through the JWST survey but not included in the sample analyzed in previous papers in the series. This system was excluded from previous analyses because our inference pipeline could not resolve the flux from individual lensed images in the merging triplet, which leads to unreliable flux ratio predictions. For this work, we have made improvements to our forward modeling pipeline that allow us to model this system. These improvements also lead to a factor of 5--10 increase in the speed with which we can forward model other lenses, as discussed in Appendix \ref{sec:appB}. The modeling of B2045 is discussed in Appendix \ref{sec:appA}, and the flux ratios measurements are presented by \citet{Keeley++25}. Third, we have measured spectroscopic redshifts for two systems in our sample which previously lacked a spec-$z$: we place the system J0608+4229 at $z_{\rm d} = 1.0$ and J0803+3908 at $z_{\rm d} = 1.12$, respectively \citep{Sheu++26}. 
	
	\subsection{Analysis method}
	Our goal is to compute the posterior distribution 
	\begin{eqnarray}
		\label{eqn:posterior}
		\nonumber p\left( \qsub | \data \right) &\propto& \pi\left(\qsub \right) \mathcal{L}\left(\data | \qsub \right) \\
		&\propto&\pi\left(\qsub \right) \prod_{{\rm n}=1}^{N} \mathcal{L}\left(\datan | \qsub \right).
	\end{eqnarray}
	where $\qsub$ represents a set of hyper-parameters that specify statistical properties of a halo population, such as the amplitude and slope of the halo mass function, the free-streaming length, a core collapse timescale, etc., and $\pi\left(\qsub\right)$ is the prior probability. The data vector is $\data = \left(\boldsymbol{d}_{1}, \boldsymbol{d}_2, ..., \boldsymbol{d}_N\right)$, where $\datan$ refers to the dataset for the n-th strong lens system. As in our previous analyses of this sample, for each lens we measure the relative image positions, the flux ratios, and the lensed emission from the quasar host galaxy, or the lensed arcs, which we use to constrain the lens macromodel. 
	
	We compute the likelihood function, $\mathcal{L}\left(\datan | \qsub \right)$, by marginalizing over many possible \textit{realizations} $\rsub$ of each strong lens system, given the model specified by $\qsub$
	\begin{equation}
		\label{eqn:likelihoodfull}
		\nonumber  \mathcal{L}\left(\datan | \qsub \right) =  \int p\left(\datan | \rsub, \boldsymbol{n} \right)  p\left(\rsub | \qsub \right) p\left(\boldsymbol{n}\right)  d \boldsymbol{n} d \rsub.
	\end{equation}
	A single realization includes thousands of parameters that specify the masses, positions, and density profiles of individual perturbers. We include (sub)halos in the main lens plane and along the line of sight, as well as globular clusters. The vector of nuisance parameters, $\boldsymbol{n}$, specifies the lens mass model, the source light model, and the size of the emission region surrounding the background quasar. 
	
	We follow the methodology outlined in Section II of \citet{Gilman++25a} to compute the likelihood function. This approach involves reconstructing each lens system in the presence of different realizations $\rsub$. We simultaneously vary the dark matter hyper-parameters of interest, $\qsub$, alongside the nuisance parameters, ${\boldsymbol{n}}$, generating lens systems that include different configurations of small-scale structure and different configurations of the main deflector mass profile. The models for the light profiles and the main deflector mass profile are the same as those used in \citet{Gilman++25a}, and we refer to Section V A of \citet{Gilman++25a} for details. 
	
	As in \citet{Gilman++25a}, we jointly reconstruct the extended lensed arcs with the quasar flux ratios. The lensed arcs constrain the main deflector mass profile on angular scales comparable to the size of a pixel in the image plane, or $\sim 30 - 100$ m.a.s. depending on the instrument used to observe the lensed arcs. The flux ratios are sensitive to milli-arcsecond scale perturbations to the lens model from dark matter substructure and globular clusters. Both the imaging data and flux ratios constrain global deformation to the main deflector mass profile. Following the approach discussed in detail by \citet{Gilman++25a}, we include substructure in the lens model when reconstructing the imaging data, and marginalize over the halo population to derive importance sampling weights $w\left(\boldsymbol{n} | \datan \right)$ that depend only on a subset of the macromodel parameters. We then propagate these importance weights into the likelihood function for the dark matter inference, down-weighting lens models that fail to reproduce the morphology of the lensed arcs. These large-scale constraints on the macromodel lead to more-precise model-predicted flux ratios, which increases the constraining power on small-scale structure. We refer to Section II of \citet{Gilman++25a} for additional details on the lens modeling methodology. 
	
	We perform gravitational lensing calculations, including PSF modeling, multi-plane ray tracing, and joint lens and source reconstruction, using {\tt{lenstronomy}}\footnote{\url{https://github.com/lenstronomy/lenstronomy}} \citep{Birrer++18,Birrer++21}. We generate substructure realizations in SIDM using {\tt{pyHalo}}\footnote{\url{https://github.com/dangilman/pyHalo}} \citep{Gilman++20}. The open-source software {\tt{samana}}\footnote{\url{https://github.com/dangilman/samana}} carries out the forward modeling simulations using {\tt{pyHalo}} and {\tt{lenstronomy}}.
	
	\section{Models for halo substructure in SIDM}
	\label{sec:structureformationmodel}
	This section discusses the modeling of halo abundance and internal structure in SIDM. We account for properties of field halos along the line of sight and the subhalo population around the main deflector. Our modeling pipeline is implemented in the open source code {\tt{pyHalo}}
    \citep{Gilman++20}. We organize this section following the order of operations implemented in {\tt{pyHalo}} for generating SIDM halo populations. We discuss each key step in this process in Sections \ref{ssec:mfuncs}-\ref{ssec:collapsedsidm}: 
	
	\begin{itemize}
		\item Step 1 and Section \ref{ssec:mfuncs}: In the first step of our modeling pipeline we generate a population of dark matter subhalos and field halos with abundance and density profiles as predicted by CDM. Section \ref{ssec:cdmhalos} presents the halo mass function, spatial distribution, and concentration--mass relation used for this process. Section \ref{ssec:globularclusters} presents an improved modeling framework for globular clusters.
		\item Step 2 and Section \ref{ssec:timescale}: We assign each halo in the lens system generated in Step 1 a characteristic core collapse timescale, $t_c$, and identify halos that have core collapsed by $t / t_c > 1$, where $t$ represents the age of the halo. Section \ref{ssec:timescale} discusses how we define the age of a halo, with a separate treatment applied to field halos and subhalos to account for the effects of tidal stripping on the collapse time. 
		\item Step 3 and Section \ref{ssec:collapsedsidm}: We transform the density profiles of halos identified to have core collapsed in Step 2 into a form that matches the density profiles of deeply collapsed halos predicted by the simulations of core collapse presented by \citet{Gurian++25}. Section \ref{ssec:collapsedsidm} presents our model for the internal structure of core collapsed halos, and discusses two limiting cases that bracket the range of systematic uncertainty regarding the internal structure of deeply collapsed objects.
	\end{itemize}

	We conclude in Section~\ref{ssec:lensingsignal} by illustrating the lensing perturbations caused by core collapsed halos, which may aid in building physical intuition for how quadruply imaged quasars can test SIDM predictions. 
	
	\subsection{Mass functions, concentration-mass relations, and spatial distribution}
	\label{ssec:mfuncs}
	This section describes how we model dark matter substructure in CDM. We begin with the halo mass function, concentration-mass relation, and spatial distribution of dark matter subhalos in Section~\ref{ssec:cdmhalos}, and discuss our modeling of globular clusters in Section~\ref{ssec:globularclusters}.
	
	\subsubsection{Structure in cold dark matter}
	\label{ssec:cdmhalos}
	Our modeling pipeline begins by generating a population of CDM halos, which we will later transform into a population of SIDM halos based on a core collapse timescale assigned to each halo in the lens system. Throughout this work, we assume a mass definition for field halos $m=m_{200}$, or the mass inside a sphere with a mean density $200 \rho_{\rm{crit}}\left(z\right)$. We compute halo concentrations using the median concentration-mass-redshift relation of \citet{DiemerJoyce19} with 0.2 dex scatter. For field halos, we evaluate $\rho_{\rm{crit}}\left(z\right)$ and the concentration-mass relation at the halo redshift. For subhalos, we define their mass at the infall redshift, $z_{\rm{infall}}$, evaluate the concentration-mass relation at $z_{\rm{infall}}$, and define structural parameters with respect to $\rho_{\rm{crit}}\left(z_{\rm{infall}}\right)$. 
	
	The parametric form for the halo and subhalo mass functions are the same as those presented by \citet{Gilman++25a}. We model the field halo mass function as 
    \begin{equation}
		\label{eqn:losmfunc}
		\frac{d^2 N}{dm dV}\left(m,z\right) = \delta_{\rm{LOS}} \left(1+\xi\left(m_{\rm{host}},z\right)\right) \frac{d^2 N_{\rm{ST}}}{dm dV} \left(m,z\right),
	\end{equation}
	where $\frac{d^2 N_{\rm{ST}}}{dm dV}$ represents the mass function model presented by Sheth-Tormen \citep{Sheth++01}, and $\delta_{\rm{LOS}}$ allows for an overall scaling in the amplitude of the line-of-sight mass function. $\xi\left(m_{\rm{host}},z\right)$ is the two-halo term, which adds a $\sim 20\%$ boost to the number of halos near the main deflector as a result of clustering around the host \citep{Gilman++19,Lazar++21}. 
    
    We model the subhalo mass function as
	\begin{equation}
		\label{eqn:subhalomfunc}
		\frac{d^2 N}{dm dA} = \frac{\Sigma_{\rm{sub}}}{m_0}\left(\frac{m}{m_0}\right)^{\alpha} \mathcal{F}\left(m_{\rm{host}},z_{\rm{d}}\right).
	\end{equation}
	The last term accounts for the evolution of the projected number density with host halo mass and redshift \citep{Gilman++20,Gannon++25}, $\alpha\sim -1.9$ is the logarithmic slope of the subhalo mass function predicted by both CDM and SIDM \citep{Springel++08a,Nadler++25a,Nadler++25b}, and $m_0 = 10^8 M_{\odot}$. $\Sigma_{\rm{sub}}$ sets the normalization of the subhalo mass function at infall, or the total number of objects that fall into the host halo that eventually appear in projection within $30 \ \rm{kpc}$ of the host halo center at the lens redshift $ z_{\rm{d}}$. The CDM predictions for $\Sigma_{\rm{sub}}$ from $N$-body simulations and the semi-analytic model {\tt{galacticus}} correspond to $\Sigma_{\rm{sub}}=0.1 \ \rm{kpc^{-2}}$ and $\Sigma_{\rm{sub}}=0.16 \ \rm{kpc^{-2}}$, respectively \citep{Gannon++25,Du++25,Gilman++25a}.

	When generating subhalo populations, we first draw subhalo infall masses using Equation \ref{eqn:subhalomfunc}. We then sample the infall redshift of each subhalo from a distribution $p\left(z_{\rm{infall}}|m/m_{\rm{host}}\right)$ derived from the semi-analytic model {\tt{galacticus}} \citep{Benson2012}. Next, we use the tidal evolution model presented by \citet{Du++25} to assign each subhalo a bound mass fraction $f_{\rm{b}}=m_{\rm{bound}}/m$ drawn from a distribution $p\left(f_{\rm{b}}|c,z_{\rm{infall}},c_{\rm{host}}\right)$. The distribution of bound mass fractions depends on the infall concentration, the infall time, and the host concentration, with scatter around the mean driven by different orbital and mass loss histories. We refer to \citet{Du++25} and Section II of Paper IV for additional details regarding this tidal evolution model. 
	
	The susceptibility of cored halos to complete tidal disruption can alter subhalo abundance in SIDM relative to CDM, even for a CDM-like power spectrum and initial conditions \citep{Penarrubia++10,Errani++23,Du++24}. Our tidal evolution model predicts the bound mass function using relations calibrated for CDM, and does not explicitly account for the suppression of the subhalo mass function that results from cored halos becoming completely disrupted. We will instead examine how a suppression of the subhalo mass function impacts our inferences on SIDM properties by jointly constraining $\Sigma_{\rm{sub}}$, the normalization of the subhalo mass function, with other quantities, and examine how our inferences on SIDM parameters depend on subhalo abundance by assigning different importance sampling weights to $\Sigma_{\rm{sub}}$. 
	
	Each non-collapsed field halo and subhalo rendered in our simulations has a truncated NFW (hereafter TNFW) profile \citep{Baltz++09}
	\begin{equation}
		\label{eqn:nfwprofile}
		\rho_{\rm{tnfw}} = \frac{f_t \rho_\mathrm{s}}{x\left(1+x\right)^2} \frac{\tau^2}{\tau^2+x^2},
	\end{equation}
	with $x\equiv r/r_\mathrm{s}$ and $\tau\equiv r_\mathrm{t}/r_\mathrm{s}$. For field halos, we set $r_t = r_{\rm{200}}$ and $f_t = 1$. For subhalos, we compute $r_t$ and $f_t$ from the bound mass fraction based on the tidal tracks \citep{Errani++21,Du++24}. We note that cored SIDM halos, with core sizes typically $r_c / r_s \lesssim 0.5$, produce lensing perturbations that are nearly indistinguishable on the population level from NFW profiles \citep{Gilman++21,Gilman++23}. Appendix \ref{sec:appC} demonstrates this by computing flux ratio statistics with populations of cored halos.  

    The geometry of the lensing volume is that of a double-cone that opens towards the lens and closes at the source redshift. We render subhalos in the main lens plane within a disk of radius $3 R_{\rm{E}}$, where $R_{\rm{E}} \sim 1 \ \rm{arcsec}$ is the Einstein radius. Simulations of dark matter substructure predict that the projected spatial distribution of subhalos is approximately uniform on these scales \citep{Xu++15,Fiacconi++16,Gannon++25}, so we distribute subhalos uniformly within this disk. CDM also predicts that field halos cluster. However, the effect leads to correlations on physical scales significantly larger than the transverse size of each lens plane. Clustering effects also become diluted from projecting each line of sight volume element, with a redshift spacing $\Delta z = 0.02$, corresponding to a comoving distance $\sim 50  \ \rm{Mpc}$, into two dimensions. These effects are therefore negligible on the scales relevant to our analysis, and we render field halos uniformly in each lens plane.  

    We generate halo masses in the range $10^6 - 10^{10.7} M_{\odot}$. The lower bound corresponds to the minimum halo mass sensitivity of our data, which is set by the angular size of the background source \citep{Dobler++06,Nierenberg++24}. We assume that halos more massive than $10^{10.7} M_{\odot}$ would host a galaxy luminous enough that we would detect it in space-based imaging. When we detect a luminous satellite galaxy in the imaging data we include it explicitly in the lens model \citep{Gilman++25a}. Therefore, the range $10^6 - 10^{10.7} M_{\odot}$ covers the halo mass scales that can affect our data without double-counting the more massive objects. We note that many subhalos with infall masses in the range $10^6 - 10^{10.7} M_{\odot}$ appear in our lens models with bound masses below $10^6 M_{\odot}$. Despite the bound mass falling below $10^6 M_{\odot}$, flux ratios mainly probe the dense central regions of the halo, meaning a subhalo stripped to below $10^6 M_{\odot}$ can sometimes impart a significant perturbation \citep{Du++25}.  
    
	Some classes of SIDM theories predict a suppression of the linear matter power spectrum or acoustic oscillations below some scale \citep{Vogelsberger++16,Nadler++25b}. This occurs, for example, in models where dark matter is coupled to a new light relativistic species. Whether this occurs alongside a given SIDM cross section depends on the details of the particle physics model. For this work, we assume the SIDM models in question leave the matter power spectrum unchanged on the scales probed by our data: halo masses $10^6 - 10^{10.7} M_{\odot}$ and $P\left(k\right)$ down to $k \sim 200 \ \rm{Mpc^{-1}}$. 
	
	\subsubsection{Globular clusters}
	\label{ssec:globularclusters}
	Globular clusters (hereafter GCs) can have masses of $10^5 - 10^6 M_{\odot}$ with high central densities relative to CDM halos in this mass range. We base our model for GCs around strong lenses on the observed population of GCs around nearby massive elliptical galaxies. We model the mass function as a log-normal distribution 
	\begin{equation}
		\label{eqn:gcmf}
		\frac{d^2N}{d \ln m_{\rm{gc}} \, dA} = \frac{\Sigma_{\rm{gc}}}{\langle m_{\rm{gc}} \rangle \, \sigma \sqrt{2\pi} \, \ln 10} \exp\left[-\frac{\left(\log_{10} m_{\rm{gc}} / \mu\right)^2}{2\sigma^2}\right]
	\end{equation}
	where $\mu = 2 \times 10^5 M_{\odot}$ is the median GC mass, we assume a standard deviation of $\sigma = 0.6$ dex \citep{Jordan++07}, and $\langle m_{\rm{gc}} \rangle \approx 2.6 \mu$ is the corresponding mean GC mass. With these definitions $\Sigma_{\rm{gc}}$ is then the projected mass density in GCs integrated over the mass function. 
    
    We use scaling relations measured for GC systems around massive ellipticals in the Virgo cluster to determine the expected abundance of GCs around strong lenses. In particular, the total mass of the GC population, $m_{\rm{gc,tot}}$, scales linearly with host halo mass as $m_{\rm{gc,tot}} = \eta \, m_{\rm{host}}$, with $\eta \sim 3 \times 10^{-5}$ \citep{Hudson++14,Harris++17,Forbes++18}. The radial profile of the GC systems follows a S{\'e}rsic profile with S{\'e}rsic index $n=2$, and the effective radius scales with the host halo virial radius, $r_{\rm{eff,gc}} = f \, r_{200}$ with $f \sim 0.06$ \citep{Forbes17,Lim++24}. Applying these relations to a typical lens host halo mass of $2\times10^{13} M_{\odot}$ \citep{Lagattuta++10} at $z=0.5$ gives $\Sigma_{\rm{gc}} \sim 10^{5.4} M_{\odot} \ \rm{kpc^{-2}}$, corresponding to a projected number density of $\sim 0.5 \ \rm{kpc^{-2}}$. Given the scatter in these relations and the range of deflector masses in our sample, we expect a factor of $\sim 2$ uncertainty in this estimate, and we make the conservative choice of taking $\Sigma_{\rm{gc}} \sim 10^{5.6} M_{\odot} \ \rm{kpc^{-2}}$ as the expected GC abundance. Our estimate is consistent with the GC abundance computed by \citet{He++18} for Virgo ellipticals. 
    
    We model the density profiles of GCs using the empirical King \citep{King62} profile
	\begin{equation}
		\Sigma_{\rm{king}}(r) =
		\begin{cases}
			\Sigma_{0} \left( \dfrac{r_c}{\sqrt{r_c^2 + r^2}} - \dfrac{r_c}{\sqrt{r_c^2 + r_t^2}} \right)^2 & r < r_t \\[8pt]
			0 & r \geq r_t
		\end{cases}.
	\end{equation}
    The total mass of each GC is
    \begin{equation}
    \label{eqn:gcmass}
	   m_{\rm{gc}} = 2 \pi \Sigma_{0} r_c^2 F(r_t / r_c)
    \end{equation}
    where
    \begin{equation}
    	F(x) = \frac{1}{2} \ln\left(1 + x^2\right)
    	       - 2 a \left(\sqrt{1 + x^2} - 1\right)
    	       + \frac{1}{2} a^2 x^2
    \end{equation}
    and $a = \left(1 + \left(r_t/r_c\right)^2\right)^{-1/2}$. 

    We parameterize the profile in terms of quantities that are robustly measured for nearby GCs. These include the GC concentration $c_{\rm{gc}} = \log_{10}\left(r_t/r_c\right)$ and the projected half-mass radius $r_{1/2}$, which satisfies
    \begin{equation}
        F\left(\frac{r_{1/2}}{r_c}\right) = \frac{1}{2} F\left(\frac{r_t}{r_c}\right).
    \end{equation}
    We solve this equation numerically to obtain $r_c$ from $r_{1/2}$ and $c_{\rm{gc}}$.  

    \citet{Gilman++25a} modeled GCs as point masses with a mass function given by Equation~\ref{eqn:gcmf} and a fixed normalization $\Sigma_{\rm{gc}} =  10^{5.6} \ M_{\odot} \rm{kpc^{-2}}$. We note that the point-mass assumption is likely a very good approximation to the true lensing signal from GCs since their Einstein radii exceed their half-mass radii for $m_{\rm{gc}} \gtrsim 10^5 M_{\odot}$. In this work, because we consider a dark matter model with a lensing signal that we expect will resemble the kinds of perturbations caused by GCs, we have implemented a more careful treatment of the GC population. We expect the modeling in this work will better capture the lensing signal from the less massive GCs, whose half-mass radii are comparable to their Einstein radii. 
    \begin{figure}
		\centering
		\includegraphics[trim=0cm 0.5cm 0cm
		0.cm,width=0.48\textwidth]{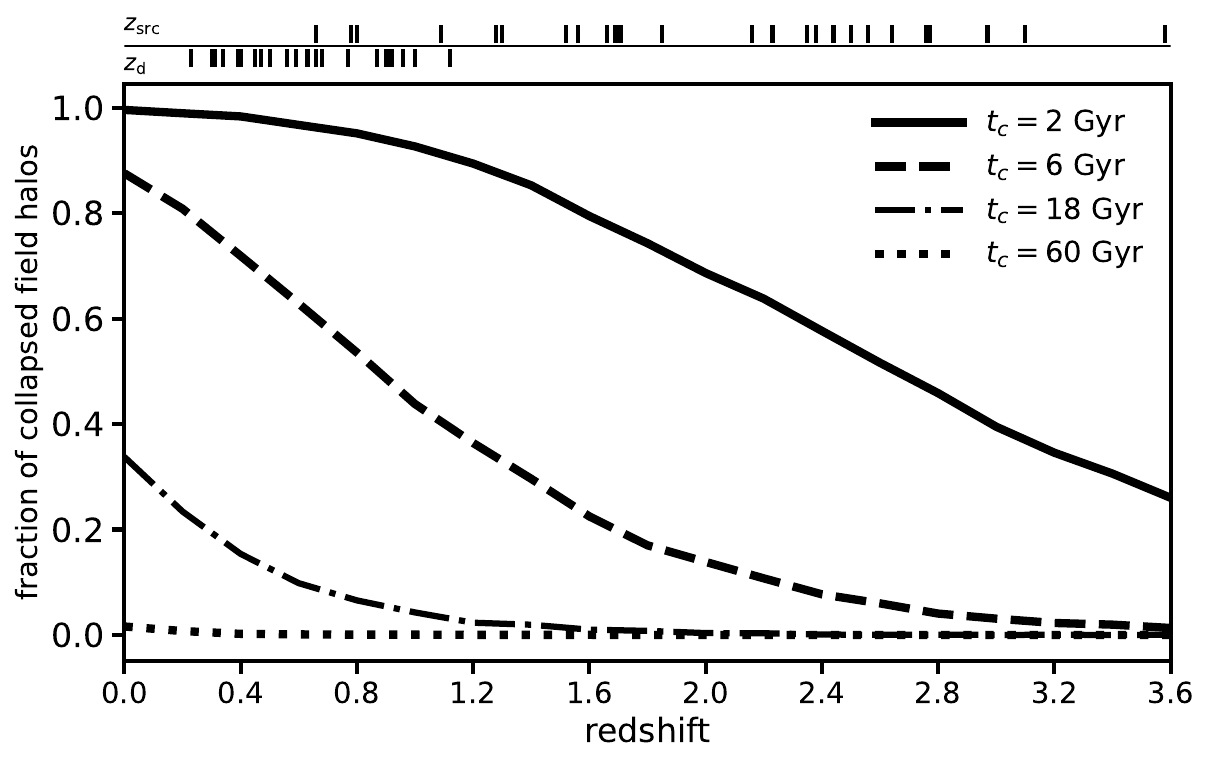}
		\caption{\label{fig:fieldhalocollapse} The fraction of core collapsed field halos as a function of redshift for the core collapse timescales $t_{6/8} = t_{8/10} = t_c$, and $t_c$ values spanning 2 - 60 Gyr (see Section \ref{ssec:timescale} for a discussion of the collapse timescales). The lens and source redshifts for the 29 systems in our sample are marked in the upper x-axis.}
	\end{figure}
    \begin{figure}
	\centering
	\includegraphics[trim=0cm 1cm 0cm
	0.cm,width=0.48\textwidth]{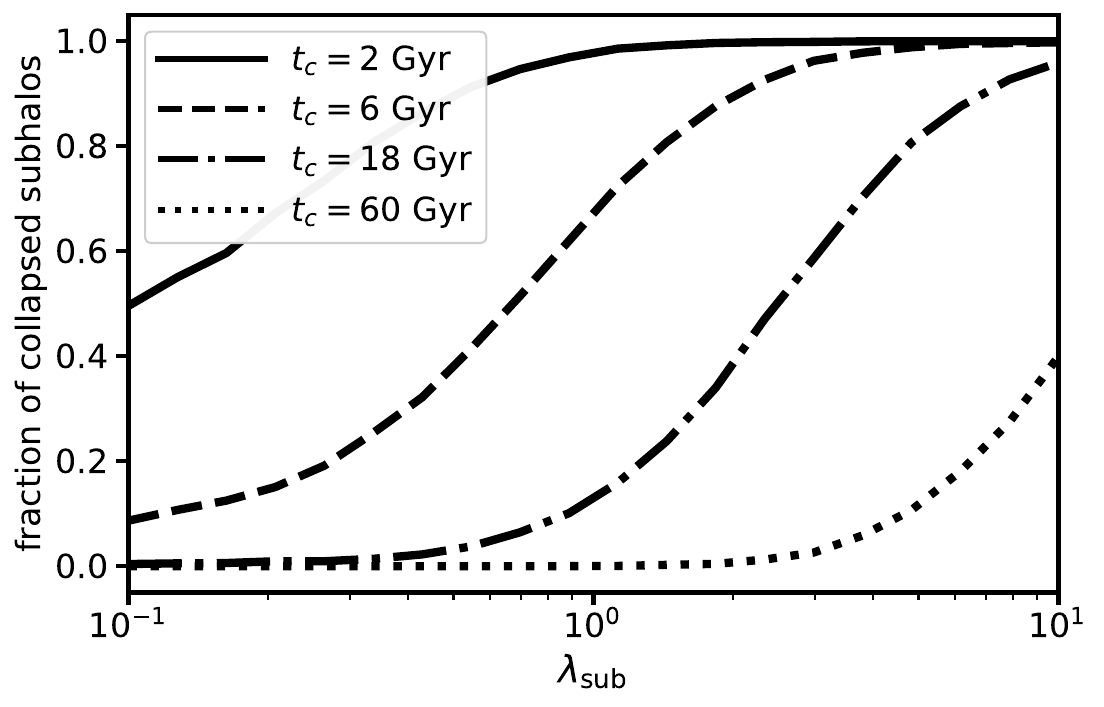}
	\caption{\label{fig:subhalocollapse} The fraction of core collapsed subhalos at $z=0.5$ for different core collapse timescales, as a function of $\lambda_{\rm{sub}}$, the linear rescaling of the subhalo collapse timescale (Equation \ref{eqn:timesub}). $\lambda_{\rm{sub}}>1$ corresponds to faster core collapse as a result of tidal stripping \citep{Nishikawa++20}, while $\lambda_{\rm{sub}} < 1$ could result from delayed collapse from environmental effects such as evaporation \citep[e.g.,][]{Zeng++22}.}
\end{figure}
    
    We draw the half-mass radius $\log_{10} \left(r_{1/2} / \rm{pc}\right) \sim \mathcal{N}\left(0.5,0.2\right)$, and the GC concentration from a Gaussian distribution $c_{\rm{gc}}\sim\mathcal{N}\left(1.5, 0.1\right)$ \citep{Harris96,McLaughlin++05}. We distribute GCs with a uniform spatial distribution inside a circular aperture with radius of $0.2$ arcsecond around each lensed quasar image. We draw $\log_{10} \Sigma_{\rm{gc}}$ from a sampling distribution $\mathcal{U}\left(5.3, 7.0\right)$, varying this parameter simultaneously with the dark matter parameters of interest. The sampling distribution on $\Sigma_{\rm{gc}}$ spans a factor of 50 in GC abundance, and allows for 25 times more GCs than expected for a typical lens host halo mass. When quoting our main results in Section \ref{ssec:results}, we use an informative prior on $\Sigma_{\rm{gc}}$ based on the expected GC population around lens host halos with a factor of 2 uncertainty $\log_{10} \Sigma_{\rm{gc}} \sim \mathcal{N}\left(5.6, 0.3\right)$. By extending the sampling distribution to significantly higher $\Sigma_{\rm{gc}}$, we can quantify the GC abundance required to explain our data without invoking SIDM. 
	 
	\subsection{The core collapse timescale}
	\label{ssec:timescale}
	In SIDM, scattering between dark matter particles drives a dynamic evolution of dark matter halo profiles. The central density evolves with a characteristic collapse timescale, $t_c$, that one can connect directly to the SIDM cross section. For example, in the long mean free path limit, $t_c \propto \left(\rho_s v_s \sigma\right)^{-1}$, where $v_s \sim \sqrt{G \rho_s r_s^2}$ and $\sigma$ typically represents a thermally averaged cross section \citep{YangYu22,Yang++23}. However, the precise relation between $\sigma$ and $t_c$ can change based on environment, such as the accelerated collapse of subhalos due to tidal stripping \citep{Nishikawa++20,Sameie++20,Turner++21,ShahAdhikari24}, the delay of core collapse from evaporation \citep{Zeng++22,Slone++23}, major mergers \citep{Silverman++26}, or the presence of baryons inside more massive halos \citep{Zhong++23}. Dissipative or inelastic scattering processes can also decrease the collapse timescale \citep{Essig++19,ONeil++23,Roy++25,Shen++24}. Moreover, the $t_c \propto \sigma^{-1}$ scaling only holds in the long mean free path limit, transitioning to $t_c \propto \sigma$ in the short mean free path limit \citep{Balberg++02,KodaShapiro11}. Given the dependence of $t_c$ on environment and on the particle physics model, in this analysis we work directly in terms of $t_c$ as the quantity inferred from the data, rather than the cross section itself. We present a particle physics interpretation of our results in a companion paper, where we assume an SIDM cross section from a long-range interaction and connect the mediator mass, dark matter particle mass, and interaction strength to the collapse timescale. 
	
	The halo mass sets a characteristic velocity scale for self-interactions. Given that many SIDM theories predict a velocity-dependent cross section \citep[e.g.][]{Tulin++13}, we introduce two timescales, $t_{6/8}$ and $t_{8/10}$, that set the median value of $t_c$ for halos (subhalos) with masses (infall masses) in the range $10^6 M_{\odot}-10^8 M_{\odot}$ and $10^8 M_{\odot}-10^{10.7} M_{\odot}$, respectively. For a given timescale $t_{6/8}$ or $t_{\rm{8/10}}$, we assign each halo a $t_c$ drawn from a log-normal distribution centered at $t_{6/8}$ or $t_{8/10}$ with a scatter of 0.3 dex. The 0.3 dex scatter is tied to the scatter in the concentration-mass relation through the Knudsen number \citep{Balberg++02,KodaShapiro11}, which depends on the combination $\rho_s r_s$. The Knudsen number delineates the transition between the long and short mean free path regimes, where scatter in $\rho_s r_s$ propagates differently onto $t_c$. As we interpret this inference in terms of particle physics models that span the long and short mean free path regimes \citep{Gilman++26toappear}, we do not explicitly up or down-scatter $t_c$ based on the halo concentration, and instead treat the scatter as a random draw per halo\footnote{We note that introducing scatter in $t_c$ avoids unphysical behavior, such as an abrupt jump from 0$\%$ core collapse to $100 \%$ core collapse at the redshift where the typical halo age exceeds $t_{6/8}$ or $t_{8/10}$.}. Appendix \ref{sec:appC} discusses the effect of including directional scatter in the core collapse timescale based on halo concentration. 

	Halos enter the deeply collapsed regime when $t / t_c > 1$, where $t$ represents the age of a halo, or the time elapsed since the dense center of a halo forms in the early Universe and self-interactions begin to thermalize the profile. For halos in the field at redshift $z$, we evaluate their age according to
	\begin{equation}
		t_{\rm{field}} \equiv t\left(z_{\rm{form}}\right) - t\left(z\right),
	\end{equation}
	  where $t\left(z\right)$ is the lookback time to a redshift $z$. We assign all halos a formation redshift $z_{\rm{form}} =10$, approximately corresponding to the time when halos more massive than $10^6 M_{\odot}$ have built their dense central cusps \citep{Bullock++01,Wechsler++02}. 

     Given $t_{6/8}$ and $t_{8/10}$, our model predicts the fraction of collapsed field halos as a function of redshift. Figure~\ref{fig:fieldhalocollapse} shows these fractions assuming $t_{6/8} = t_{8/10} = t_c$ for $t_c$ values of 2, 6, 18, and 60 Gyr. The abundance of collapsed field halos of a given mass increases at lower redshift, reflecting the fact that these halos have had more time to evolve since formation, on average. The colored ticks above the figure mark the deflector and source redshifts for the 29 lenses in our sample, indicating the redshift range probed by our data. We note that the sensitivity to field halos out to the source redshift, typically $z\sim 2-3$, gives strong lensing the distinct capability to probe the evolution of halo properties over cosmic time. 
	
	For subhalos, various environmental factors can affect the progression towards core collapse, relative to objects in the field. For example, tidal stripping can remove material from the halo, accelerating core collapse \citep{Sameie++20,Nishikawa++20}, while interactions between subhalo particles and host halo particles can heat subhalos and delay collapse \citep{Zeng++22,Slone++23}. Baryons can accelerate the onset of core collapse, although on the halo mass scales probed by our data this effect alters the collapse time by only $10-20\%$ \citep{Zhong++23}. We allow for either accelerated or decelerated collapse in subhalos relative to field halos by introducing a free parameter, $\lambda_{\rm{sub}}$, which rescales the collapse timescale for subhalos. The parameter $\lambda_{\rm{sub}}$ determines an ``effective'' age for subhalos
	\begin{equation}
		\label{eqn:timesub}
		t_{\rm{sub}} = t\left(z_{\rm{form}}\right) - t\left(z_{\rm{infall}}\right) + \lambda_{\rm{sub}}\left[t\left(z_{\rm{infall}}\right) - t\left(z_{\rm{d}}\right)\right],
	\end{equation} 
	where  $z_{\rm{d}}$ is the main deflector redshift, and $z_{\rm{infall}}$ is the unique infall redshift of each subhalo.  
	\begin{figure}
		\centering
		\includegraphics[trim=0cm 0.5cm 0cm
		0.cm,width=0.48\textwidth]{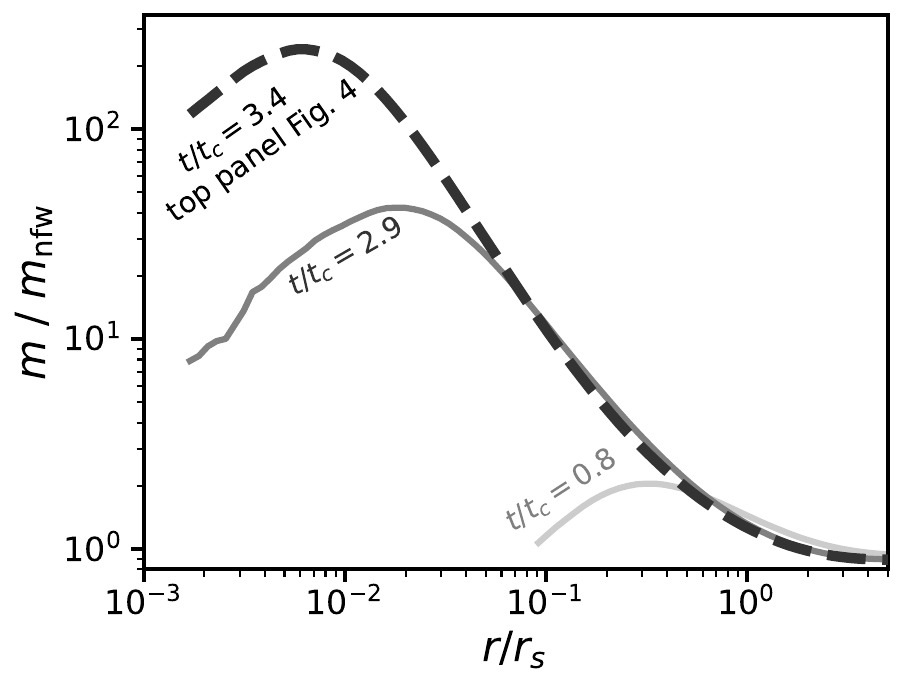}
		\caption{\label{fig:earlymediumlate} The enclosed mass of a collapsing halo as predicted by the simulations presented by \citet{Gurian++25}, which track halos into the deeply core collapsed regime. Our core collapse model is based on the most deeply collapsed profile, shown as a dashed black curve (see also Figure~\ref{fig:sidmdensity}). For this figure we define $t_c$ with the expression in the long mean free path regime $t_c = 200 \left(\sigma \rho_s \sqrt{4 \pi G \rho_s r_s^2} \right)^{-1} $\citep{Balberg++02} with $\sigma = 50 \ \rm{cm^2} \ \rm{g^{-1}}$.}
	\end{figure}
    
	For halos and subhalos at the same redshift and for a fixed $t_c$, $\lambda_{\rm{sub}}=1$ gives $t_{\rm{sub}}=t_{\rm{field}}$, and thus gives the same fraction of collapsed halos (or the fraction of halos with $t / t_c > 1$). In contrast, $\lambda_{\rm{sub}} > 1$ corresponds to accelerated subhalo core collapse, while $\lambda_{\rm{sub}}<1$ corresponds to delayed subhalo collapse. This empirical treatment allows significant flexibility in our implementation of subhalo core collapse, allowing us to explore the associated systematic uncertainty. In Section~\ref{sec:structureinference}, we quantify how our inferences on $t_{6/8}$ and $t_{8/10}$ depend on assumptions related to subhalo core collapse by assigning different importance sampling weights to this parameter. 
	
	Figure~\ref{fig:subhalocollapse} shows the fraction of collapsed subhalos for the same $t_c$ values considered in Figure~\ref{fig:fieldhalocollapse}, for different values of $\lambda_{\rm{sub}}$. For $\lambda_{\rm{sub}} > 1$, corresponding to accelerated core collapse due to tidal stripping, our model predicts that nearly $100 \%$ of subhalos should core collapse for $t_c < 5 \ \rm{Gyr}$, and roughly $50 \%$ of field halos collapse between $z=0$ and $z=2$. On the other hand, if we set $t_c = 18 \ \rm{Gyr}$ with $\lambda_{\rm{sub}}=5$, we obtain substructure realizations in which only $\sim 20 \%$ of field halos core collapse at $z < 0.4$, but more than $80\%$ of subhalos collapse at $z=0.5$. These examples illustrate the diversity of outcomes we can obtain by simultaneously sampling $t_{6/8}$, $t_{8/10}$, and $\lambda_{\rm{sub}}$.  
    
    From both Figures~\ref{fig:fieldhalocollapse} and~\ref{fig:subhalocollapse} we see that only a small number of halos collapses when $t_c$ becomes much longer than the age of the Universe. We can therefore associate models with $t_{6/8}> 80 \ \rm{Gyr}$ and $t_{8/10} > 80 \ \rm{Gyr}$ as indistinguishable from CDM, given our data. 
	
	Once the collapse phase begins, the central density of the halo increases exponentially (or super-exponentially) with time. Prior to collapse, the lensing cross section of core forming SIDM halos is very similar to that of an NFW profile \citep{Gilman++23}. These factors  motivate our modeling of SIDM halo evolution as a binary state of pre-collapse and post-collapse, with collapse occurring when $t_{\rm{field}} / t_c = 1$ for field halos, and at $t_{\rm{sub}} / t_c = 1$ for subhalos. Once we have selected the subset of halos and subhalos that have core collapsed based on these criteria, we transform their density profiles into a core collapsed representation.

	\subsection{Density profiles of collapsed SIDM halos}
	\label{ssec:collapsedsidm}
	We base our model for collapsed halo profiles on the simulations of deeply core collapsed objects presented by \citet{Gurian++25}. Figure \ref{fig:earlymediumlate} shows the enclosed mass of the collapsed profile in their simulations, relative to the enclosed mass of an NFW profile, as a function of radius. The three curves span the early to late collapse regimes, with the density profile beginning to collapse around $t / t_c \sim 0.8$ and entering the deeply collapsed stage by $t / t_c \sim 2.9$. We note that most numerical simulations of SIDM core collapse stop around $t / t_c \sim 1$ due to the numerical challenges involved with evolving the halo further into the collapse phase \citep[e.g.][]{Mace++24,Fischer++25}. In addition to the work by \citet{Gurian++25}, other simulations have begun addressing this challenge \citep[e.g.][]{Fischer++26,Zier++26}, although most of these new approaches have not yet evolved subhalos deep into the collapse regime while also accounting for processes such as tidal stripping. Parametric models for SIDM subhalo evolution calibrated to $t/t_c \sim 1$ \citep{YangD++24,Yang++25} evolve halos along tidal mass loss histories computed for CDM halos, which does not self-consistently capture the correlated effects of tidal stripping, host--subhalo particle interactions, and core collapse. We note that most subhalos around a strong lens have lost a significant fraction of their mass since infall \citep{Du++25}, so future work that develops a more detailed understanding of how core collapse proceeds in heavily stripped subhalos with bound mass fractions $f_{\rm{b}}\sim 0.05$ would significantly improve our ability to forward model SIDM halo populations. 
    \begin{figure}
		\centering
		\includegraphics[trim=0cm 0.cm 0cm
		0cm,width=0.48\textwidth]{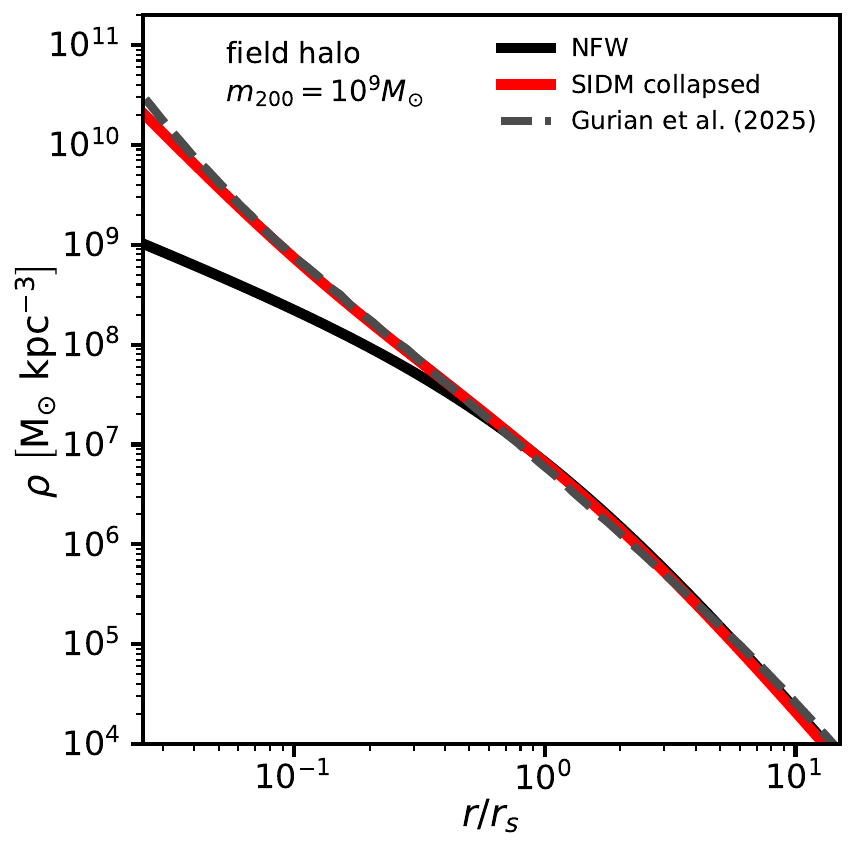}
		\includegraphics[trim=0cm 0.cm 0cm
		0cm,width=0.48\textwidth]{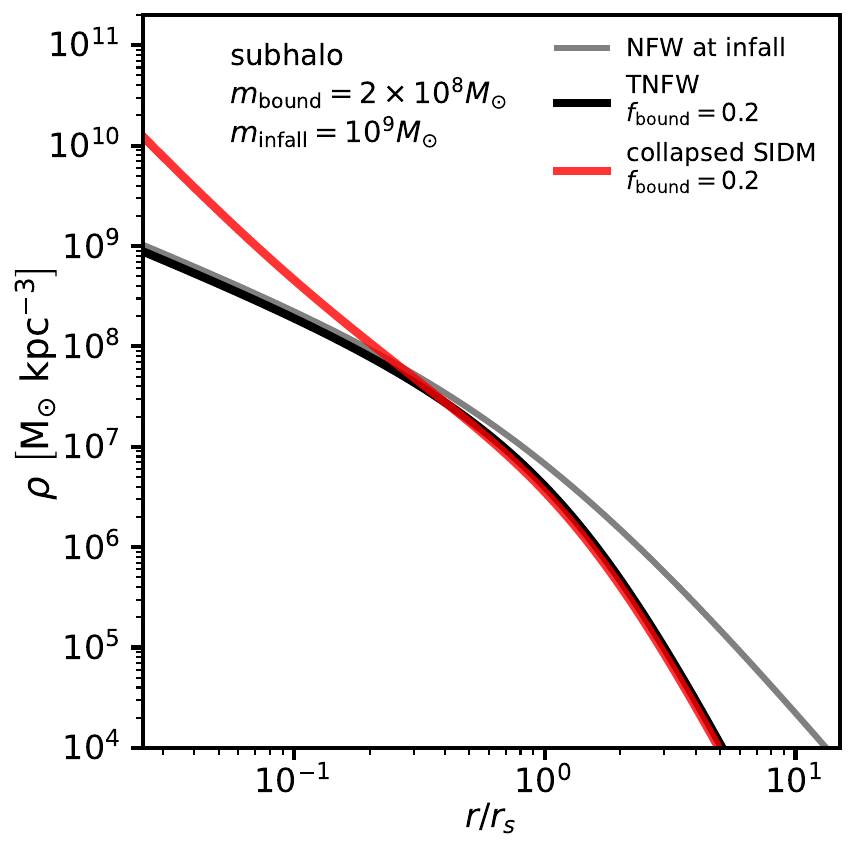}
		\caption{\label{fig:sidmdensity} The density profiles of a core collapsed field halo (top) and subhalo (bottom). The black line in each panel shows the reference TNFW profile, and the gray line in the bottom panel shows the profile at infall, before tidal stripping. The dashed black curve in the upper panel shows the halo simulated by \citet{Gurian++25}, and the red curves show the power law model for the density profile (Equation \ref{eqn:rhosidm1}).}
	\end{figure}
    
    Figure \ref{fig:sidmdensity} shows the density profile snapshot at $t / t_c = 3.4$ as a dashed black curve. The red curve in Figure \ref{fig:sidmdensity} shows a parametric fit to the collapsed profile
	\begin{equation}
		\label{eqn:rhosidm}
		\rho_{\rm{SIDM}} = \rho_{\rm{central}} + A_{\rm{m}} \rho_{\rm{tnfw}},
	\end{equation}
	where $\rho_{\rm{tnfw}}$ is a truncated NFW profile, $A_{\rm{m}}$ is an overall rescaling factor introduced to conserve mass, and $\rho_{\rm{center}}$ represents the collapsing central part of the halo. We model the central region with
	\begin{equation}
		\label{eqn:rhosidm1}
		\rho_{\rm{central}} = \frac{\rho_0}{\left(r/s\right)^{p} \left(1+r^2/s^2\right)^{\left(6-p\right)/2}}.
	\end{equation}
	This profile has a central cusp with logarithmic slope $p$ that transitions to $r^{-6}$ beyond a transition radius $s$. The $r^{-6}$ scaling outside $s$ ensures the profile transitions to a TNFW profile at larger radii. 
    
	To construct collapsed representations of halos and subhalos we parameterize
	the collapsed profile using only the logarithmic slope of the reference CDM halo and its enclosed mass, so that our model contains no fixed physical scale. We define the transition radius $s$ in Equation~\ref{eqn:rhosidm1} as the radius at which the TNFW profile that the halo would have had in CDM reaches a logarithmic slope $\gamma_{s}$
	\begin{equation}
		\label{eqn:slopematch}
		\gamma_{s} \equiv \frac{\partial \log \rho_{\rm{tnfw}}}{\partial \log r} \Big |_{r=s}.
	\end{equation}
	Next, we partition the total mass of a halo into the collapsed region, and into the outer envelope. We conserve mass interior to $s$ through 
	\begin{equation}
		\label{eqn:massconserve1}
		M_{\rm{central}}\left(s\right) + A_{\rm{m}} M_{\rm{tnfw}}\left(s\right)=B_{\rm{m}} M_{\rm{tnfw}}\left(s\right).
	\end{equation}
	and conserve mass within the virial radius $r_{200}$
	\begin{equation}
		\label{eqn:massconserve2}
		M_{\rm{central}}\left(r_{\rm{200}}\right) + A_{\rm{m}} M_{\rm{tnfw}}\left(r_{200}\right)= M_{\rm{tnfw}}\left(r_{200}\right)
	\end{equation}
	where $A_{\rm{m}}$ is the same as in Equation \ref{eqn:rhosidm} and $B_{\rm{m}}$ is an additional free parameter. We note that for subhalos, conserving mass within $r_{200}$ is the same as conserving the bound mass of the subhalo, i.e. $M_{\rm{tnfw}}\left(r_{200}\right)=f_{\rm{b}} m $.
	
	The parameters $p$, $s$, and $B_{\rm{m}}$ have a clear physical interpretation as the logarithmic profile slope of the deeply collapsed region, the radius where the thermalized region of the SIDM halo transitions to the CDM envelope, and the mass enclosed inside the collapsed central region, relative to the mass of the reference CDM halo, respectively. We determine values for $p$, $\gamma_{s}$, and $B_{\rm{m}}$ by matching the inner structure of the halo simulated by \citet{Gurian++25}, shown as the dashed black curve in Figure \ref{fig:sidmdensity}. We reproduce their density profile with $p=2.6$, $\gamma_{s} = -1.9$ and $B_{\rm{m}}=1.6$, which fixes 3 of the 5 free parameters. We solve for the remaining two, $\rho_0$ and $A_{\rm{m}}$, using Equations \ref{eqn:massconserve1} and \ref{eqn:massconserve2}. The resulting parametric approximation to the halo simulated by \citet{Gurian++25} is shown as a red curve in Figure \ref{fig:sidmdensity}. 

	We follow the procedure described in the previous paragraph to construct core collapsed density profiles for field halos, solving Equations \ref{eqn:slopematch}, \ref{eqn:massconserve1}, and \ref{eqn:massconserve2} with parameters $p=2.6$, $\gamma_{s} = -1.9$ and $B_{\rm{m}}=1.6$. Because our model does not have a fixed physical scale, the resulting profiles are approximately self-similar when lengths and densities are scaled by $r_s$ and $\rho_s$, respectively. This self-similarity is partially motivated by the gravothermal fluid formalism, which predicts a self-similar halo evolution in the long mean free path regime when length scales and densities are expressed in terms of the properties of an NFW profile \citep{Balberg++02,KodaShapiro11,Yang++23}. Like the fluid model, the simulations by \citet{Gurian++25} predict an approximately self-similar halo shape, which motivates our calibration of halo density profiles at different masses and concentrations using the shape of the collapsed halo shown in Figure \ref{fig:sidmdensity}. We note, however, that \citet{Gurian++25} consider a velocity-independent cross section, and find that the density profile may deviate from self-similarity as the halo transitions from the long to short mean free path regimes.
    
    We also use Equations~\ref{eqn:slopematch}-\ref{eqn:massconserve2} to construct physically plausible representations of deeply core collapsed subhalos. We use the same set of parameters $B_{\rm{m}}=1.6$, $\gamma_s = -1.9$ and $p=2.6$, such that the collapsed central region of the subhalo retains approximately the same shape as a field halo. We define the transition radius $s$ relative to the logarithmic slope of the tidally-stripped CDM subhalo, and conserve (bound) mass using Equations~\ref{eqn:massconserve1} and \ref{eqn:massconserve2}. An example density profile for a collapsed subhalo is shown in the bottom panel of Figure \ref{fig:sidmdensity}. We emphasize that a tidal evolution model for SIDM subhalos that accounts for the correlated effects of mass loss, core expansion, and eventual collapse would improve the fidelity of our structure formation model. Lacking this capability, we explore how assumptions related to subhalo abundance and internal structure affect our results through different importance sampling weights on $\Sigma_{\rm{sub}}$, and by introducing additional flexibility in the collapsed halo density profile, as discussed below. 
	 
	\begin{figure}
		\centering
		\includegraphics[trim=0cm 0.5cm 0cm
		0.cm,width=0.48\textwidth]{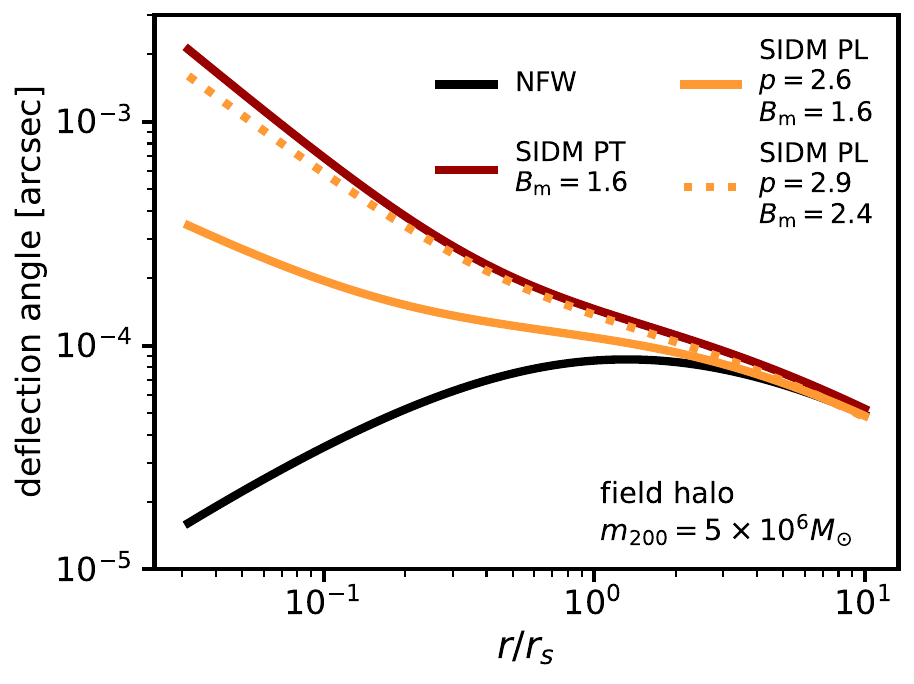}
		\caption{\label{fig:alphar} The deflection angle as a function of radius from a $5 \times 10^6 M_{\odot}$ field halo at $z=0.5$. The black curve shows the deflection angle of an NFW profile, and the orange and red curves show collapsed profiles corresponding to the power-law (PL) and point-mass (PT) models (Equations~\ref{eqn:rhosidm1} and \ref{eqn:rhosidm2}), respectively. The dashed orange curve shows a power-law profile that closely matches the deflection angle of the point mass model, which corresponds to $\xi_{\rm{core}}=0.9$ (see Equation \ref{eqn:yeffect}). Our model allows for a range of perturbation strengths from collapsed halos bounded from above by the red curve and from below by the solid orange curve.}
	\end{figure}
	\begin{figure}
		\centering
		\includegraphics[trim=0cm 0.cm 0cm
		0.cm,width=0.48\textwidth]{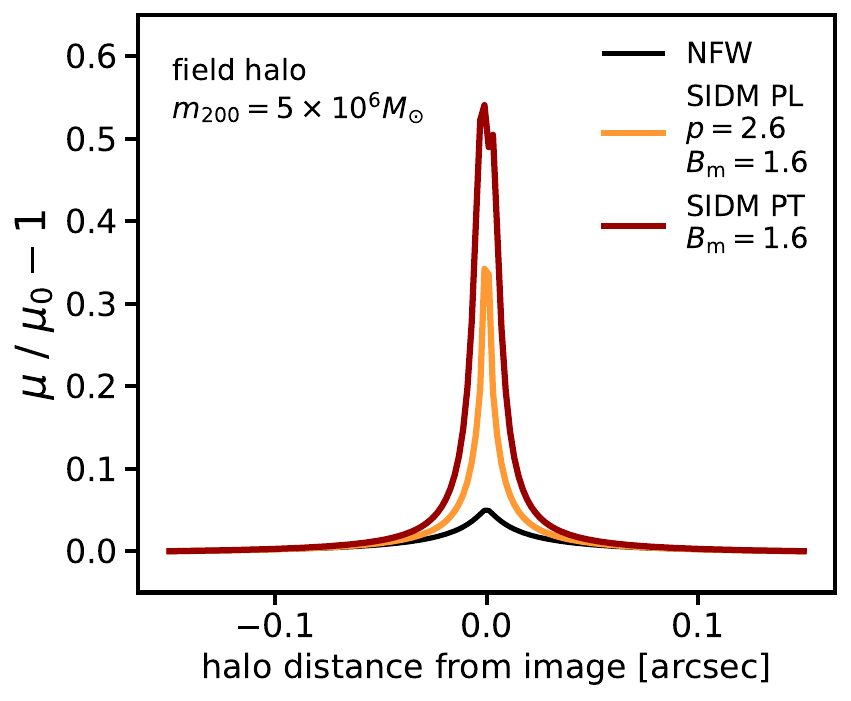}
		\includegraphics[trim=0cm 0.cm 0cm
		0.cm,width=0.48\textwidth]{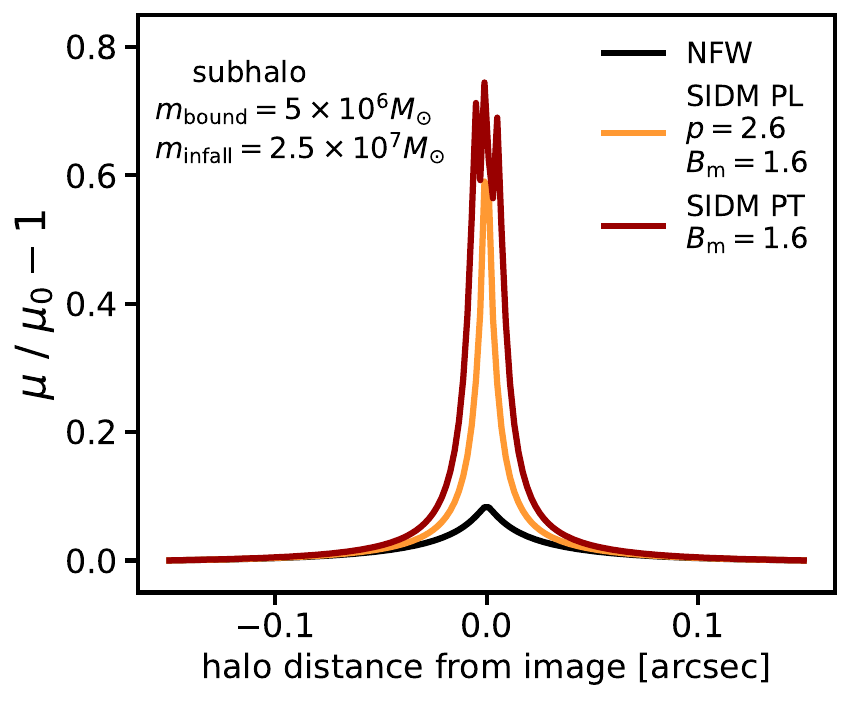}
		\caption{\label{fig:sidmmagcross} The strength of magnification perturbations produced by dark matter halos in CDM (black) and SIDM as a function of their projected offset from a lensed image. Orange curves assume the power-law (PL) model for core collapsed halos (Equation \ref{eqn:rhosidm1}), and the red curve assumes the point-mass (PM) model (Equation \ref{eqn:rhosidm2}). The reference NFW profiles have a concentration of 10, and we assume a source size of 10 pc. {\bf{Top:}} The perturbations caused by field halos of mass $m=5 \times 10^6 M_{\odot}$. The halo scale radius $r_s = 0.3 \ \rm{kpc}$ subtends 0.05 arcsec at $z=0.5$. {\bf{Bottom:}} The perturbation caused by a subhalo with infall (bound) mass $2.5 \times 10^7 M_{\odot}$ ($5 \times 10^6 M_{\odot}$) and $z_{\rm{infall}}=2$. The jagged feature produced by the point mass model occurs because the lensed image is split in two by the halo. We use a flexible model for the internal structure of collapsed halos (Equation \ref{eqn:yeffect}) with perturbation strengths bounded from below by the orange curves and from above by the red curves.}
	\end{figure}

	Had \citet{Gurian++25} run their simulation longer, for example, until $t / t_c = 10$, the central density would have increased by many additional orders of magnitude. In terms of our model parameters, for $t / t_c \gg 1$ both $p$ and $B_{\rm{m}}$ may increase with time. Pushing even further into the collapse phase, it remains an open question whether any mechanism would stop the runaway contraction, or if the central parts of an SIDM halo eventually collapse into black holes \citep{Balberg++02b,Pollack++15,Feng++21}. We note that a point mass inside an extended body of material is the best-fitting model for the extremely dense $\sim 10^6 M_{\odot}$ perturber recently detected in an extended lensed arc \citep{Powell++25,Vegetti++26}. If we adopt an SIDM explanation for this unusual object, this suggests core collapse can form structures more compact than allowed Equation \ref{eqn:rhosidm1} with $p=2.6$ and $B_{\rm{m}}=1.6$.
    \begin{figure*}
		\centering
		\includegraphics[trim=0.5cm 2.5cm 0.5cm
		0.5cm,width=0.95\textwidth]{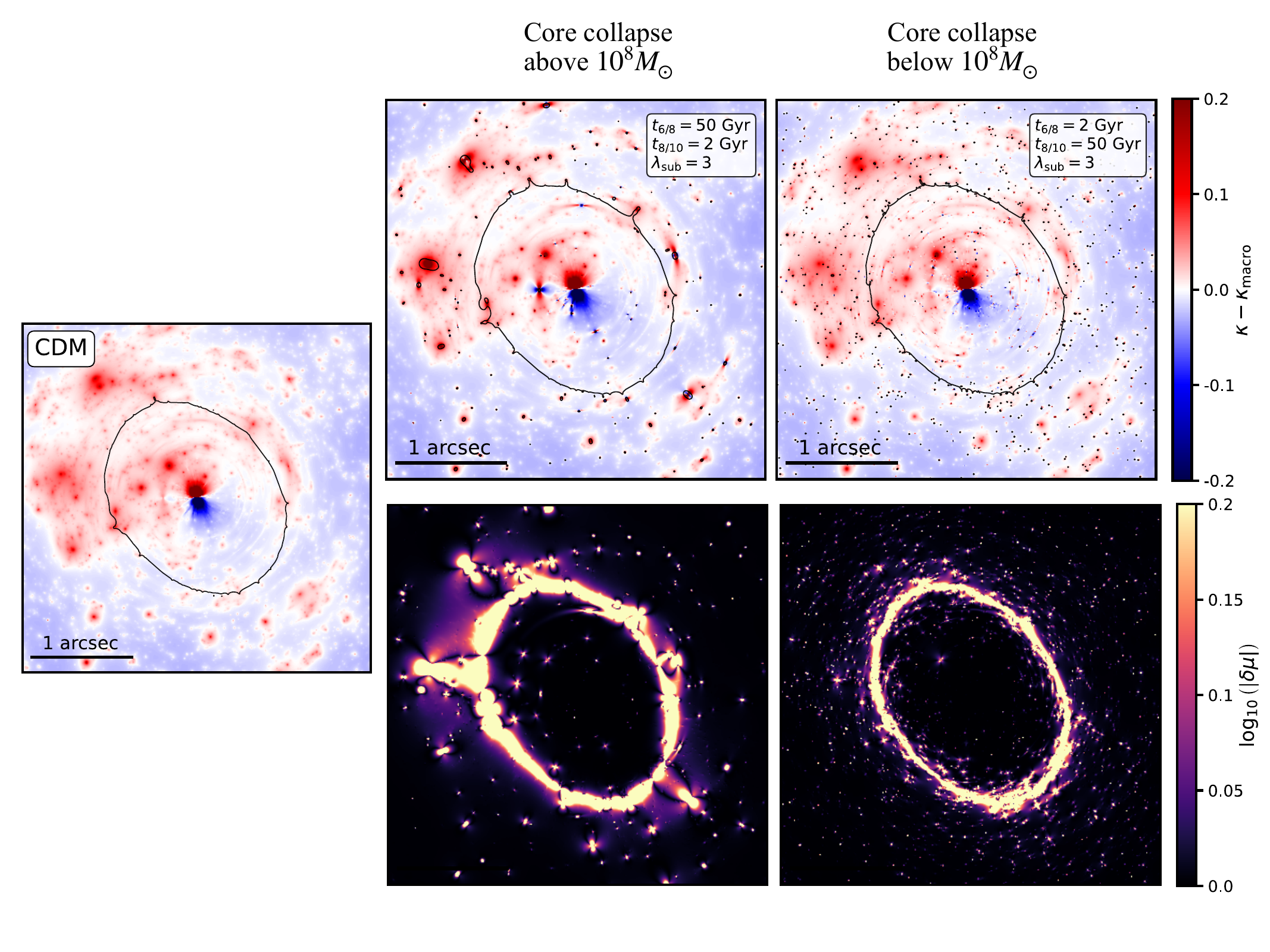}
		\caption{\label{fig:kappamaps}  Illustrations of core collapsed halo populations. {\bf{Left:}} The projected view of a population of CDM subhalos and field halos in terms of the effective multi-plane convergence (Equation \ref{eqn:kappaeff}). The color map shows the effective convergence relative to the average projected mass in dark matter. The critical curve of the lens system is shown in black. {\bf{Top row:}} Two SIDM realizations generated from the CDM halo population shown of the left, following the procedures outlined in Sections \ref{ssec:timescale} and \ref{ssec:collapsedsidm}. The center panel has median collapse timescales $t_{6/8}=50 \ \rm{Gyr}$ and $t_{8/10}=2 \ \rm{Gyr}$, driving core collapse primarily in halos more massive than $10^8 M_{\odot}$. The panel on the right has $t_{6/8}=2 \ \rm{Gyr}$ and $t_{8/10}=50 \ \rm{Gyr}$, causing the collapse of low-mass halos. Many core collapsed halos become dense enough to form their own critical curves. {\bf{Bottom row:}} The change in magnification $\delta \mu \equiv \mu / \mu_{\rm{cdm}}-1$ between the halo populations shown in the top row, and the CDM realization shown on the left. Our analysis is sensitive to the numerous localized perturbations to the magnification map from collapsed halos.}
	\end{figure*}
	
	Either a more concentrated central power-law profile or a black hole would impart stronger perturbations to lensed images than predicted by the power-law profile model with $p=2.6$ and $B_{\rm{m}}=1.6$. This motivates some additional flexibility in our model for the collapsed halo profile. A limiting case for the collapsing central region of an SIDM halo is a point mass
	\begin{equation}
		\label{eqn:rhosidm2}
		\rho_{\rm{central}} = M_{\rm{central}}\left(s\right) \, \delta^{(3)}\left(\boldsymbol{r}\right)
	\end{equation}
	where $M_{\rm{central}}\left(s\right)$ is the mass of the central collapsed profile interior to $s$ obtained by solving Equations~\ref{eqn:slopematch}-- \ref{eqn:massconserve2} for $\gamma_s=-1.9$, $p=2.6$, and $B_{\rm{m}}=1.6$. We note that our model gives $M_{\rm{central}}\left(s\right) / m \sim 0.03$--$0.08$ for field halos. For subhalos, $M_{\rm{central}}\left(s\right) / m_{200}$ depends on the truncation radius and bound-mass fraction. 

	As we discuss further in the next section, the deflection angle we obtain from modeling the collapsed region as a point mass can be closely approximated by a power-law profile described by Equation~\ref{eqn:rhosidm1} with larger $p$ and $B_{\rm{m}}$. In particular, $p=2.95$ and $B_{\rm{m}}=2.5$ produce the point-mass deflection across the radial scales relevant for our analysis (see Figure \ref{fig:alphar}). We implement a flexible model for the collapsed halo center by allowing $p$ and $B_{\rm{m}}$ to vary between the power-law and point-mass limits through an additional parameter $\xi_{\rm{core}}$
	\begin{equation}
	\label{eqn:yeffect}
	\begin{split}
		p &= 2.6 + 0.35 \, \xi_{\rm{core}} \\
		B_{\rm{m}} &= 1.6 + 0.9 \, \xi_{\rm{core}}.
	\end{split}
    \end{equation}
	At $\xi_{\rm{core}}=0$ this recovers the power-law fit to \citet{Gurian++25} shown in Figure \ref{fig:sidmdensity}, while at $\xi_{\rm{core}}=1$ it reproduces the point-mass deflection from Equation \ref{eqn:rhosidm2}. In Section~\ref{sec:structureinference} we quantify how our inference on $t_{6/8}$ and $t_{8/10}$ depends on the internal structure of collapsed halos through importance sampling weights on $\xi_{\rm{core}}$. As a single value of $\xi_{\rm{core}}$ is applied to every collapsed halo generated in a realization, it sets the population-average degree of central mass concentration, and hence the deflection angles produced by collapsed halos relative to their CDM counterparts. Finally, we note that increasing the density of collapsed halos through $\xi_{\rm{core}}$ approximates the effect of including directional scatter in the core collapse timescale based on halo concentration. We discuss this topic further in Appendix \ref{sec:appC}. 
    
	 \subsection{The lensing signal of SIDM halos}
	 \label{ssec:lensingsignal}
    Figure~\ref{fig:alphar} compares the deflection angle of an NFW field halo with a mass of $5 \times 10^6 M_{\odot}$ and a concentration $c=10$ (black), the power law model for collapsed halos (orange), and that in a scenario where $\rho_{\rm{central}}$ is modeled as a point-mass model (red) according to Equation \ref{eqn:rhosidm2}. The dotted orange curve, which uses $\xi_{\rm{core}} = 0.9$ in Equation \ref{eqn:yeffect}, shows our approximation of the lensing signal from the point-mass model (Equation \ref{eqn:rhosidm2}) in terms of a power-law profile. Relative to the NFW profile, the deflection angle increases for collapsed SIDM halos at $r / r_s = 0.1$ by a factor of 10-100. 
 
	Flux ratios are sensitive to the derivatives of the deflection field across a lensed image with an angular sensitivity scale determined by the source size \citep{Dobler++06}. Figure~\ref{fig:sidmmagcross} compares the strength of a perturbation to an image magnification from halos in CDM and SIDM. We show both the power-law and point-mass representations of the core-collapsed density profile (Equations \ref{eqn:rhosidm1} and \ref{eqn:rhosidm2}), and assume a background source size of 10 pc. The x-axis shows the projected angular separation between the halo and a simulated lensed image, and the y-axis shows the magnification, $\mu$, relative to the magnification produced by the macromodel without a perturber included, $\mu_0$. In the top panel, we consider a field halo with a total mass $5 \times 10^6 M_{\odot}$, the same as in Figure~\ref{fig:alphar}. For the objects causing the perturbation shown in upper panel $r_s \sim 0.05 \ \rm{arcsec}$, and the collapsed halos produce stronger magnifications than an NFW halo by a factor of $5-10$ when they have an angular separation from the lensed image $\lesssim 0.5 r_s$. 
    
    The lower panel of Figure \ref{fig:sidmmagcross} shows the perturbation caused by a subhalo accreted at $z_{\rm{infall}}=2$ with an infall mass of $2.5 \times 10^7 M_{\odot}$ that is tidally stripped down to $5 \times 10^6 M_{\odot}$, matching the virial mass of the field halos shown in the upper panel. In the case of the subhalo perturber, the point-mass model (and our power-law approximation to this profile) causes the lensed image to split in two, producing a triple-peak structure in the magnification curve. For both sets of assumptions regarding the collapsed halo internal structure, the subhalo produces a stronger perturbation than the field halo of the same bound mass because the subhalo's characteristic density, $\rho_s$, is $\sim 5$ times higher. Because $\rho_s \propto \rho_{\rm{crit}}\left(z_{\rm{infall}}\right)$ at fixed mass and concentration, halos accreted earlier are denser. We emphasize that predicting the strength of lensing perturbations by subhalo populations, or interpreting a single-halo detection in a lensed arc in terms of a subhalo perturber \citep[e.g.][]{Minor++21,He++25,Enzi++25,Vegetti++26}, necessarily involves assumptions about $z_{\rm{infall}}$. As discussed in Section \ref{ssec:cdmhalos} in relation to the subhalo tidal evolution model, we use a distribution of infall redshifts calibrated for CDM using {\tt{galacticus}}, and assume the mass accretion history of SIDM subhalos follows a similar distribution. 
    
    Although the strength of a perturbation caused by a halo depends on the magnification produced by the underlying macro lens model and the background source size, the qualitative conclusions we can draw from Figure~\ref{fig:sidmmagcross} do not strongly depend on these details. Collapsed halos imparts significantly stronger perturbations to a lensed image than a CDM halo of the same total mass. The difference in lensing efficiency between collapsed halos and their CDM counterparts becomes most apparent when the collapsed halo makes a ``direct hit'' on a lensed image. In Section \ref{ssec:lensreconstructions} we give several examples of how direct hits by compact structures can explain some of the more extreme flux ratio anomalies in a subset of lenses in our sample. 

    Although lens models that reproduce extreme flux ratio anomalies often have a single dominant perturber near a lensed image, the frequency with which these configurations arise depends on the properties of the full population. Our analysis therefore constrains the collective properties of the population, rather than the properties of any individual perturber. As an example, in the left panel of Figure~\ref{fig:kappamaps} we show an illustration of one such halo population as it would appear in CDM. The two panels to the right of the CDM illustration depict SIDM models with different $t_{6/8}$ and $t_{8/10}$. These panels show the effective multiplane convergence in substructure,
 	\begin{equation}
 		\label{eqn:kappaeff}
 		\kappa_{\rm{sub}} = \kappa - \kappa_{\rm{macro}},
 	\end{equation}
 	where $\kappa \equiv \left(1/2\right) \nabla \cdot \boldsymbol{\alpha}$ is an effective convergence for a multiplane lens system defined in terms of the deflection field $\boldsymbol{\alpha}$, and $\kappa_{\rm{macro}}$ is the convergence from the main deflector. Subtracting off the projected mass from the macromodel reveals the population of dark subhalos and line-of-sight halos. Due to the definition of $\kappa$ in terms of the full non-linear multi-plane deflection field, field halos behind the main deflector appear sheared and distorted in a direction tangential to the critical curve, shown as a black curve in each panel. Collapsed objects appear prominently in the convergence maps, and often producing their own critical curves. 
 	\begin{table*}
		\setlength{\tabcolsep}{8pt}
		\caption{\label{tab:tableqsub} Description of the dark matter hyper-parameters introduced in Section \ref{sec:structureformationmodel} that are varied freely in our analysis of the data. The third column states the sampling distribution on the parameter implemented in the forward modeling procedure. Our analysis also marginalizes over parameters unique to each lens, including a uniform prior on background source size between $1-10 \ \rm{pc}$, and parameters specifying the lens macromodel, as discussed by \citet{Gilman++25a}.}
		\begin{tabular}{cccc}
			\hline
			Hyper-parameter & Description & Sampling distribution &  Remarks\\
			\hline
			$\Sigma_{\rm{sub}} \ \left[\rm{kpc^{-2}}\right]$ & amplitude of the differential subhalo & $\log_{10} \mathcal{U}\left(-1.8, 0.0\right)$ & enhanced tidal stripping of\\ 
			& mass function at infall at $10^8 M_{\odot}$ & & surviving subhalos captured \\ 
			& & & by rescaling $\Sigma_{\rm{sub}}$ (Equation \ref{eqn:subhalomfunc})  \\ \\
			$t_{6/8} \ \left[\rm{Gyr}\right]$ & median core collapse timescale for & $\log_{10} \mathcal{U}\left(-0.3, 2.3\right)$ & individual halo $t_c$ with \\
			& halos with masses $10^{6}$--$10^{8} \ \mathrm{M}_{\odot}$ & & 0.3 dex scatter \\ \\
			$t_{8/10} \ \left[\rm{Gyr}\right] $ & median core collapse timescale for & $\log_{10} \mathcal{U}\left(-0.3, 2.3\right)$ & individual halo $t_c$ with \\
			& halos with masses $10^{8}$--$10^{10.7} \ \mathrm{M}_{\odot}$ & & 0.3 dex scatter \\ \\
			$\lambda_{\rm{sub}}$ & linear rate of accelerated or & $\log_{10} \mathcal{U}\left(-1, 1\right)$ & allows faster/slower collapse \\
			& decelerated subhalo core collapse relative& & due to tidal stripping or heating \\ & to field halos at $r_{2d}< 30 \ \rm{kpc}$ & & (Equation \ref{eqn:timesub}) \\ \\
			$\Sigma_{\rm{gc}} \ \left[M_{\odot} \rm{kpc^{-2}}\right]$ & projected surface mass density & $\log_{10} \mathcal{U}\left(5.3, 7.0\right)$ & expected abundance  \\
			& of globular clusters& & corresponds to $\log_{10} \Sigma_{\rm{gc}} \sim 5.6$ \\ & & & (Equation \ref{eqn:gcmf}) \\ \\
			$\xi_{\rm{core}}$ & sets the mean lensing efficiency of & $\mathcal{U}\left(0, 1\right)$ & interpolates between power-law\\
			& deeply collapsed halos & & and point-mass (Equation \ref{eqn:yeffect}) \\ \\
			$\delta_{\rm{LOS}}$ & rescales the amplitude of the& $\mathcal{U}\left(0.9, 1.1\right)$ & $\delta_{\rm{LOS}}=1$ corresponds to the \\
			& field halo mass function  & & Sheth--Tormen prediction\\ \\
			$\alpha$ & logarithmic slope of the & $\mathcal{U}\left(-1.95, -1.85\right)$ & CDM predicts $\alpha \sim -1.9$\\
			& subhalo mass function at infall & & \\ \\
			\hline
		\end{tabular}
	\end{table*}
    
 	The bottom panels of Figure~\ref{fig:kappamaps} show the change in the magnification surface, relative to CDM. The ring-like feature around the critical curve in the bottom panels is an expected feature that stems from tiny fractional differences in the deflection field that become amplified in the immediate vicinity of the critical curve. The signal we can extract from the data corresponds to the numerous localized perturbations to the magnification surface. The strength of these perturbations and their number density depend on the collapse timescales $t_{6/8}$ and $t_{8/10}$.   

    	\begin{figure*}
		\centering
		\includegraphics[trim=0.cm 0.0cm 0.0cm
		0.0cm,width=0.5\textwidth]{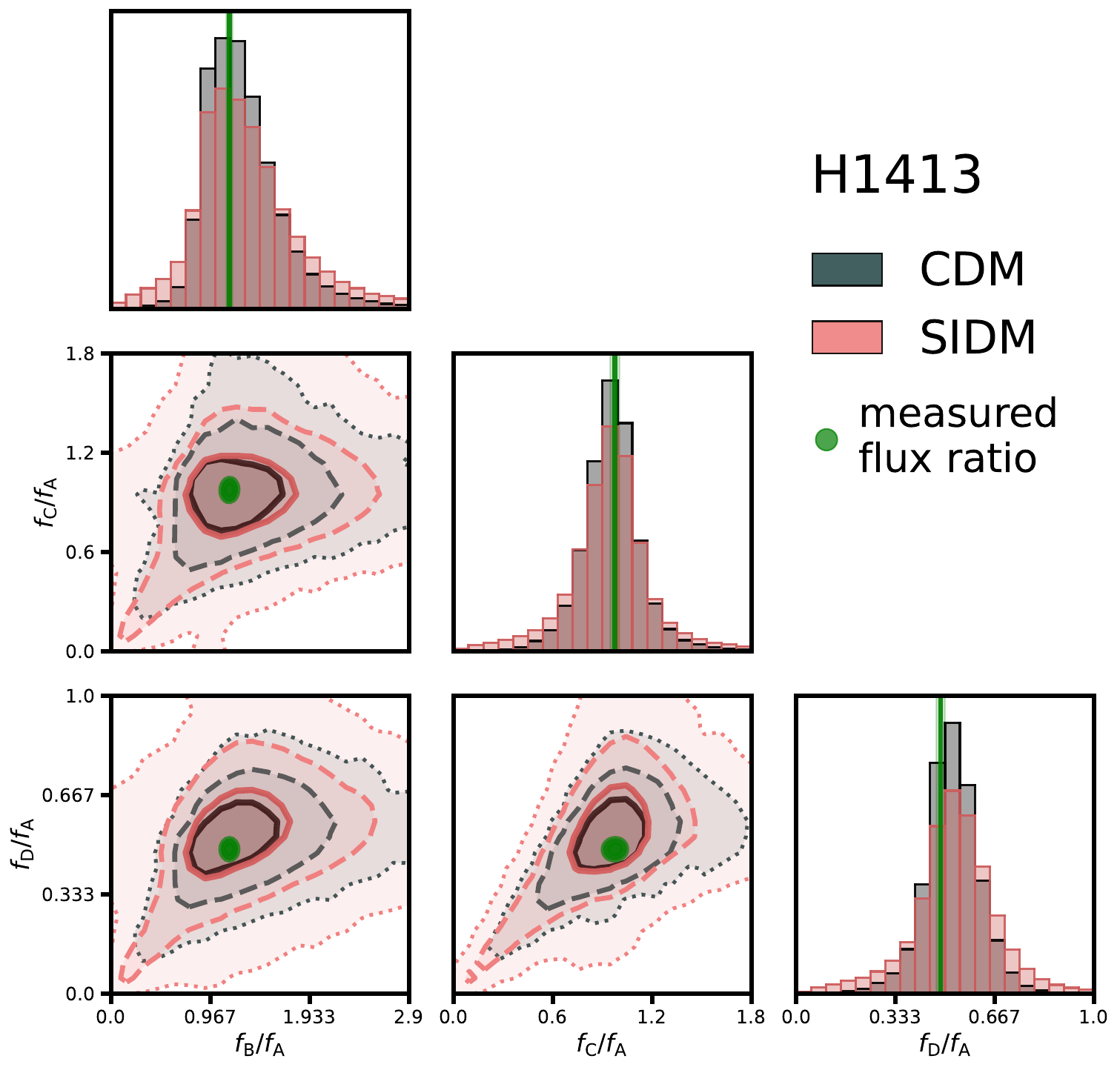}
		\includegraphics[trim=0.cm 0.0cm 0.0cm
		0.0cm,width=0.48\textwidth]{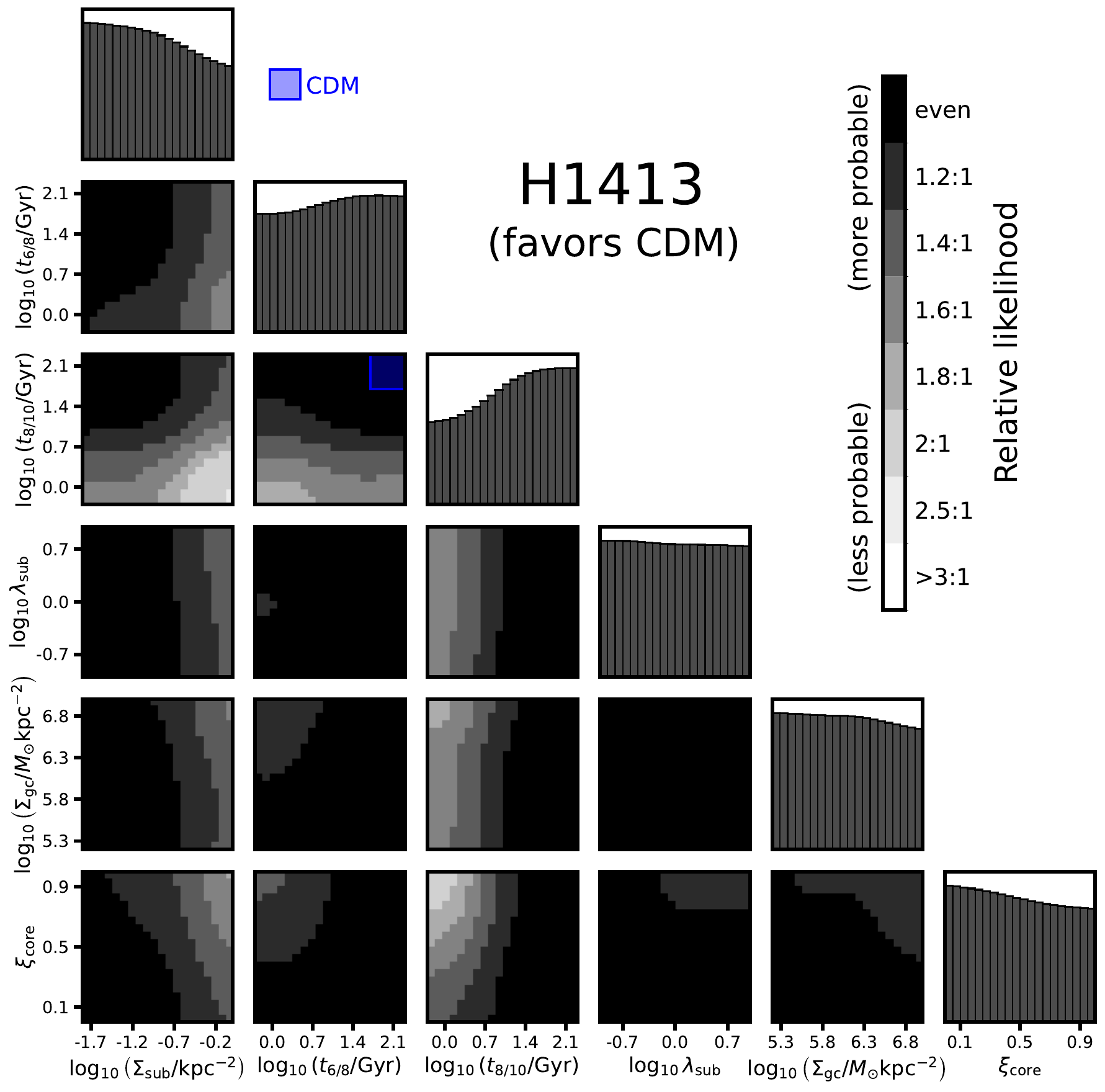}
		\includegraphics[trim=0.0cm 0.0cm 0.0cm
		0.0cm,width=0.5\textwidth]{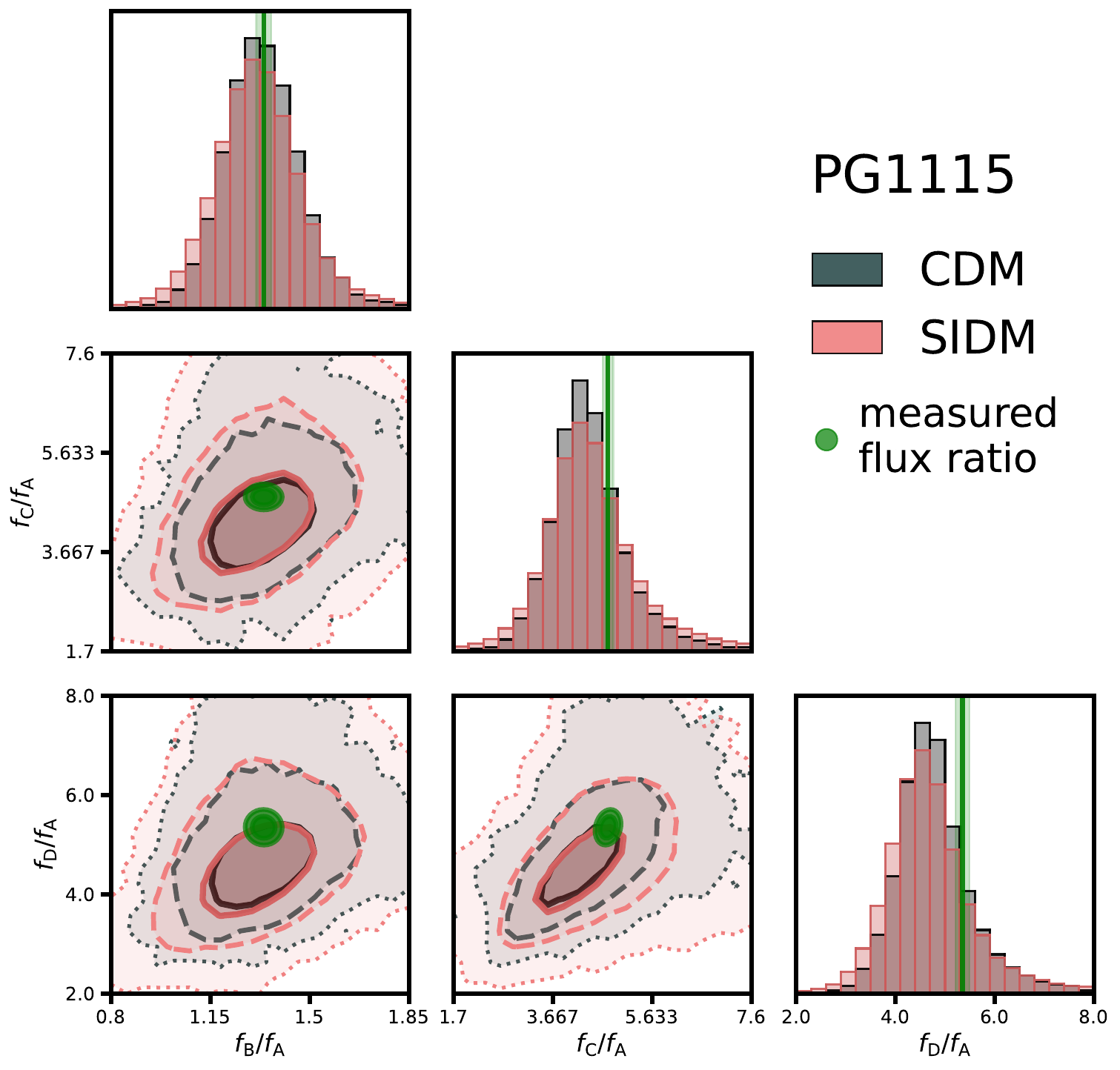}
		\includegraphics[trim=0.0cm 0.0cm 0.0cm
		0.0cm,width=0.48\textwidth]{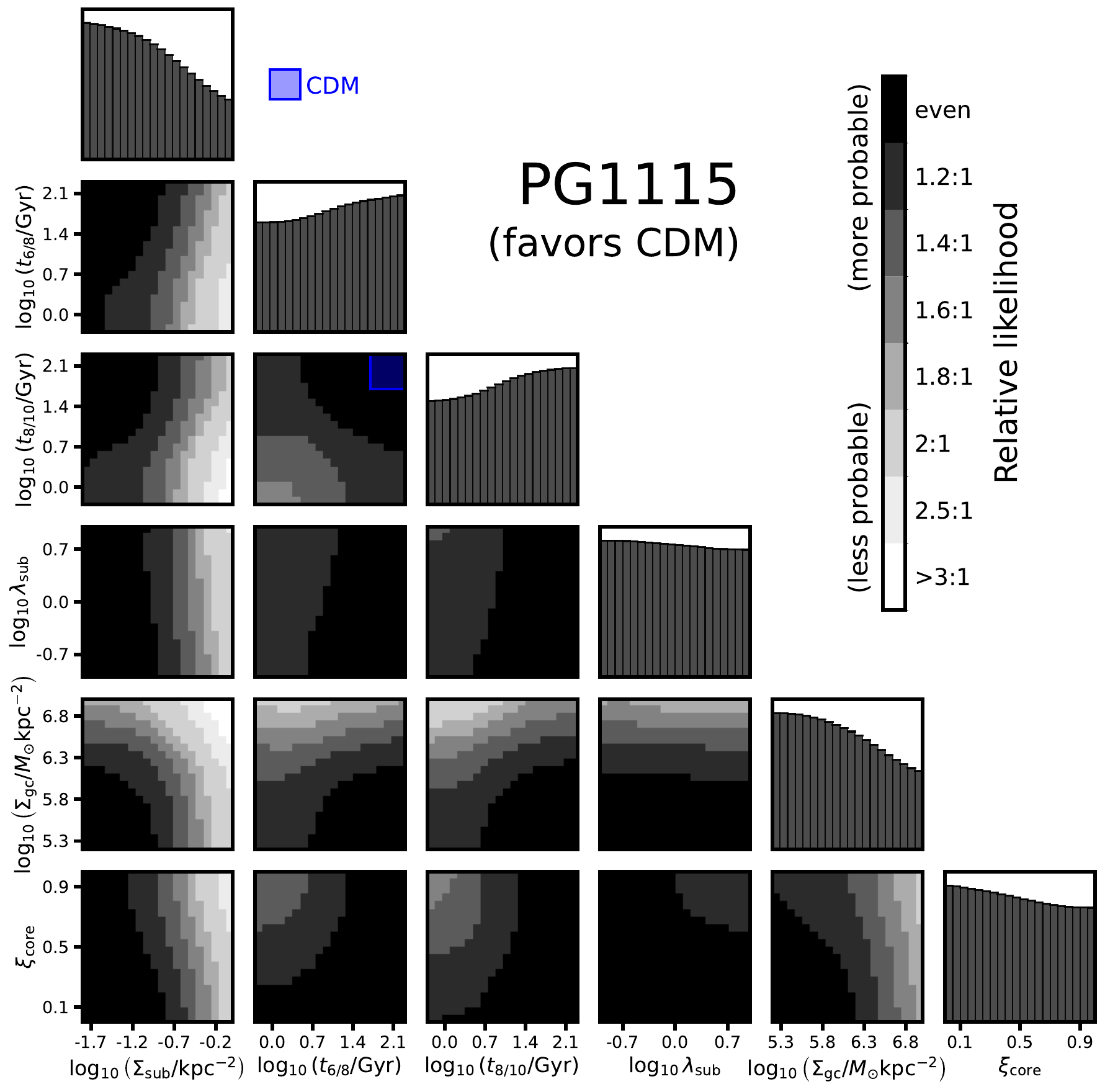}
		\caption{\label{fig:fluxratios1} {\bf{Left:}} The joint distribution of image flux ratios in CDM and SIDM predicted by our model for H1413 and PG1115. To clearly illustrate the different lensing signals between the models, we define CDM models as realizations with $t_{6/8}>80 \ \rm{Gyr}$ and $t_{8/10} > 80 \ \rm{Gyr}$, such that very few halos core collapse, and SIDM realizations as those with $t_{6/8}<5 \ \rm{Gyr}$ and $t_{8/10} < 5 \ \rm{Gyr}$, such that many halos collapse. The green ellipse and shaded bands represents the measured flux ratio and $2 \sigma$ uncertainty. The solid, dashed, and dotted contours correspond to $68 \%$, $95\%$, and $99.5 \%$ confidence regions. {\bf{Right:}} The likelihood functions for H1413 and PG1115. The color scale indicates likelihood of each point in the marginal 2D parameter space relative to the most probable point in each marginal 2D parameter space. We identify the region of $t_{6/8}$ and $t_{8/10}$ parameter space consistent with CDM with a blue square in the joint distribution of $t_{6/8}$ and $t_{8/10}$.}
	\end{figure*}
	
	\begin{figure*}
		\centering
		\includegraphics[trim=0.cm 0.0cm 0.0cm
		0.0cm,width=0.5\textwidth]{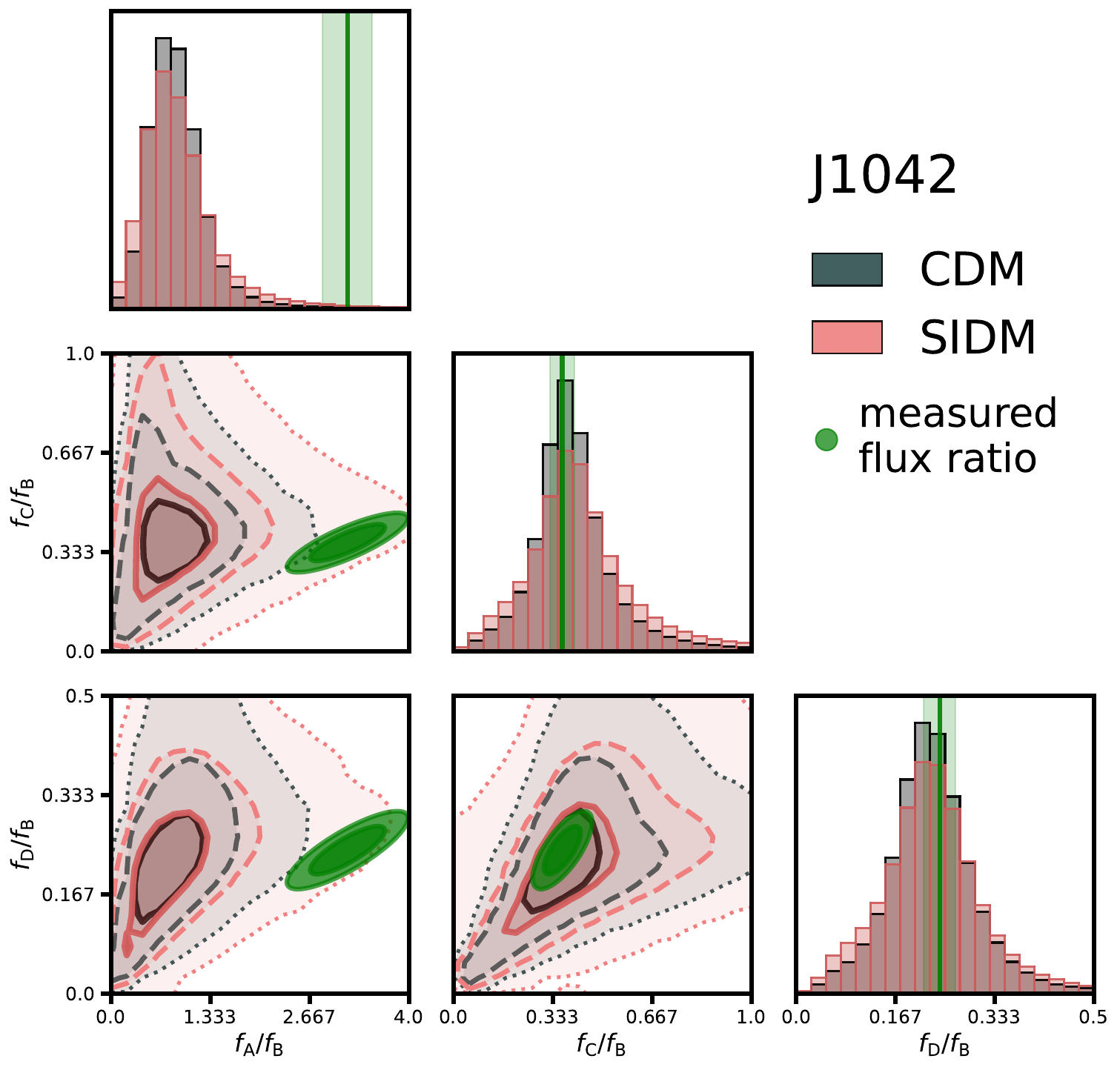}
		\includegraphics[trim=0.cm 0.0cm 0.0cm
		0.0cm,width=0.48\textwidth]{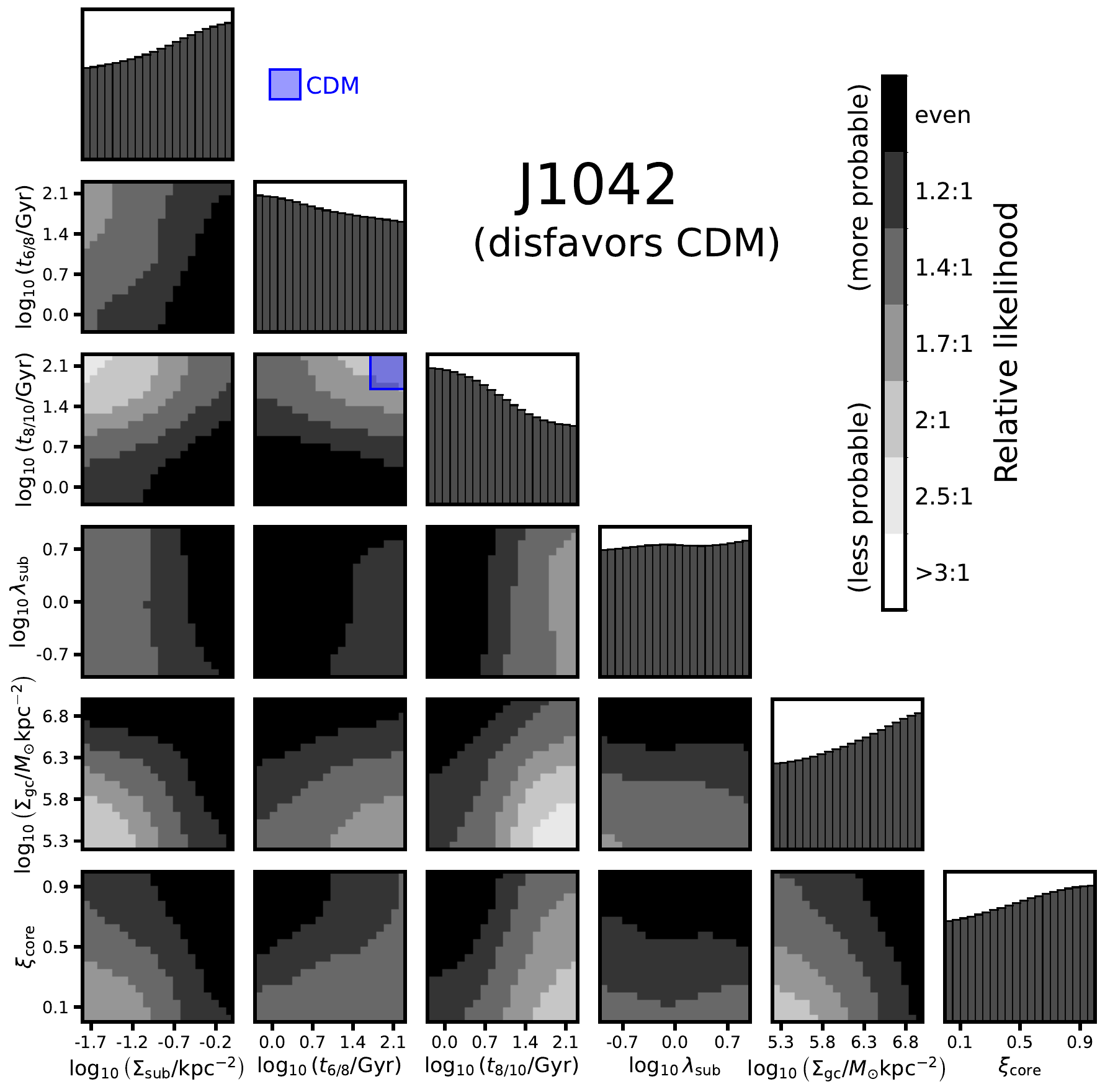}
		\includegraphics[trim=0.0cm 0.0cm 0.0cm
		0.0cm,width=0.5\textwidth]{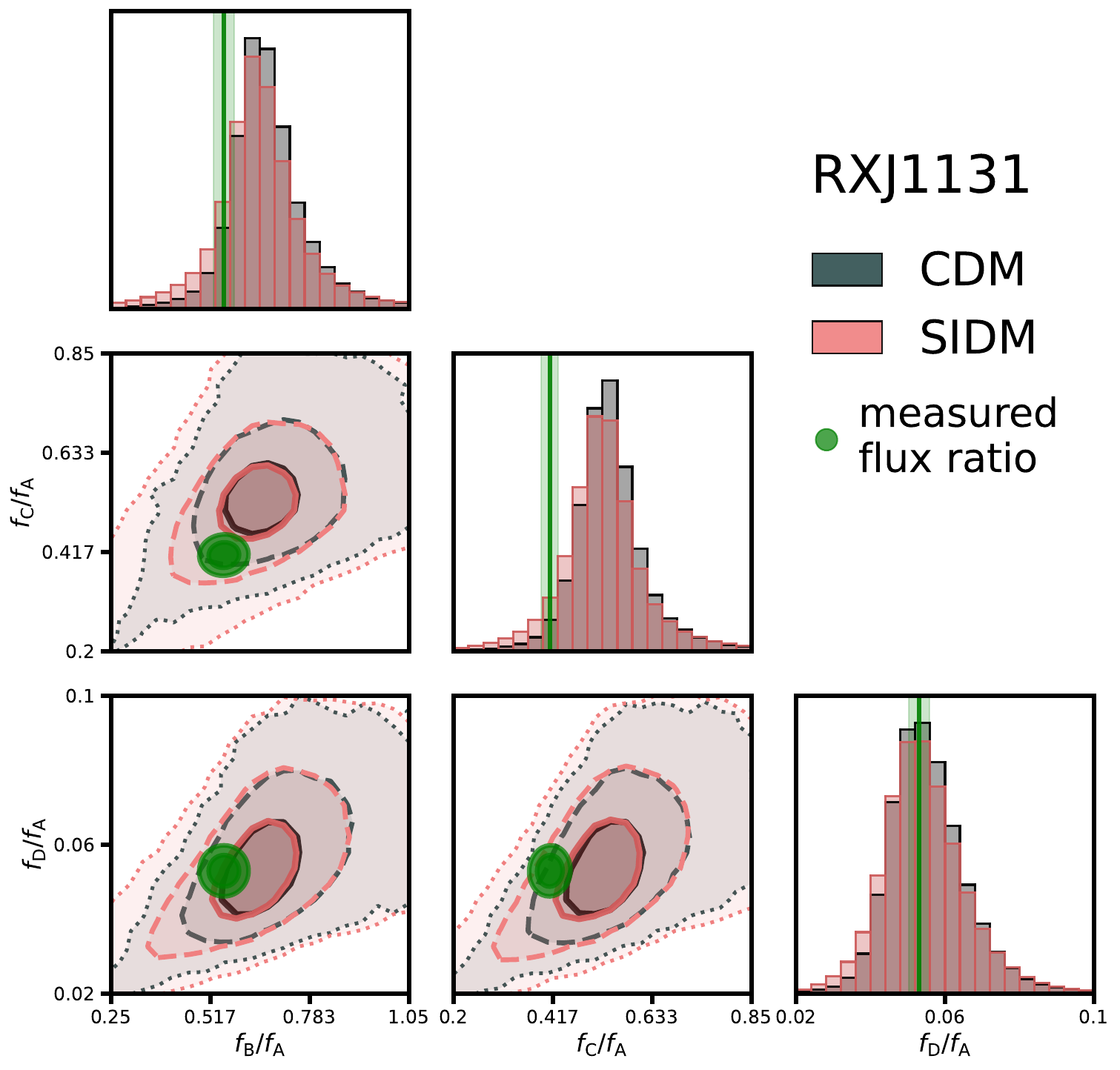}
		\includegraphics[trim=0.0cm 0.0cm 0.0cm
		0.0cm,width=0.48\textwidth]{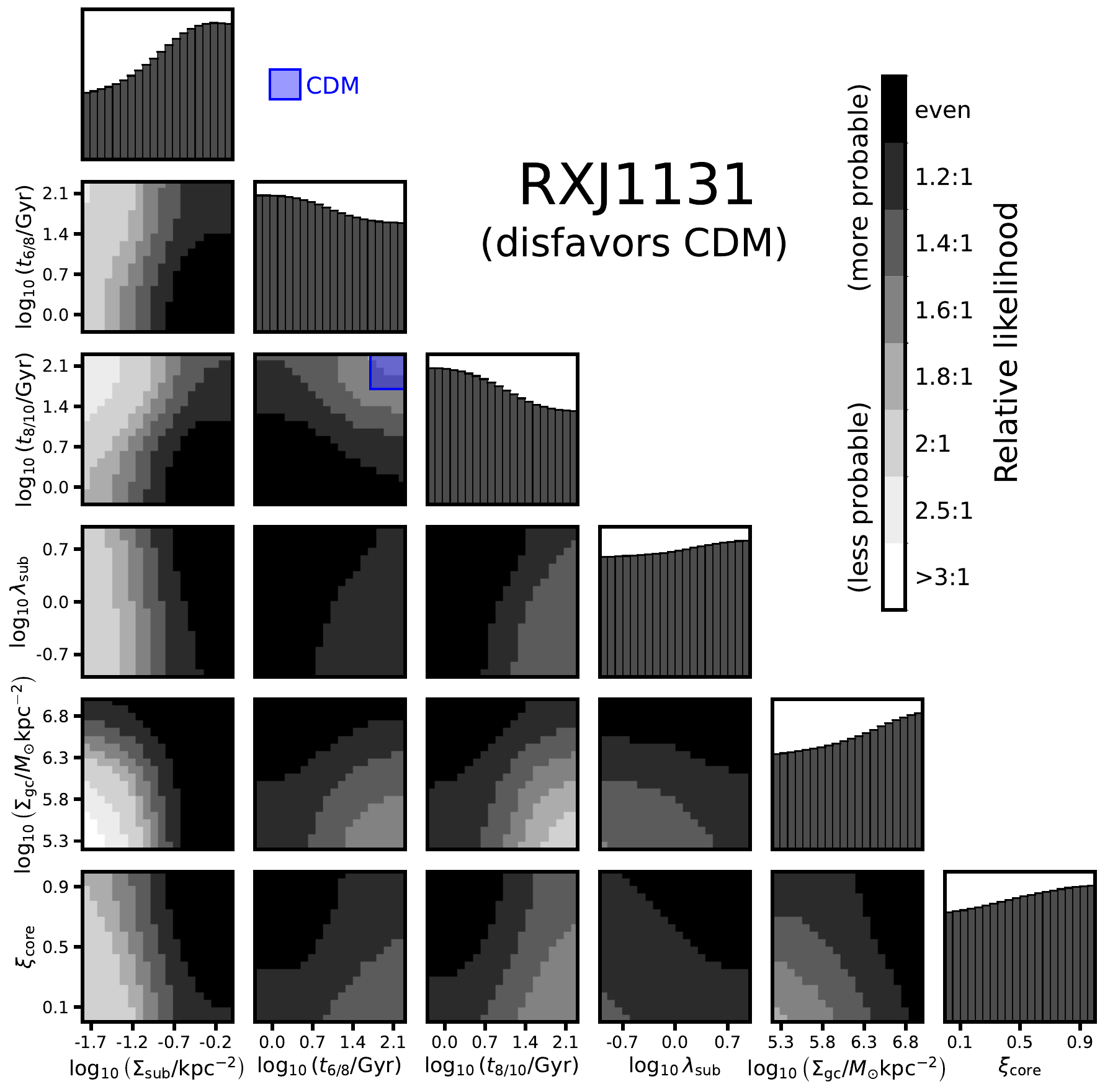}
		\caption{\label{fig:fluxratios2} The same as Figure \ref{fig:fluxratios1}, but we show the model-predicted flux ratio distributions (left) and likelihood functions (right) for J1042 and RXJ1131, two lens systems that disfavor CDM relative to SIDM models in which some fraction of halos core collapse. As in Figure \ref{fig:fluxratios1}, solid, dashed, and dotted contours in the flux ratio distributions correspond to $68 \%$, $95\%$, and $99.5 \%$ confidence regions, respectively. Lens systems for which SIDM is favored over CDM typically favor higher subhalo abundance ($\Sigma_{\rm{sub}}$), shorter core collapse timescales ($t_{6/8}$ and $t_{8/10}$), and a higher projected mass density in globular clusters ($\Sigma_{\rm{gc}}$).}
	\end{figure*}
	\begin{figure}
		\centering
		\includegraphics[trim=0.cm 0.0cm 0.0cm
		0.0cm,width=0.48\textwidth]{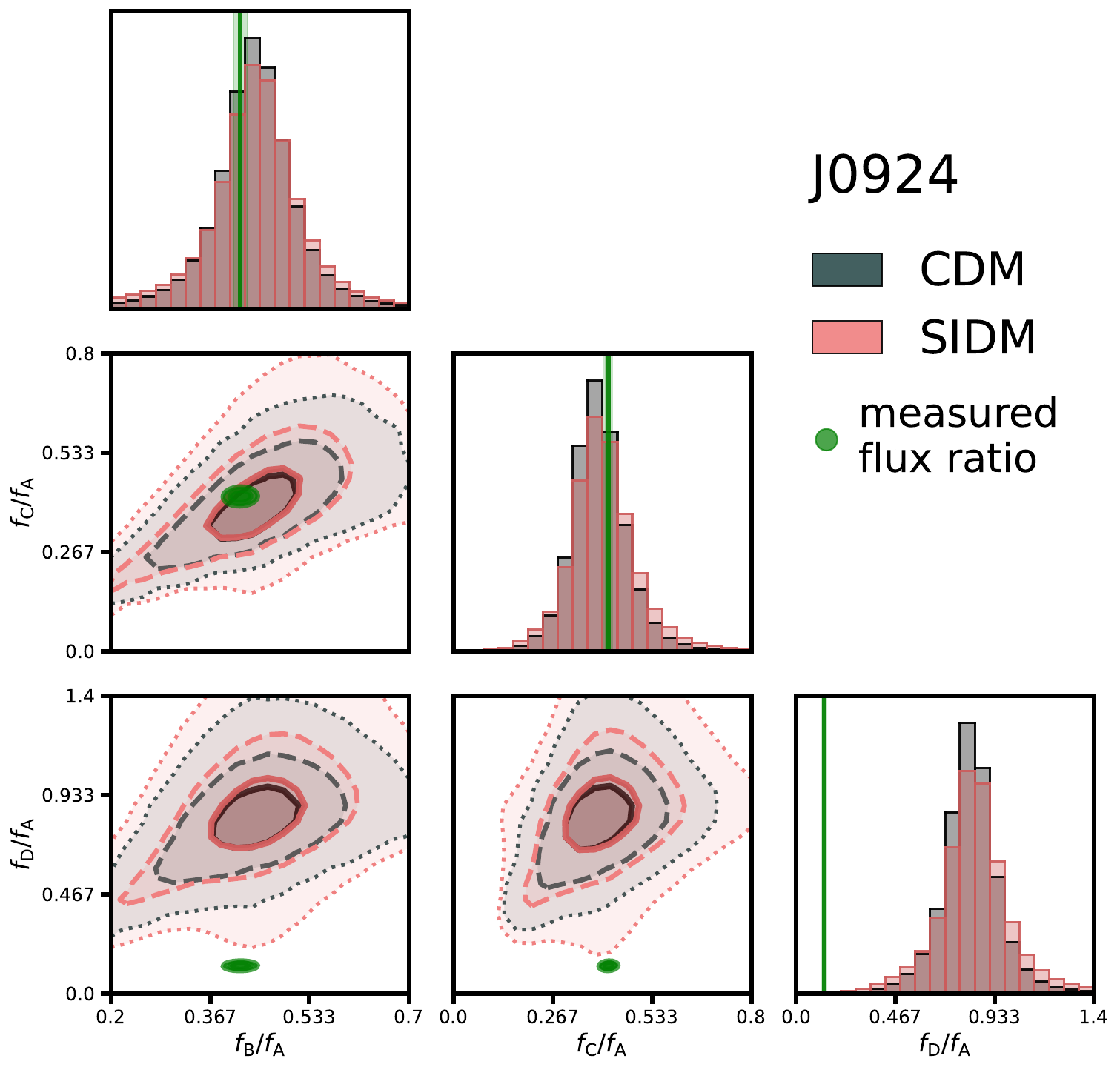}
		\includegraphics[trim=0.0cm 1cm 0.0cm
		0.0cm,width=0.3\textwidth]{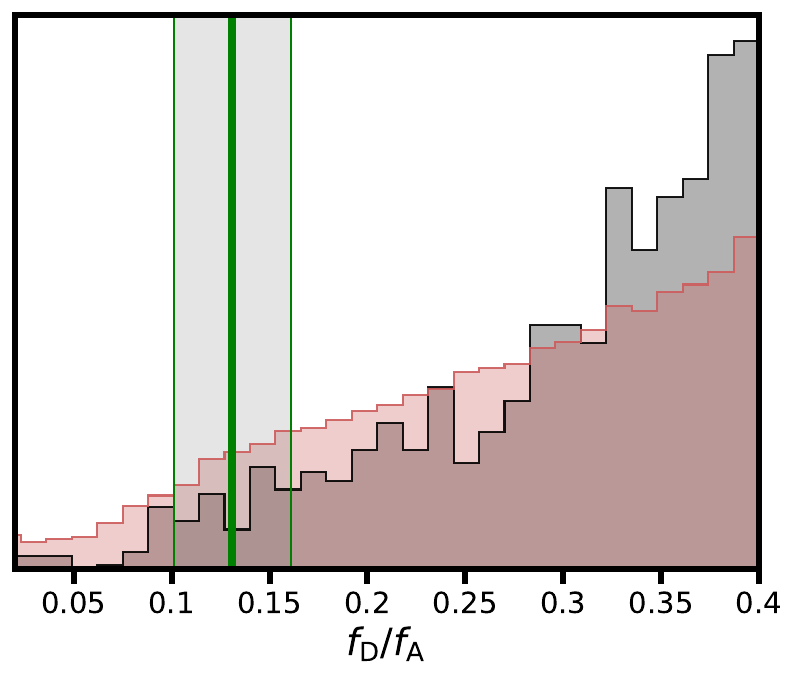}
		\caption{\label{fig:fluxratiosj0924} The model-predicted ratio distributions for the lens system J0924, which exhibits a large flux ratio anomaly (demagnification) in image D. The upper panel shows the joint distribution of the three flux ratios, and the lower panel zooms in on the tail of the $f_{\rm{D}}/f_{\rm{A}}$ ratio. As in Figure \ref{fig:fluxratios1} that solid, dashed, and dotted contours in the flux ratio distributions correspond to $68 \%$, $95\%$, and $99.5 \%$ confidence regions, respectively.}
	\end{figure}
    \begin{figure}
	 	\centering
	 	\includegraphics[trim=0.cm 0.0cm 0.0cm
	 	0.0cm,width=0.48\textwidth]{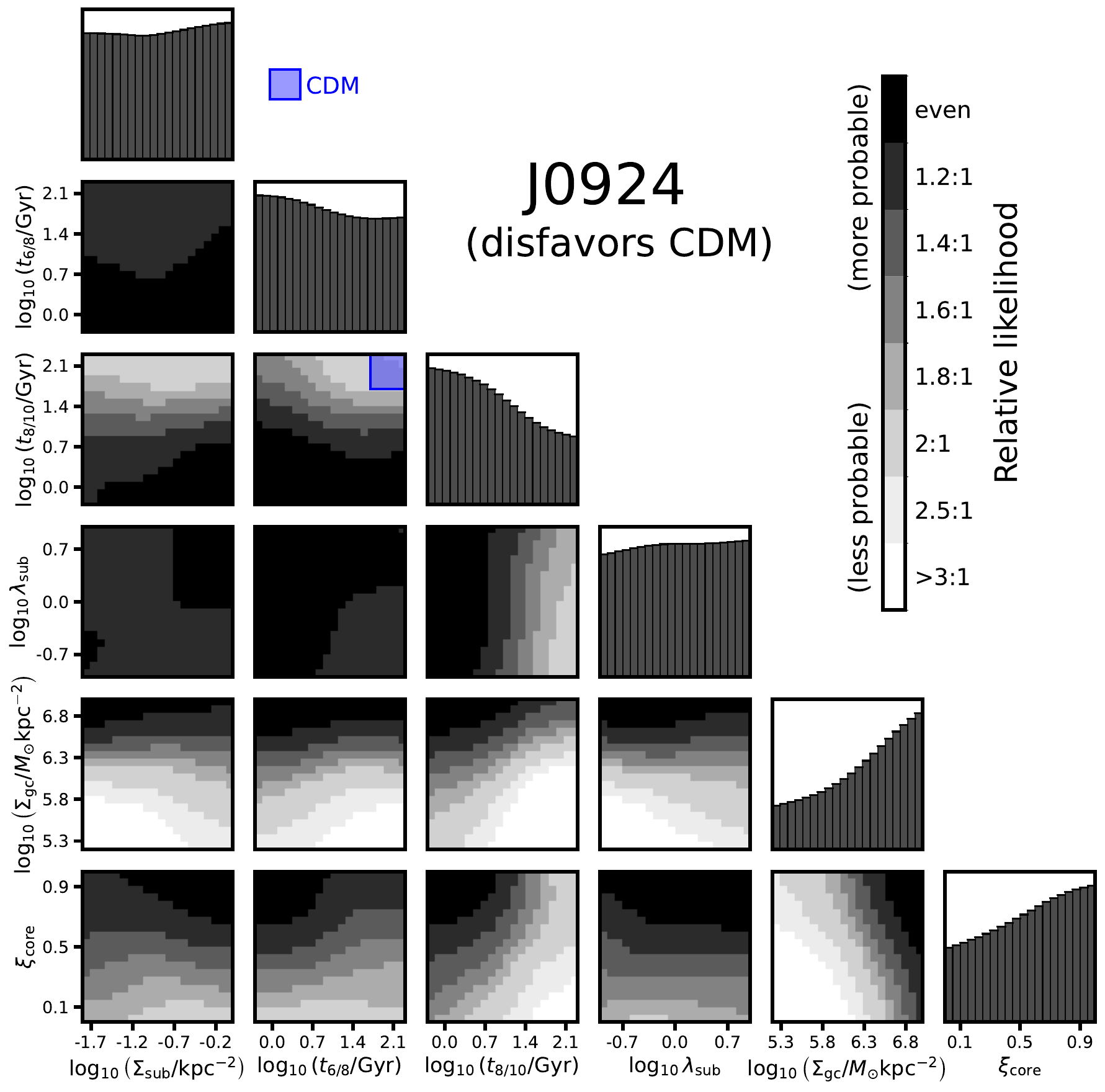}
	 	\caption{\label{fig:likelihood0924} The likelihood function for J0924. Due to the large flux anomaly in image D, this system exhibits the strongest relative likelihood disfavoring CDM among the lenses in our sample, with a penalty of $\sim$ 3:1. Enforcing a prior on the surface mass density of globular clusters based on observations of nearby elliptical, the relative likelihood increases to $\sim$ 5:1.}
	 \end{figure}
     
	\section{Results}
	\label{sec:structureinference}
	This section presents the results of our analysis, in which we compute the likelihood function of the data given the core collapse timescales, and the other dark matter parameters summarized in Table \ref{tab:tableqsub}. We begin in Section \ref{ssec:sidmstats} by discussing flux ratio statistics in SIDM models. We show model-predicted flux ratio distributions and likelihood functions for individual systems. These illustrations demonstrate the typical level of constraining power of an individual lens system, and help build intuition for how population-level inferences account for many different sources of small-scale perturbation, for example, from dark substructure, globular clusters, and angular structure in the main deflector. Section \ref{ssec:structureinf} presents constraints on the core collapse timescales under different sets of assumptions for the effect of tidal stripping on core collapse, and the abundance of globular clusters and dark subhalos.  
	
	\subsection{Flux ratio statistics in SIDM models}
	\label{ssec:sidmstats}
	\subsubsection{Model-predict flux ratio distributions and individual likelihood functions}
	SIDM and CDM predict different flux ratio statistics on a per-lens basis. In the left panels of Figure \ref{fig:fluxratios1}, we show examples of the model-predicted flux ratio distributions for two systems, H1413 and PG1115. We distinguish the black and red probability densities based on the core collapse timescale: the black distribution includes only flux ratios corresponding to realizations with $t_{6/8}$ and $t_{8/10}$ both longer than 80 Gyr and $\lambda_{\rm{sub}} < 1$, while the red distribution includes only models with $t_{6/8}$ and $t_{8/10}$ both shorter than 5 Gyr and $\lambda_{\rm{sub}} > 1$. The black distribution therefore corresponds to CDM, because these realizations contain a negligible fraction of collapsed halos in the lens model. Conversely, the red distribution corresponds to SIDM models in which many halos collapse. We marginalize each probability density over other sources of flux ratio perturbation, including globular clusters and angular structure from the $m=1$, $m=3$, and $m=4$ elliptical multipoles. 

	The width of model-predicted flux ratio distributions, such as the one shown in Figure \ref{fig:fluxratios1}, depends on the degree to which the image positions and extended lensed arcs constrain the main deflector mass profile, and also on the amount of small-scale structure in the lens system. The distributions peak near the median smooth-model prediction. Substructure imparts perturbations to the flux ratios, broadening the distributions. SIDM models with core collapsed halos result in broader distributions because collapsed halos impart stronger perturbations than NFW profiles.
	
	The likelihood functions for H1413 and PG1115 are shown in the right panels of Figure \ref{fig:fluxratios1}. In these figures and throughout this section, we use a color scale based on the relative likelihood taken with respect to the most probable point in each projected 2D parameter space. In Figure \ref{fig:fluxratios1}, black regions are all within a factor 1.2 in relative likelihood with respect to the most probable point, while white regions are disfavored by more than 2.5:1. We highlight a region of parameter space consistent with CDM with a blue blox. In this region, both core collapse timescales exceed $80 \ \rm{Gyr}$, and thus a negligible fraction of halos undergo core collapse according to our model. 
    
    For H1413 and PG1115, the measured flux ratios sit near the peak of the flux ratio distributions, which coincides with the flux ratio predictions from a smooth lens model. The data therefore disfavors models with increased subhalo abundance and in which a significant fraction of halos undergo core collapse. We emphasize that a single lens system, with a factor of $2$--$3$ in relative likelihood between two extreme ends of parameter space, does not on its own rule out any particular model. The constraining power comes from the population-level inference, where a relative likelihood of $2$--$3$ across the 29 lenses in our sample could yield a strong constraint. 
	
	Figure \ref{fig:fluxratios2} shows examples of lens systems where the data favors models with additional small-scale perturbation. For these two cases, J1042 and RXJ1131, SIDM realizations produce flux ratios that match the data more often than CDM realizations. As a result, the likelihood functions for these systems, shown in the right panels of Figure \ref{fig:fluxratios2}, disfavor core collapse timescales much longer than a Hubble time with a relative likelihood of $\sim$2:1. 
    
    In both Figure \ref{fig:fluxratios1} and \ref{fig:fluxratios2} the core collapse timescales are strongly correlated with subhalo and globular cluster abundance. As discussed in Section \ref{sec:structureformationmodel}, we expect a projected mass density in globular clusters at a projected distance from the central galaxy of $6-12 \ \rm{kpc}$ of $\Sigma_{\rm{gc}} \sim 10^{5.6} M_{\odot} \rm{kpc^{-2}}$. By matching the amplitude of the bound mass function predicted by our tidal evolution model to numerical simulations \citep{Du++25,Gannon++25}, we calculate a subhalo abundance corresponding to $\Sigma_{\rm{sub}}=0.1 - 0.16 \ \rm{kpc^{-2}}$ in CDM, and a possible suppression in SIDM due to the increased susceptibility of SIDM halos to tidal disruption. In the following section, we use an informative prior on $\Sigma_{\rm{sub}}$ and $\Sigma_{\rm{gc}}$ to disentangle the relative contribution of increased subhalo and globular clusters abundance from the number of core-collapsed halos.  
	
	Two lens systems in our sample, J0924 and B2045, have flux ratios that deviate strongly (by more than $5$ standard deviations) from the predictions of a smooth lens model. Figure~\ref{fig:fluxratiosj0924} shows the model-predicted flux ratios for J0924, the more extreme of the two (we discuss B2045 below). J0924 appears in a fold image configuration, with images A and D separated by approximately $3/4$ the Einstein radius. For a smooth lens model, the flux ratio of these two merging images should be close to unity, but in J0924 it is $0.13 \pm 0.01$. We note that this flux ratio anomaly is isolated to the D/A ratio, which points towards a local perturbation to the lens model near image D, rather than a global perturbation to the main deflector mass profile, which would likely affect all four images.  
     
	Figure~\ref{fig:fluxratiosj0924} shows that the strength of the perturbation to image D is an outlier, even in SIDM. Nonetheless, as shown in the lower panel of Figure \ref{fig:fluxratiosj0924}, which zooms in on the tail of the $f_{\rm{D}} / f_{\rm{A}}$ distribution, SIDM models produce a flux ratio $f_{\rm{D}} / f_{\rm{A}} \sim 0.13$ more frequently than CDM models. As a result, the likelihood function for J0924, shown in Figure \ref{fig:likelihood0924}, disfavors models with long core collapse timescales. We emphasize that the data disfavor CDM even when we marginalize over the prior volume associated with globular clusters, which allows a number density exceeding the expected abundance around lens host halos by a factor of 25. 
	
	Many previous studies of J0924 have treated it as a hallmark example of stellar microlensing because the flux anomaly in image D is also detected at optical wavelengths \citep[e.g.,][]{Keeton++06,Floyd++09}. In our data the dust sublimation zone around the quasar, whose size depends on the black hole mass and the luminosity of the AGN, sets a minimum size for the warm dust region of $\sim 1 \ \rm{pc}$, $100$ times larger than a typical stellar Einstein radius. This region is therefore too extended to experience detectable microlensing \citep{Sluse++13}. On the other hand, \citet{Badole++20} measure a flux ratio near unity between images $A$ and $D$ from radio emission emanating from a kpc-scale region around the background quasar, a region too extended to be perturbed by a low-mass halo. The near-unity flux ratio disfavors a perturbation caused by a very massive object, such as an intervening undetected dwarf or satellite galaxy. These considerations support an alternative hypothesis: there is a compact object, or multiple objects, near image D massive enough to de-magnify the warm dust region but not massive enough to affect the radio measurements of \citet{Badole++20}. It is possible that the same perturber demagnifies image D at optical wavelengths, contributing to the flux anomaly sometimes attributed to microlensing.  
	\begin{figure*}
		\centering
		\begin{minipage}[t]{0.32\textwidth}
			\centering
			\includegraphics[width=\linewidth]{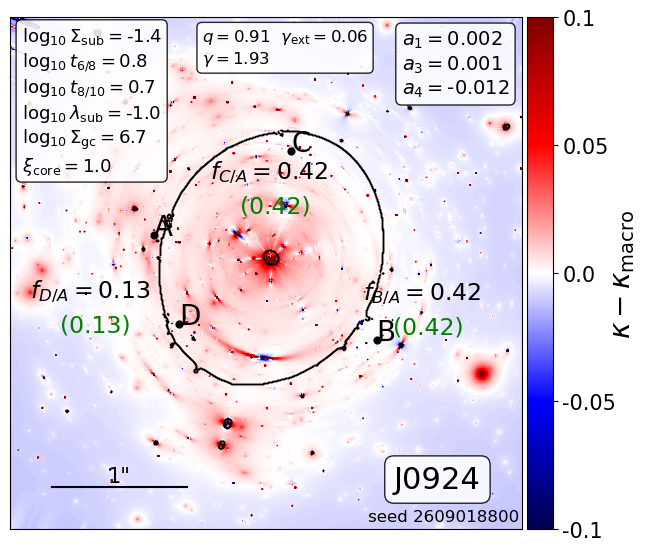}\\[0.3em]
			\begin{tabular}{@{}c@{\hspace{0.5em}}c@{}}
				\includegraphics[trim=0.0cm 1cm 0.0cm 0.0cm,width=0.48\linewidth]{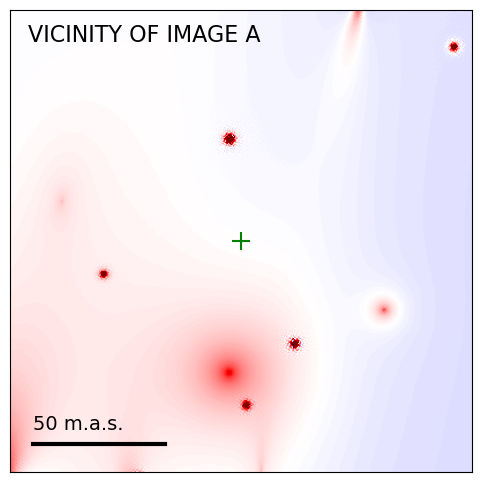} &
				\includegraphics[trim=0.0cm 1cm 0.0cm 0.0cm,width=0.48\linewidth]{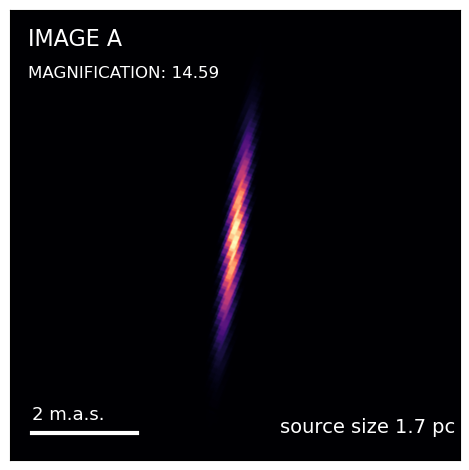} \\
				\includegraphics[trim=0.0cm 1cm 0.0cm 0.0cm,width=0.48\linewidth]{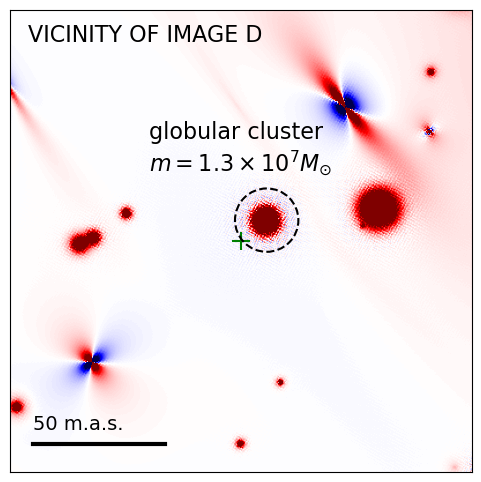} &
				\includegraphics[trim=0.0cm 1cm 0.0cm 0.0cm,width=0.48\linewidth]{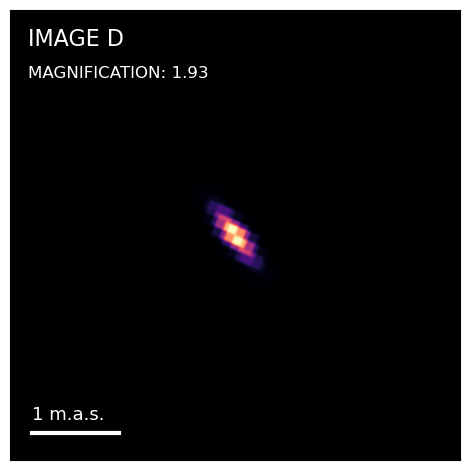}
			\end{tabular}
		\end{minipage}\hfill
		\begin{minipage}[t]{0.32\textwidth}
			\centering
			\includegraphics[width=\linewidth]{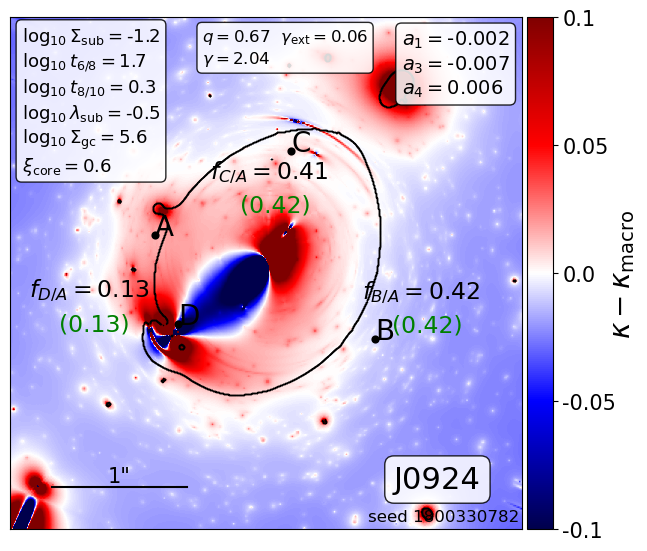}\\[0.3em]
			\begin{tabular}{@{}c@{\hspace{0.5em}}c@{}}
				\includegraphics[trim=0.0cm 1cm 0.0cm 0.0cm,width=0.48\linewidth]{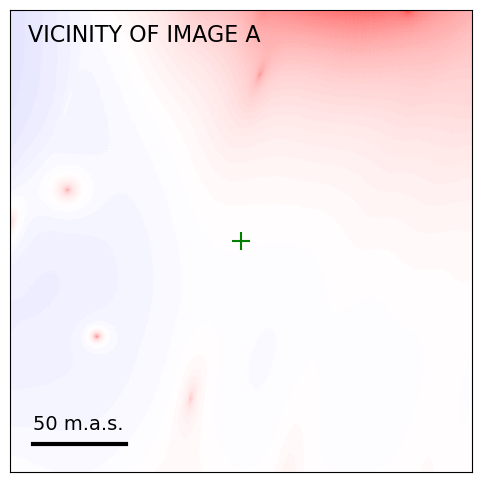} &
				\includegraphics[trim=0.0cm 1cm 0.0cm 0.0cm,width=0.48\linewidth]{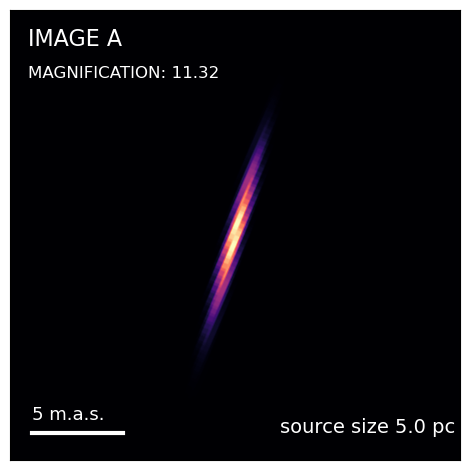} \\
				\includegraphics[trim=0.0cm 1cm 0.0cm 0.0cm,width=0.48\linewidth]{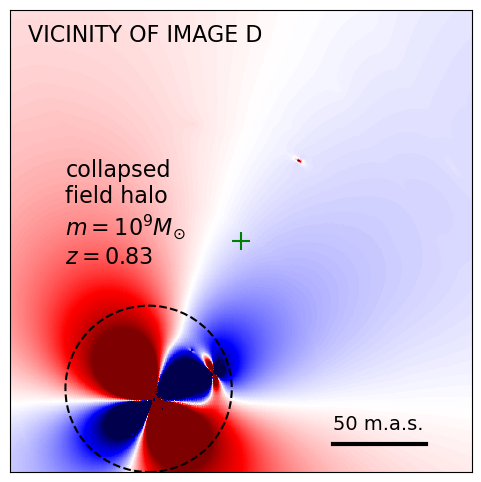} &
				\includegraphics[trim=0.0cm 1cm 0.0cm 0.0cm,width=0.48\linewidth]{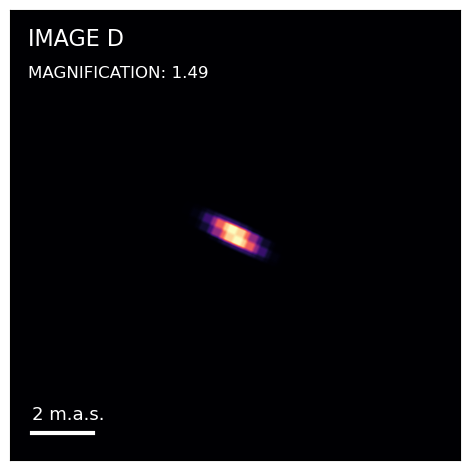}
			\end{tabular}
		\end{minipage}\hfill
		\begin{minipage}[t]{0.32\textwidth}
			\centering
			\includegraphics[width=\linewidth]{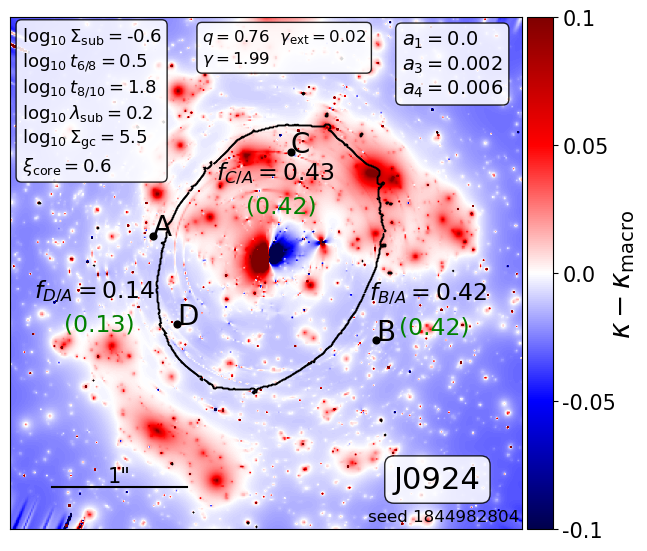}\\[0.3em]
			\begin{tabular}{@{}c@{\hspace{0.5em}}c@{}}
				\includegraphics[trim=0.0cm 1cm 0.0cm 0.0cm,width=0.48\linewidth]{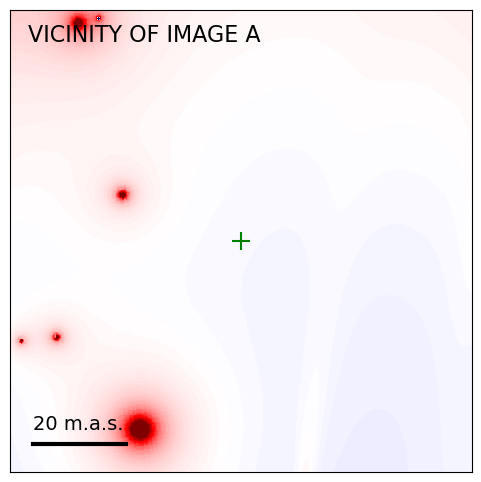} &
				\includegraphics[trim=0.0cm 1cm 0.0cm 0.0cm,width=0.48\linewidth]{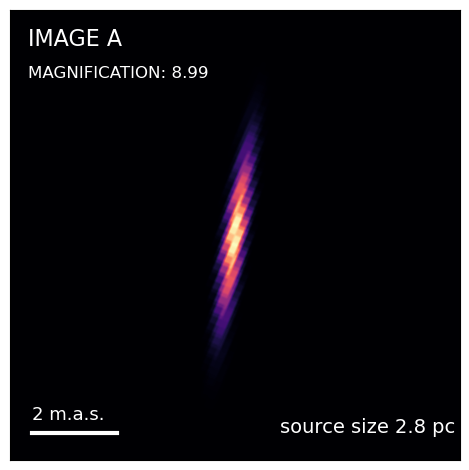} \\
				\includegraphics[trim=0.0cm 1cm 0.0cm 0.0cm,width=0.48\linewidth]{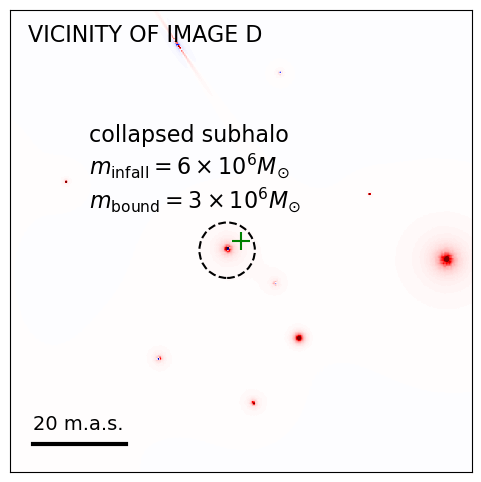} &
				\includegraphics[trim=0.0cm 1cm 0.0cm 0.0cm,width=0.48\linewidth]{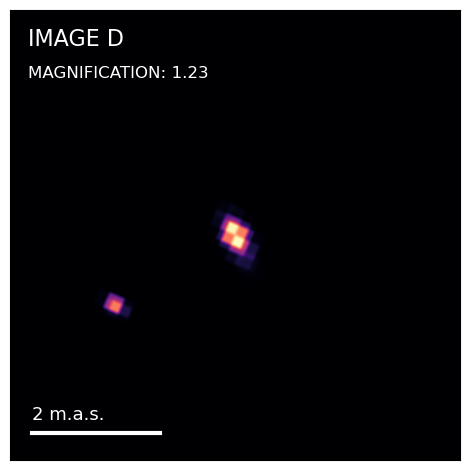}
			\end{tabular}
		\end{minipage}
        \caption{\label{fig:j0924models} Lens models that reproduce the flux ratios of J0924, a system with a strong flux anomaly (demagnification) in image D. We highlight these three particular examples out of millions of lens models generated for this system in our analysis to illustrate different possible explanations for the data. Our forward modeling approach quantifies the relative likelihood of these possible solutions. {\bf{Top panels:}} The substructure convergence map of the lens system. Dark matter and globular cluster parameters are listed in the top left inset, macromodel parameters in the top center inset, and the strengths of $m=1$, $m=3$, and $m=4$ multipole perturbations to the main deflector mass profile are shown in the top right. Image positions are labeled A-D, with model-predicted (black) and measured (green) flux ratios listed alongside each image. {\bf{Lower panels:}} The four panels below each convergence map show zoomed-in regions of the convergence map around images A and D, and the lensed quasar images computed by ray tracing through the lens systems. Globular clusters, subhalos, and line-of-sight halos produce the various small-scale features visible in the convergence maps. Line-of-sight halos appear sheared and distorted in the convergence maps due to coupling between their deflection angles and the main deflector. In the lower panels, we highlight the perturber that most strongly affects the magnification of image D.}
    \end{figure*}
	\begin{figure*}
	 	\centering
	 	\begin{minipage}[t]{0.32\textwidth}
	 		\centering
	 		\includegraphics[width=\linewidth]{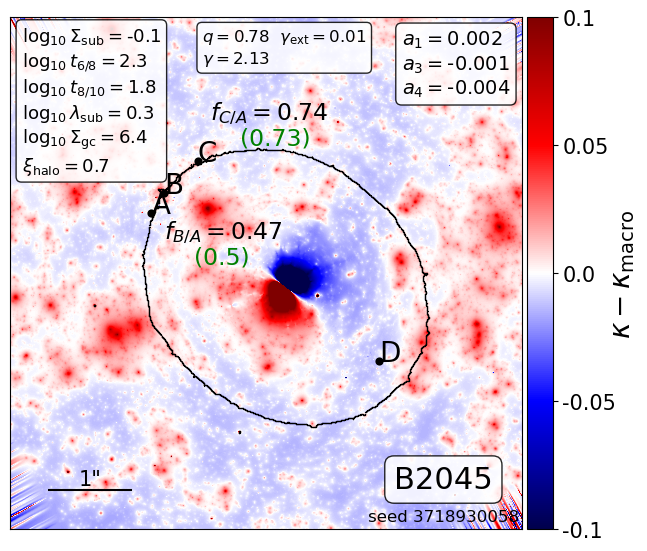}\\[0.3em]
	 		\begin{tabular}{@{}c@{\hspace{0.5em}}c@{}}
	 			\includegraphics[trim=0.0cm 1cm 0.0cm 0.0cm,width=0.48\linewidth]{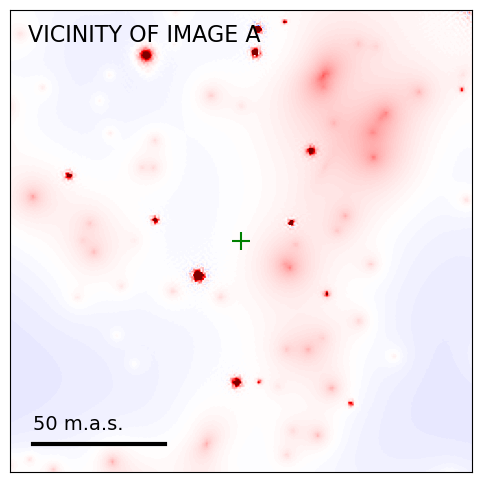} &
	 			\includegraphics[trim=0.0cm 1cm 0.0cm 0.0cm,width=0.48\linewidth]{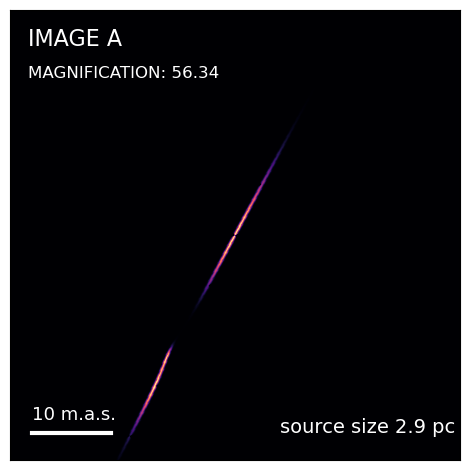} \\
	 			\includegraphics[trim=0.0cm 1cm 0.0cm 0.0cm,width=0.48\linewidth]{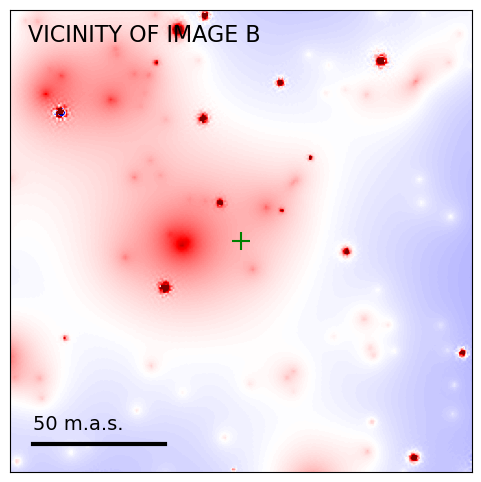} &
	 			\includegraphics[trim=0.0cm 1cm 0.0cm 0.0cm,width=0.48\linewidth]{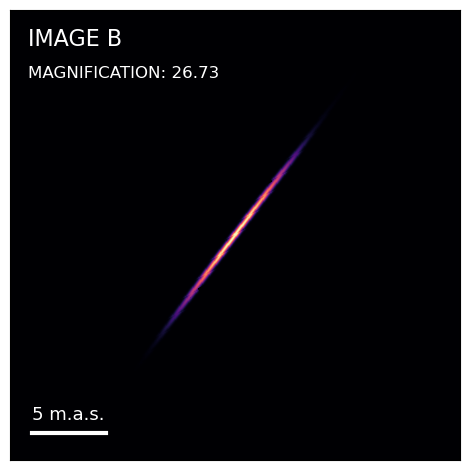}
	 		\end{tabular}
	 	\end{minipage}\hfill
	 	\begin{minipage}[t]{0.32\textwidth}
	 		\centering
	 		\includegraphics[width=\linewidth]{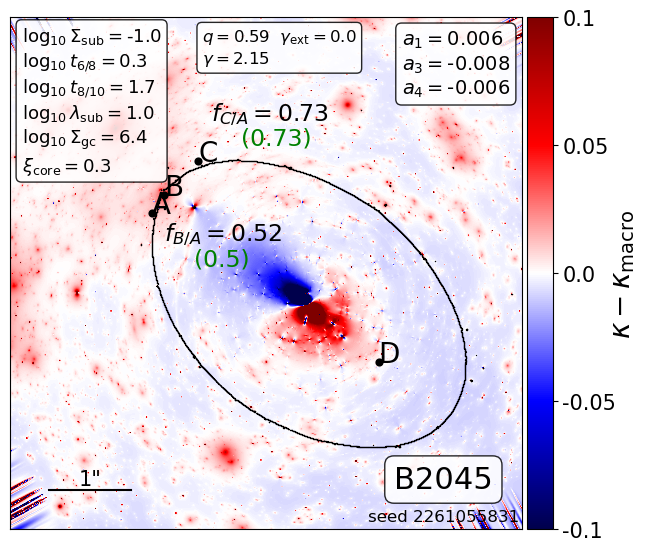}\\[0.3em]
	 		\begin{tabular}{@{}c@{\hspace{0.5em}}c@{}}
	 			\includegraphics[trim=0.0cm 1cm 0.0cm 0.0cm,width=0.48\linewidth]{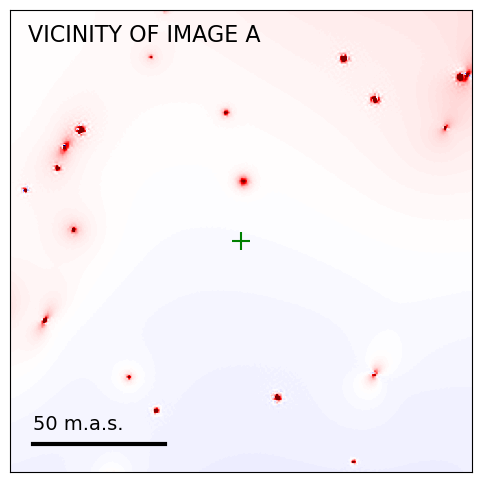} &
	 			\includegraphics[trim=0.0cm 1cm 0.0cm 0.0cm,width=0.48\linewidth]{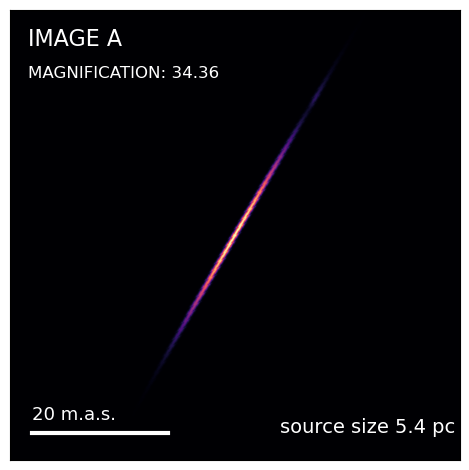} \\
	 			\includegraphics[trim=0.0cm 1cm 0.0cm 0.0cm,width=0.48\linewidth]{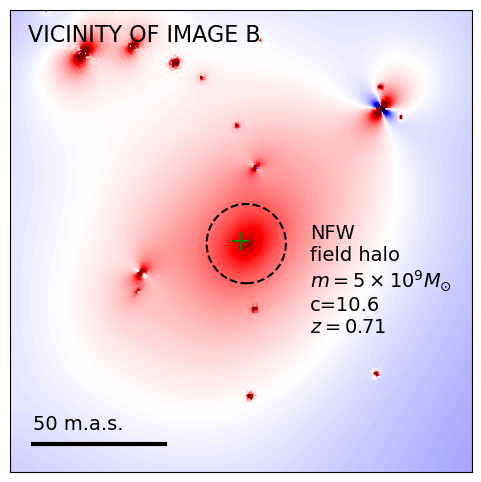} &
	 			\includegraphics[trim=0.0cm 1cm 0.0cm 0.0cm,width=0.48\linewidth]{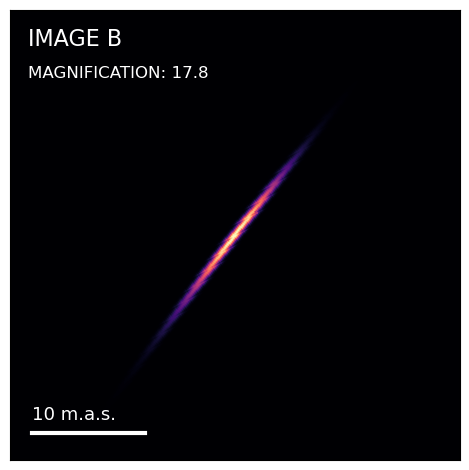}
	 		\end{tabular}
	 	\end{minipage}\hfill
	 	\begin{minipage}[t]{0.32\textwidth}
	 		\centering
	 		\includegraphics[width=\linewidth]{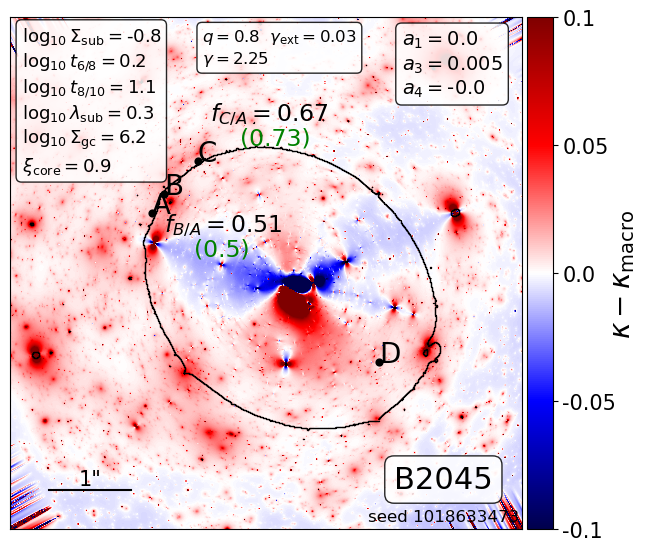}\\[0.3em]
	 		\begin{tabular}{@{}c@{\hspace{0.5em}}c@{}}
	 			\includegraphics[trim=0.0cm 1cm 0.0cm 0.0cm,width=0.48\linewidth]{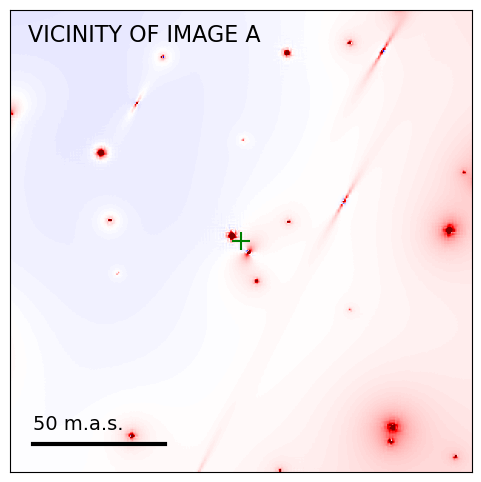} &
	 			\includegraphics[trim=0.0cm 1cm 0.0cm 0.0cm,width=0.48\linewidth]{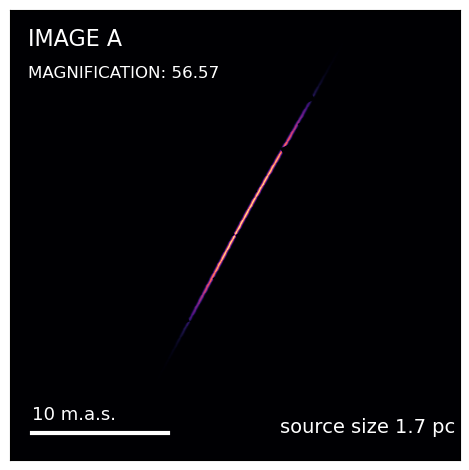} \\
	 			\includegraphics[trim=0.0cm 1cm 0.0cm 0.0cm,width=0.48\linewidth]{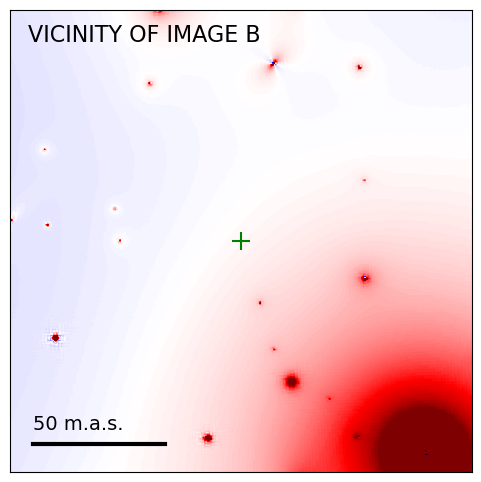} &
	 			\includegraphics[trim=0.0cm 1cm 0.0cm 0.0cm,width=0.48\linewidth]{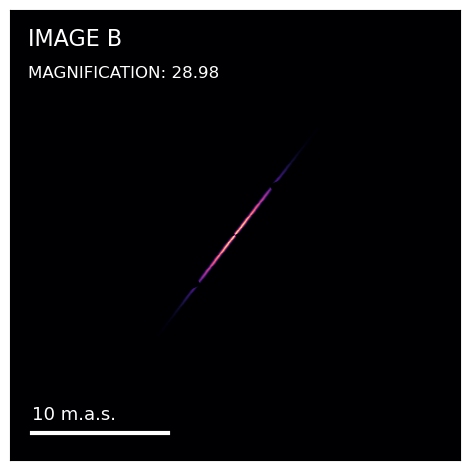}
	 		\end{tabular}
	 	\end{minipage}
	 	\caption{\label{fig:b2045models} Lens models that reproduce the flux ratios of B2045, which has strong flux ratio anomalies in the merging triplet that are also present in radio observations \citep{McKean++07}. The panel layout is the same as Figure \ref{fig:j0924models}. In the central panel we highlight a lens model that includes a foreground line of sight halo with a mass of $5 \times 10^9 M_\odot$ that causes a strong flux perturbation to image B. The lens systems shown on the left and right reproduce the observed flux ratios through the collective effect of several low-mass halos and globular clusters. The lens model on the far left includes a globular cluster that splits image A. }
	 \end{figure*}
	 \begin{figure*}
	\centering
	\begin{minipage}[t]{0.32\textwidth}
		\centering
		\includegraphics[width=\linewidth]{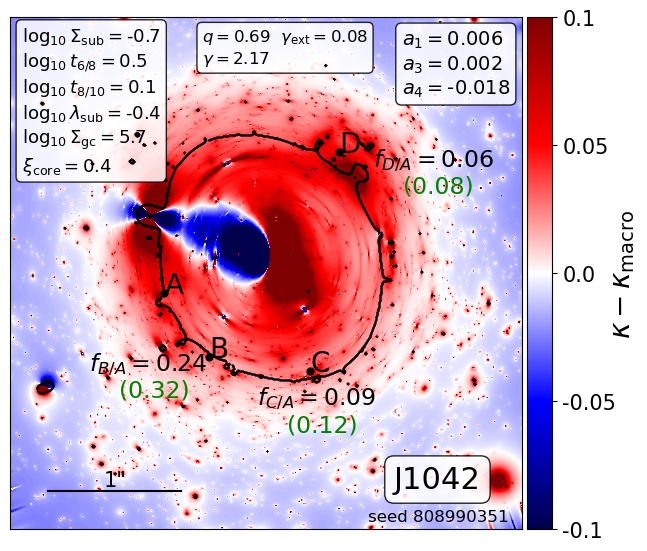}\\[0.3em]
		\begin{tabular}{@{}c@{\hspace{0.5em}}c@{}}
			\includegraphics[trim=0.0cm 1cm 0.0cm 0.0cm,width=0.48\linewidth]{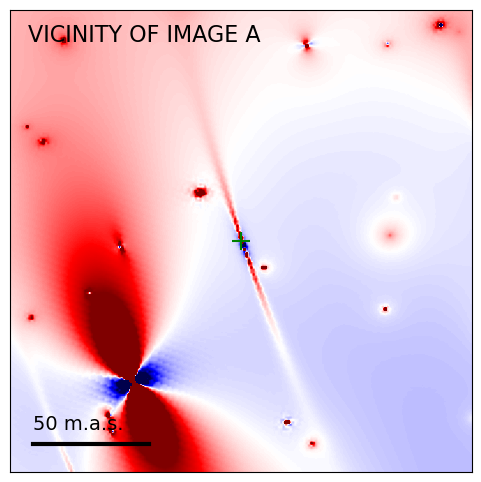} &
			\includegraphics[trim=0.0cm 1cm 0.0cm 0.0cm,width=0.48\linewidth]{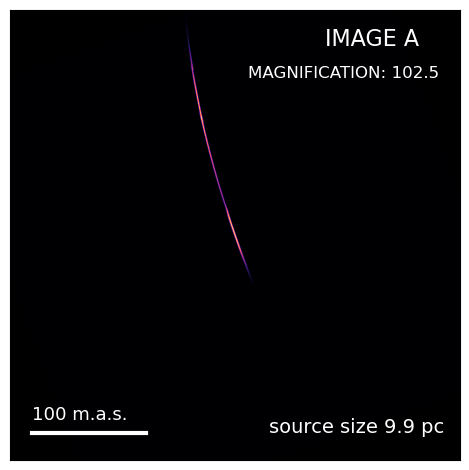} \\
			\includegraphics[trim=0.0cm 1cm 0.0cm 0.0cm,width=0.48\linewidth]{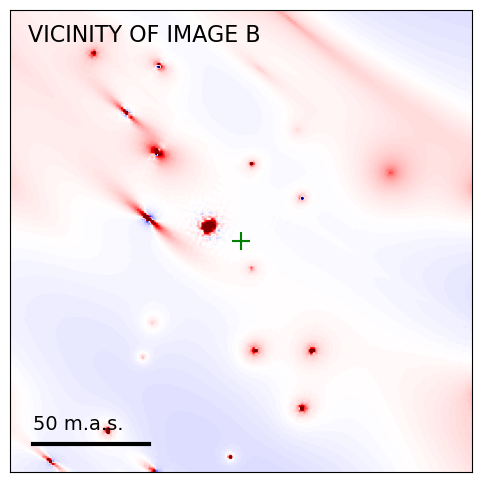} &
			\includegraphics[trim=0.0cm 1cm 0.0cm 0.0cm,width=0.48\linewidth]{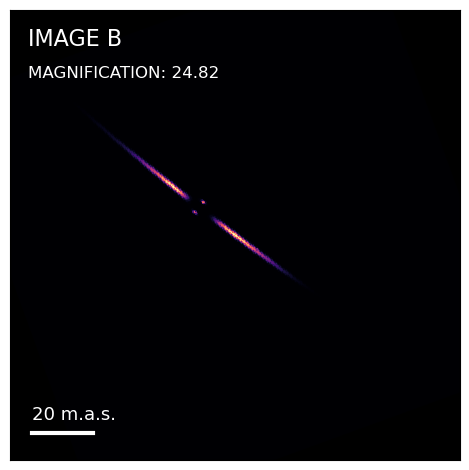}
		\end{tabular}
	\end{minipage}\hfill
	\begin{minipage}[t]{0.32\textwidth}
		\centering
		\includegraphics[width=\linewidth]{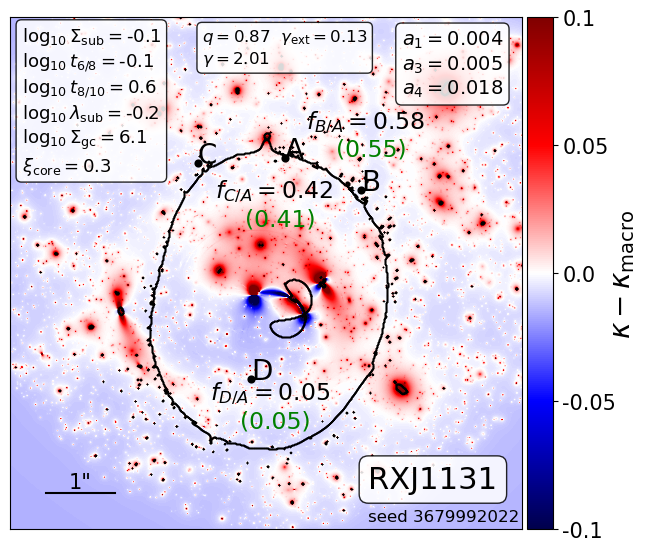}\\[0.3em]
		\begin{tabular}{@{}c@{\hspace{0.5em}}c@{}}
			\includegraphics[trim=0.0cm 1cm 0.0cm 0.0cm,width=0.48\linewidth]{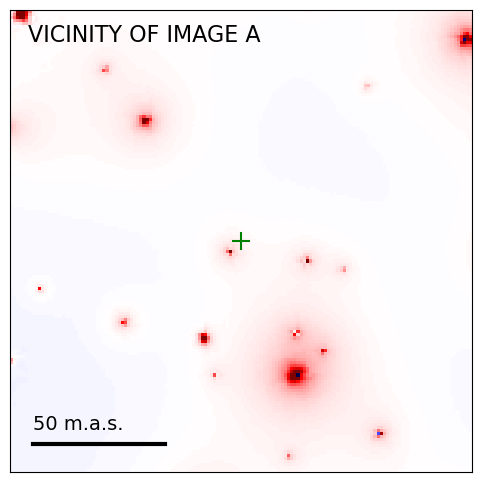} &
			\includegraphics[trim=0.0cm 1cm 0.0cm 0.0cm,width=0.48\linewidth]{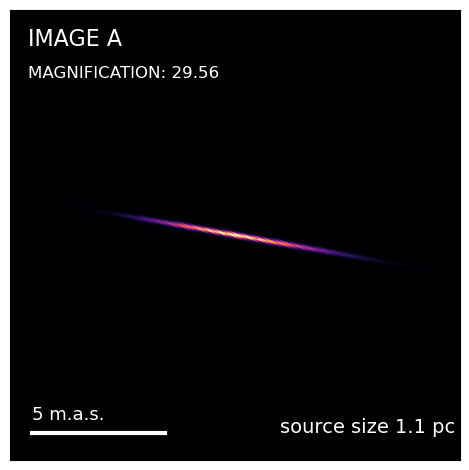} \\
			\includegraphics[trim=0.0cm 1cm 0.0cm 0.0cm,width=0.48\linewidth]{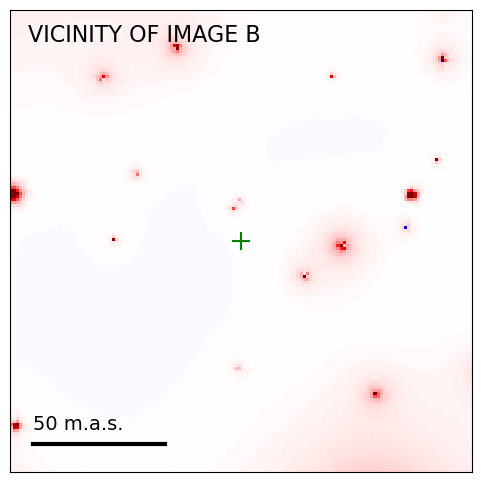} &
			\includegraphics[trim=0.0cm 1cm 0.0cm 0.0cm,width=0.48\linewidth]{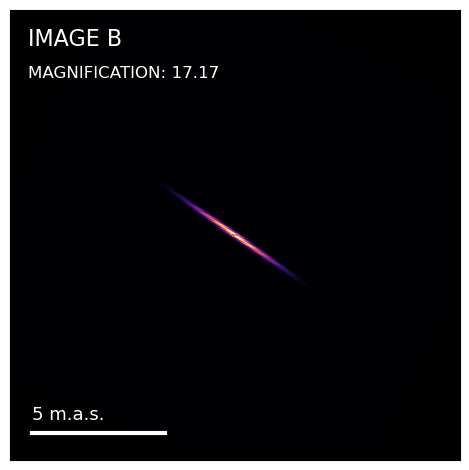}
		\end{tabular}
	\end{minipage}\hfill
	\begin{minipage}[t]{0.32\textwidth}
		\centering
		\includegraphics[width=\linewidth]{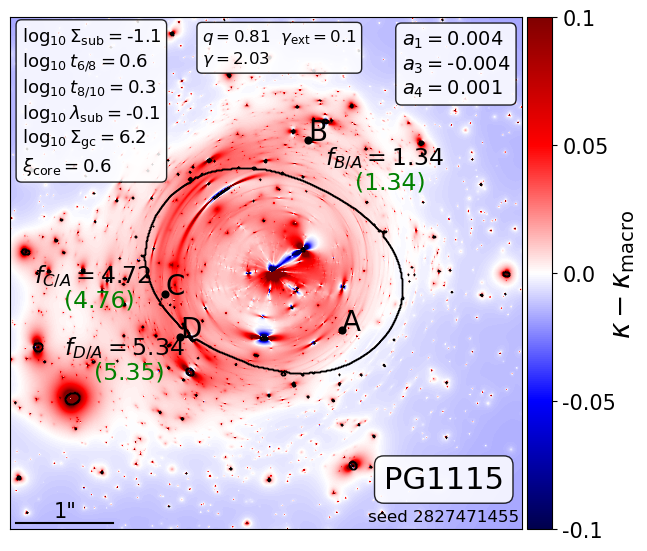}\\[0.3em]
		\begin{tabular}{@{}c@{\hspace{0.5em}}c@{}}
			\includegraphics[trim=0.0cm 1cm 0.0cm 0.0cm,width=0.48\linewidth]{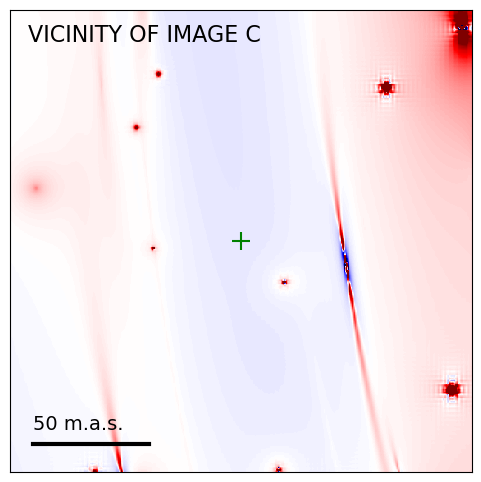} &
			\includegraphics[trim=0.0cm 1cm 0.0cm 0.0cm,width=0.48\linewidth]{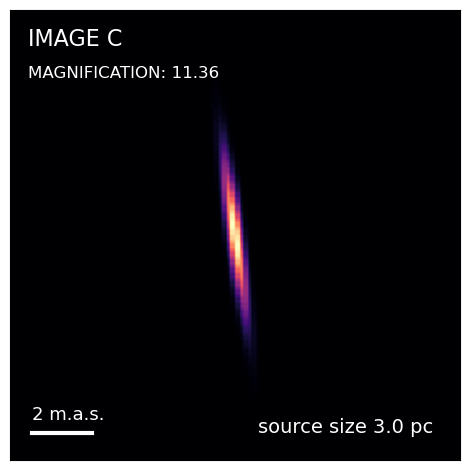} \\
			\includegraphics[trim=0.0cm 1cm 0.0cm 0.0cm,width=0.48\linewidth]{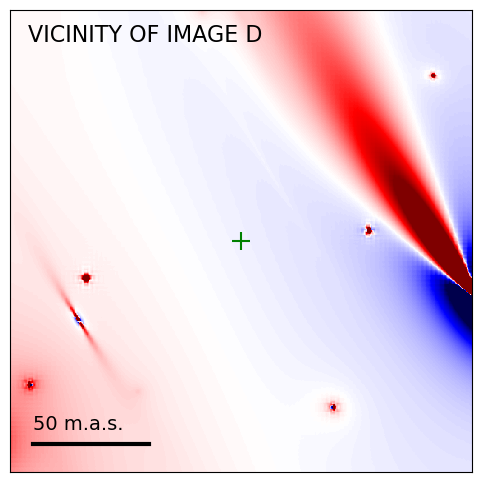} &
			\includegraphics[trim=0.0cm 1cm 0.0cm 0.0cm,width=0.48\linewidth]{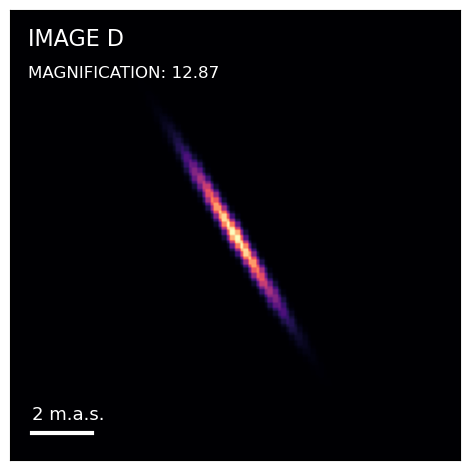}
		\end{tabular}
	\end{minipage}
	\caption{\label{fig:additionalmodels} Lens models that reproduce the flux ratios of J1042 (left), RXJ1131 (center), and PG1115 (right). The panel layout is the same as Figure \ref{fig:j0924models}. For J1042, a $ 3 \times 10^6 M_{\odot}$ collapsed field halo, and a nearby $8 \times 10^{8} M_{\odot}$ collapsed field halo, both behind the main deflector, affect the magnification of image A, while a $10^{6} M_{\odot}$ globular cluster makes a direct hit on image B and splits it into four separate components. The model for RXJ1131 includes a strong $m=4$ multipole perturbation in combination with small-scale structure near images A and B. We note that this system also exhibits an anomalous $C/A$ flux ratio that can be partially explained through a large $a_4$ term. For PG1115, the lower panels zoom in on the merging pair of images C and D, which are the most sensitive to perturbations. This particular realization has $>90 \%$ of subhalos and $>50 \%$ of field halos core collapsed, but since none of these objects appears near a lensed image the flux ratios are consistent with the predictions of a smooth lens model. The likelihood function for PG1115 disfavors scenarios in which the majority of halos core collapse (see lower-right panel Figure \ref{fig:fluxratios1}), but these solutions are possible nonetheless.}
\end{figure*}
     
	\subsubsection{Individual lens reconstructions}
    \label{ssec:lensreconstructions}
	Our analysis accounts for the collective effect of many potential sources of small-scale perturbation, including dark matter substructure, globular clusters, and angular structure in the main deflector mass profile. In this section we show reconstructions of individual lens systems that fit the image positions and flux ratios. We highlight several examples of lens systems that fit the anomalous flux ratio in J0924, where the signal often (but not always) comes from one dominant perturber. We also discuss lens model solutions derived for other systems, such as J1042, RXJ1131, and B2045, in which the collective effects of many perturbers, in combination with angular structure in the macromodel, reproduce the observations. We also present an example lens model for PG1115, which has flux ratios consistent with a smooth lens model, in which virtually all subhalos and many field halos core collapse. We emphasize that any single lens system can be explained by a variety of dark matter models, and that the constraining power derives from the full sample. We address the topic of lensing degeneracies at the end of this subsection. 
	
	Figure \ref{fig:j0924models} shows three example lens systems recovered in our analysis pipeline that fit the highly anomalous D/A flux ratio in J0924. The upper three panels show the effective multi-plane convergence in substructure for the lens system, with image positions and flux ratios labeled on top of the convergence maps. Panels on the top left, top center, and top right list the dark matter hyper-parameters, macromodel parameters (axis ratio, external shear strength, and logarithmic profile slope), and the strength of the multipole perturbations, respectively. Below each convergence map we show zoomed-in region of the convergence map near images A and D. The lensed image lies at the center of each zoomed-in convergence map and is marked with a green cross. The lower right panels show the model-predicted lensed images and their magnifications. We emphasize that the observed lensed images are unresolved, and we can only measure the total integrated flux, but we can generate the lensed images as they would appear without the effects of the PSF and with exquisite angular resolution. We note that constraints from the lensed arcs are also folded into our analysis pipeline, but the focus of this section is discussing how various sources of small-scale perturbation reproduce the flux ratios. 
	
	Each possible lens model for J0924 shown in Figure \ref{fig:j0924models} reproduces the measured image positions and flux ratios through a different mechanism. The model shown on the far left has a surface mass density in globular clusters more than 10 times higher than expected based on the globular cluster populations around nearby ellipticals. One unusually massive cluster ($m_{\rm{gc}} = 1.3 \times 10^7 M_{\odot}$) makes a direct hit on image D and causes the required demagnification. In the central panel, a core collapsed field halo at $z=0.83$ with a mass of $10^9 M_{\odot}$ causes the perturbation\footnote{We can infer that the feature in the convergence map near image D corresponds to a line-of-sight halo because of its distorted appearance in the convergence map.}. In the far right panel, a core collapsed subhalo with an infall mass of $6 \times 10^6 M_{\odot}$ appears near image D. The collapsed subhalo is dense enough to split image D and form a faint counter image.  
	
	Figure \ref{fig:b2045models} shows lens model reconstructions for B2045, a system which exhibits anomalous flux ratios between images in the merging triplet. We note that this anomalous feature is also present in radio data \citep{McKean++07} which, like the warm dust emission, should not experience stellar microlensing. Without a massive perturber near image B, the flux from images A--C becomes blended into a single elongated arc and the flux ratios stay close to unity. Lens models that reproduce the data in our modeling pipeline sometimes have a relatively massive $\sim 10^9 M_{\odot}$ line of sight perturber or subhalo that demagnifies image B, in addition to small-scale structure near images A and C. The likelihood function for this lens favors SIDM models by $\sim 1.5:1$. 
	
	Figure~\ref{fig:additionalmodels} shows three additional reconstructions for lens models of J1042, RXJ1131, and PG1115. In the case of J1042, lens models that fit the data often have one or several objects causing a magnification of image A. For the particular example shown in the left panel of Figure~\ref{fig:additionalmodels}, image A is affected by two dominant perturbers, which are core collapsed in this realization: a $3\times10^6 M_{\odot}$ field halo making a direct hit, and an $8 \times 10^8 M_{\odot}$ field halo nearby along the line of sight. Image B experiences a direct hit by a globular cluster, which introduces its own critical curve and breaks the image into four separate components. These multiple sources of perturbation, which also include a large $a_4$ multipole moment, reproduce the $f_{B/A}$ flux ratio to within $2.3 \sigma$, given the $\sim 11\%$ uncertainty on the $f_{B/A}$ ratio \citep{Keeley++25}. Note that we have written the flux ratios for J1042 in Figure~\ref{fig:fluxratios2} relative to image B to more clearly show that image A is likely the one being magnified by small-scale structure. 
     \begin{table*}
	\setlength{\tabcolsep}{8pt}
	\caption{\label{tab:priors} Importance sampling weights applied to the dark matter hyper-parameters of Table \ref{tab:tableqsub}, which lead to joint inferences on $t_{6/8}$ and $t_{8/10}$ shown in Figures \ref{fig:jointpdfbestA}-\ref{fig:marginalpdfbestGH}. Each prior is Gaussian, with $\mathcal{N}\left(\mu, \sigma\right)$ denoting the mean and standard deviation.}
	\begin{tabular}{cccc}
		\hline
		Hyper-parameter & Importance weight & Interpretation & References\\
		\hline
		$\Sigma_{\rm{sub}} \ \left[\rm{kpc^{-2}}\right]$ & $\log_{10} \mathcal{N}\left(-1.3, 0.2\right)$ & subhalo abundance suppressed by & \citep{Nadler++20b,Nadler++25a} \\
		& & $50\%$ relative to CDM $N$-body & \vspace{0.05in} \\
		& $\log_{10} \mathcal{N}\left(-1.0, 0.2\right)$ & subhalo abundance matches & \citep{Nadler++25b,Gannon++25,Du++25} \\
		& & CDM $N$-body predictions & \vspace{0.05in} \\
		& $\log_{10} \mathcal{N}\left(-0.8, 0.2\right)$ & subhalo abundance matches & \citep{Nadler++25b,Gannon++25,Du++25} \\
		& & CDM from $\tt{galacticus}$ & \\ \\
		$\lambda_{\rm{sub}}$ & $\log_{10} \mathcal{N}\left(0.5, 0.3\right)$ & subhalo core collapse accelerated &\citep{Nishikawa++20,Sameie++20,Turner++21,Zeng++22,ShahAdhikari24}\\
		& & $3\times$ relative to field halos &\vspace{0.05in}\\
		& $\log_{10} \mathcal{N}\left(0.7, 0.3\right)$ & subhalo core collapse accelerated &\citep{Nishikawa++20,Sameie++20,Turner++21,Zeng++22,ShahAdhikari24}\\
		& & $5\times$ relative to field halos & \\ \\
		$\Sigma_{\rm{gc}} \ \left[M_{\odot} \ \rm{kpc^{-2}}\right]$ & $\log_{10} \mathcal{N}\left(5.6, 0.3\right)$ & expected GC abundance in & see Section \ref{ssec:globularclusters} \\
		& & lens host halos & \\ \\
		$\xi_{\rm{core}}$ & $\mathcal{N}\left(0.0, 0.3\right)$ & deeply collapsed halos have & \citep{Gurian++25} \\
		& & $\rho_{\rm{central}} \propto r^{-2.6}$ outside core & see Section \ref{ssec:collapsedsidm} and \vspace{0.05in} \\
		& $\mathcal{N}\left(1.0, 0.3\right)$ & deeply collapsed region treated & Equation \ref{eqn:rhosidm2} \\
		& & as a point mass & \\ \\
		$t_{6/8}, \ t_{8/10} \ \left[\rm{Gyr}\right]$ & $\log_{10} \mathcal{N}\left(2.30, 0.25\right)$ & collapse timescales long compared to& \\
		& & a Hubble time (CDM limit, Figure~\ref{fig:marginalpdfcdm}) & \\ \\
		\hline
	\end{tabular}
\end{table*}

    We emphasize that our modeling pipeline also finds lens model solutions that do not invoke significant amounts of dark substructure to reproduce the measured flux ratios. \citet{Gilman++25a} present an example in the case of WFI2033 that reproduces the data with a large $a_4$ multipole amplitude. The center panel of Figure~\ref{fig:additionalmodels} shows a similar example, in this case of the system RXJ1131, where a large $a_4$ moment, in combination with small-scale structure near lensed images, reproduce the observations. As discussed in relation to Figure~\ref{fig:fluxratios2}, the likelihood function for RXJ1131 favors SIDM over CDM, indicating that models with collapsed halos match the data more frequently, or with less fine tuning, than lens models in which a particular combination of multipole moments reproduces the observations. 
	
	Finally, the right panel of Figure~\ref{fig:additionalmodels} shows a reconstruction of the lens system PG1115. As shown in Figure \ref{fig:fluxratios1}, this system has flux ratios consistent with the predictions of a smooth lens model, and disfavors models with an abundance of collapsed halos. However, the particular lens model depicted in Figure~\ref{fig:additionalmodels} shows an SIDM realization in which most subhalos and significant fraction of field halos core collapse. We can explain the properties of PG1115 with these extreme SIDM scenarios, and we emphasize that an individual lens system may be reproduced by a variety of dark matter models. A population-level inference on a sample of lenses is required to derive the meaningful constraints we present in the next section.
	
	The discussion and illustrations in this section connect to the topic of lensing degeneracies in the context of multiply-imaged quasars and gravitational imaging. These discussions are typically framed around two competing explanations of the data, for example, an explanation that invokes dark matter substructure and another that invokes angular structure in the main deflector mass profile. Population-level inferences, such as the analysis presented in this work, provide a principled way to distinguish between various competing explanations of the data. By generating millions of lens models similar to the ones shown in Figures~\ref{fig:j0924models}-\ref{fig:additionalmodels}, we assign the various competing hypotheses an appropriate statistical weight. In previous work, we demonstrated that our approach can distinguish models with only angular structure in the main deflector mass profile from lens models that include perturbations by halos \citep[see Appendix B in ][for more a discussion of this topic]{Gilman++24}. The results presented in the next section do not make definitive statements about which physical process gave rise to our dataset. Instead, we combine the independent pieces of information from each of the 29 lensed quasars in our sample to quantify which dark matter model is most likely to have explained our data, while allowing for the possibility that other sources of small-scale perturbation, such as angular structure, give rise to the observed flux ratios, either independently of or in combination with dark matter substructure. 

	\subsection{Bayesian inference on the core collapse timescales}
	\label{ssec:structureinf}
	This section presents the inference on the core collapse timescales $\boldsymbol{t} = \left(t_{6/8}, t_{8/10}\right)$ derived from the 29 lensed quasars in our sample. Given the complexity of structure formation in SIDM, and in particular the properties of subhalos and how environmental processes affect core collapse in these objects, we will present results conditioned on different sets of assumptions for subhalo abundance (determined by $\Sigma_{\rm{sub}}$, Equation \ref{eqn:subhalomfunc}) and the subhalo collapse timescale relative to field halos (determined by $\lambda_{\rm{sub}}$, Equation \ref{eqn:timesub}). Our results will also depend on the internal structure of deeply core collapsed objects (determined by $\xi_{\rm{core}}$, Equation \ref{eqn:yeffect}), and the surface mass density of globular clusters around strong lenses viewed in projection between $6-12 \ \rm{kpc}$ (determined by $\Sigma_{\rm{gc}}$, Equation \ref{eqn:gcmf}). 
	
	We begin in Section \ref{ssec:statframework} by presenting a statistical framework with which to interpret our results. As our goal involves model comparison between CDM and SIDM, we present our main results in terms of Bayes factors, which encode more information than a marginal likelihood constraint on an individual collapse timescale $t_{6/8}$ or $t_{8/10}$. Section \ref{ssec:results} presents the results of our analysis under different sets of modeling assumptions for subhalo and globular cluster populations. 
	
	\subsubsection{Statistical framework to interpret results}
	\label{ssec:statframework}
	To quantify how our inference on SIDM properties depends on the various sources of theoretical uncertainty discussed throughout this paper, we compute the marginal likelihood of the core collapse timescales conditioned on a set of importance weights, $w\left(\boldsymbol{x}\right)$: 
	\begin{equation}
			\label{eqn:effectivelikelihood}
		\mathcal{L}_{w}\left({\data} | {\boldsymbol{t}}\right) = \frac{\int w\left(\boldsymbol{x}\right) \mathcal{L}\left(\data | \boldsymbol{t}, \boldsymbol{x}\right) \pi_{\rm{samp}}\left(\boldsymbol{x}\right) d\boldsymbol{x}}{\int w\left(\boldsymbol{x}\right) \pi_{\rm{samp}}\left(\boldsymbol{x}\right) d\boldsymbol{x}}.
	\end{equation}
	Here $\boldsymbol{x} =  \left(\Sigma_{\rm{sub}}, \lambda_{\rm{sub}}, \Sigma_{\rm{gc}}, \xi_{\rm{core}} \right)$, $\pi_{\rm{samp}}\left(\boldsymbol{x}\right)$ is the sampling distribution on $\boldsymbol{x}$ (see Table \ref{tab:tableqsub}), and $w\left(\boldsymbol{x}\right)$ represent the importance weights on $\boldsymbol{x}$. Given $\mathcal{L}_{w}\left({\data} | {\boldsymbol{t}}\right)$, we compute an evidence 
	\begin{equation}
		\label{eqn:regionevidence}
		p\left({\data} | M, w\right) = \frac{\int_{\boldsymbol{t} \in M} \mathcal{L}_{w}\left({\data} | {\boldsymbol{t}}\right) \pi\left(\boldsymbol{t}\right) d\boldsymbol{t}}{\int_{\boldsymbol{t} \in M} \pi\left(\boldsymbol{t}\right) d\boldsymbol{t}},
	\end{equation}
	which is the average of the marginal likelihood over a region of parameter space $M$. Here $M$ defines a region of parameter space that corresponds to a particular dark matter model. We define the region of parameter space consistent with CDM as the region of parameter space where $t_{6/8} > 80 \ \rm{Gyr}$ and $t_{8/10} > 80 \ \rm{Gyr}$. Throughout this section, we highlight this area of parameter space in triangle plots with a shaded blue square. 
	\begin{figure*}
		\includegraphics[width=0.9\textwidth]{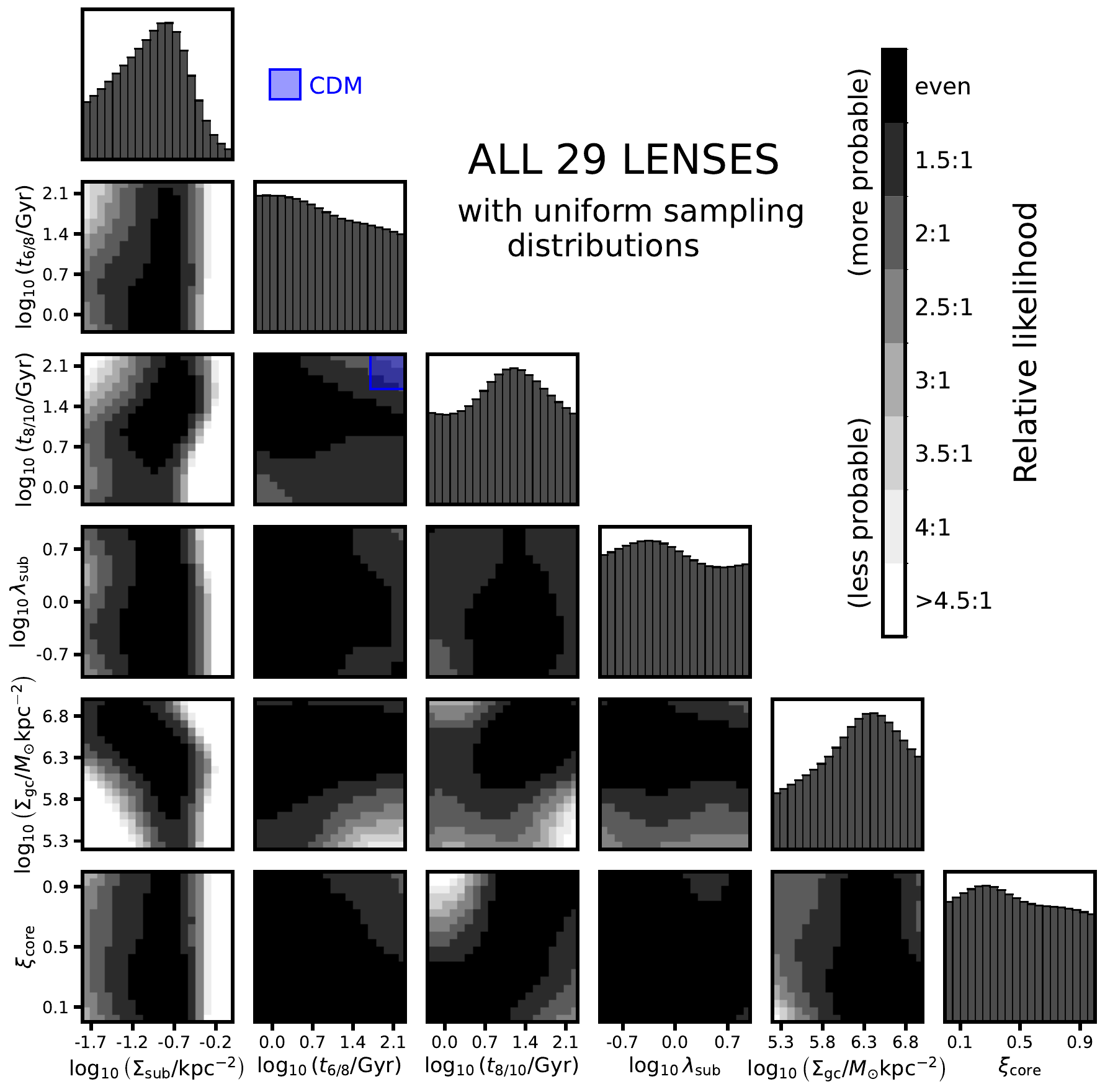}
		\caption{\label{fig:jointpdfuniform} The joint likelihood function $\mathcal{L}\left(\data | \qsub \right)$ inferred from the 29 lensed quasars in our sample. The colorscale of each panel shows the likelihood relative to the most probable point in each 2D projection of the parameter space. Here we assume a uniform sampling distribution on the dark matter parameters, meaning we draw their values from the distributions summarized in Table \ref{tab:tableqsub} and do not make any further assumption regarding subhalo abundance, the rate of subhalo core collapse, the abundance of globular clusters, or the density profile of deeply collapsed objects. We place a blue box over the region of parameter space with $t_{6/8}>80 \ \rm{Gyr}$ and $t_{8/10}> 80 \ \rm{Gyr}$. We associate this region with CDM because core collapse timescales much longer than a Hubble time result in a negligible fraction of core collapsed halos.}
	\end{figure*}
	\begin{figure*}
		\includegraphics[width=0.9\textwidth]{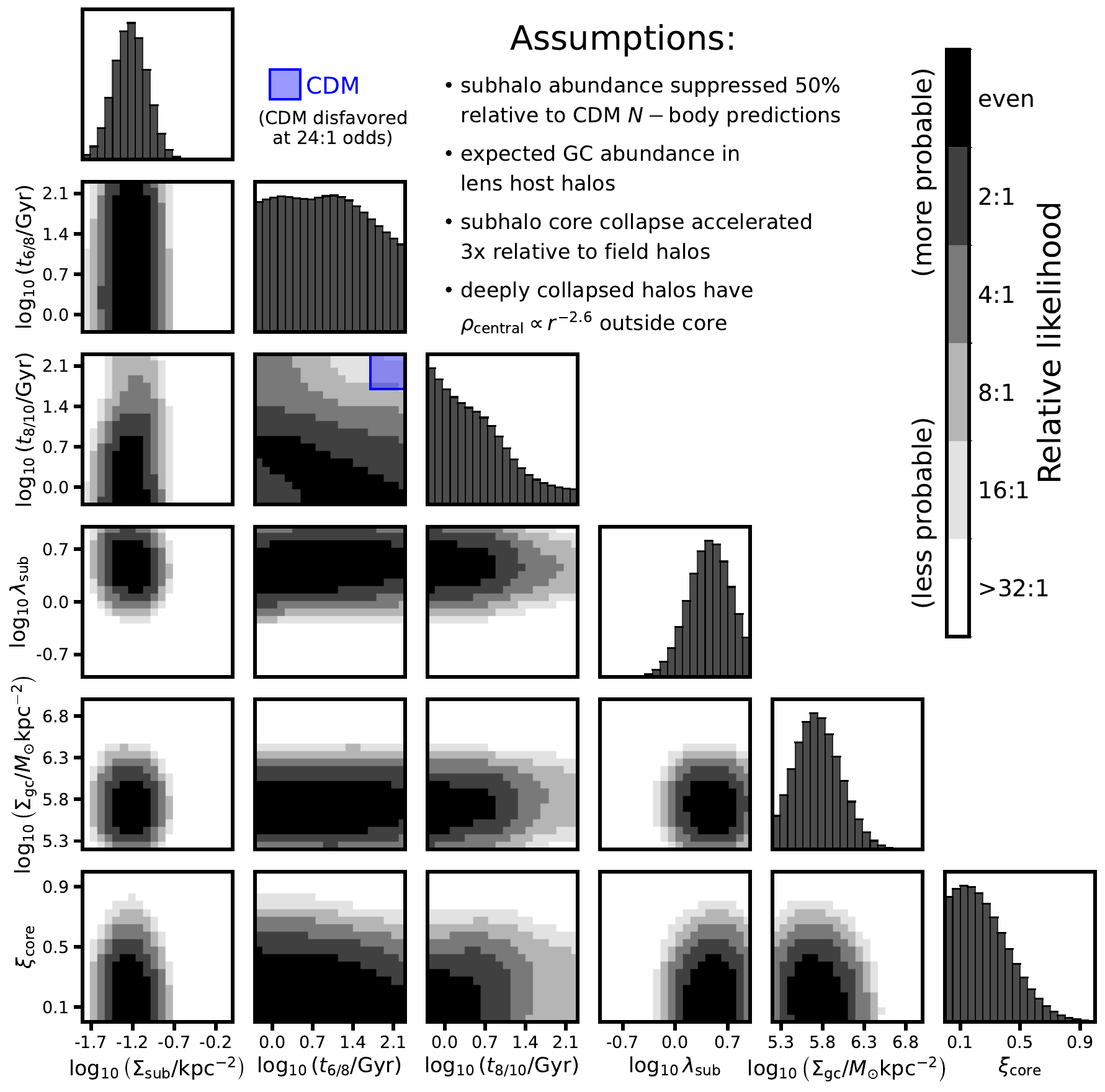}
		\caption{\label{fig:jointpdfbestA} The joint likelihood function $\mathcal{L}_{w}\left(\data | \qsub \right)$ (see Equation \ref{eqn:effectivelikelihood}) with importance weights that enforce the assumptions related to SIDM structure formation and globular cluster abundance listed above the figure. The importance weights are summarized in Table \ref{tab:priors}. As discussed in Section \ref{ssec:structureinf}, the importance weights reflect our best current understanding of SIDM structure formation, in which subhalo abundance is suppressed by $50 \%$ relative to CDM predictions, and core collapse is accelerated in subhalos by tidal stripping.}
	\end{figure*}
	\begin{figure*}
		\includegraphics[width=0.9\textwidth]{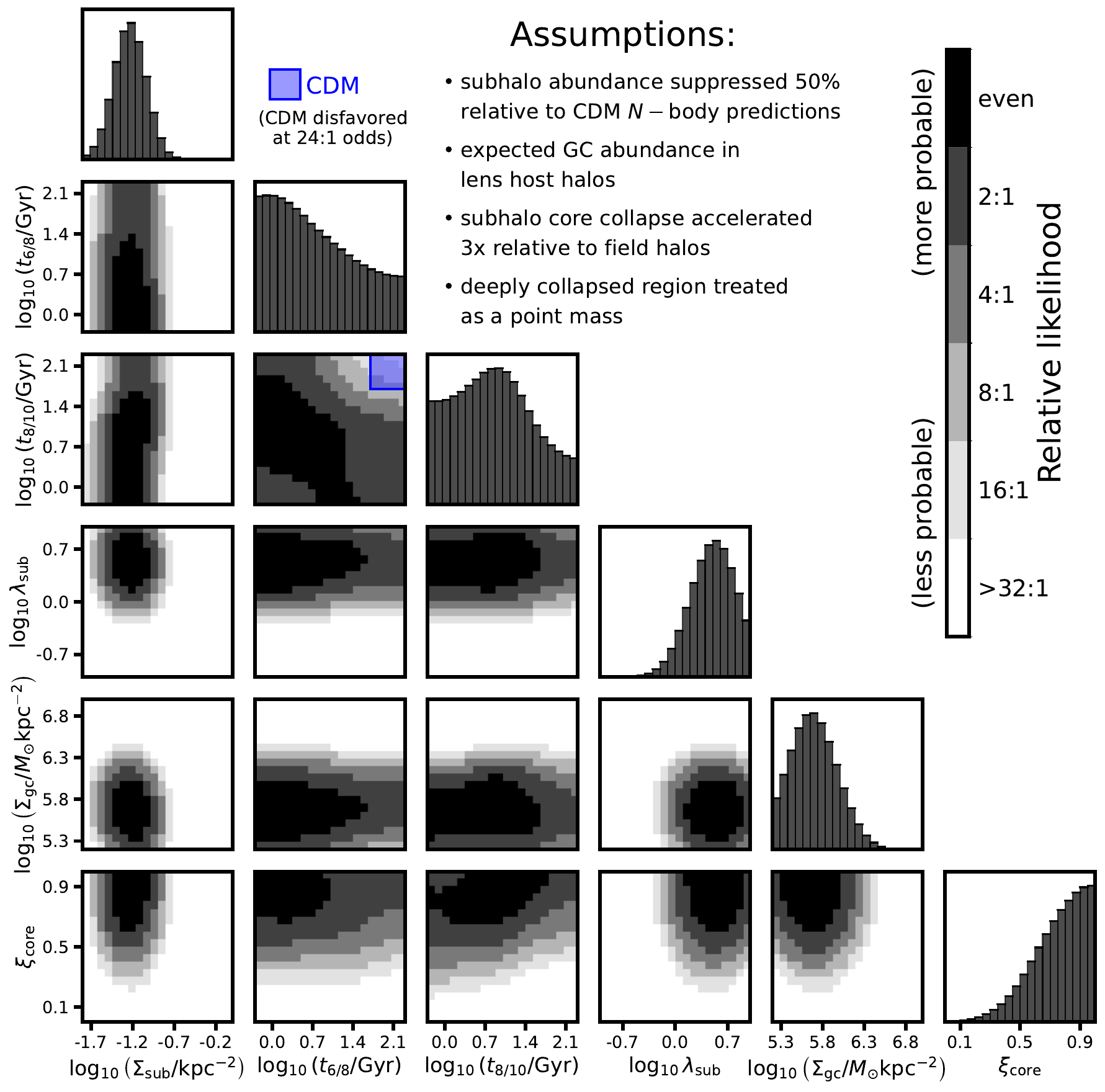}
		\caption{\label{fig:jointpdfbestB} The same as Figure \ref{fig:jointpdfbestA}, but we assume deeply collapsed halos become denser than the objects in the simulations presented by \citet{Gurian++25}. We use importance weights on $\xi_{\rm{core}} \sim \mathcal{N}\left(1,0.3\right)$, selecting realizations in which deeply collapsed halos produce a lensing signal comparable to a point mass inside of an NFW envelope (see Section \ref{ssec:collapsedsidm} and Equation \ref{eqn:rhosidm2}). Table \ref{tab:priors} summarizes the importance sampling distributions noted above the figure.}
	\end{figure*}
	\begin{figure*}
		\includegraphics[width=0.48\textwidth]{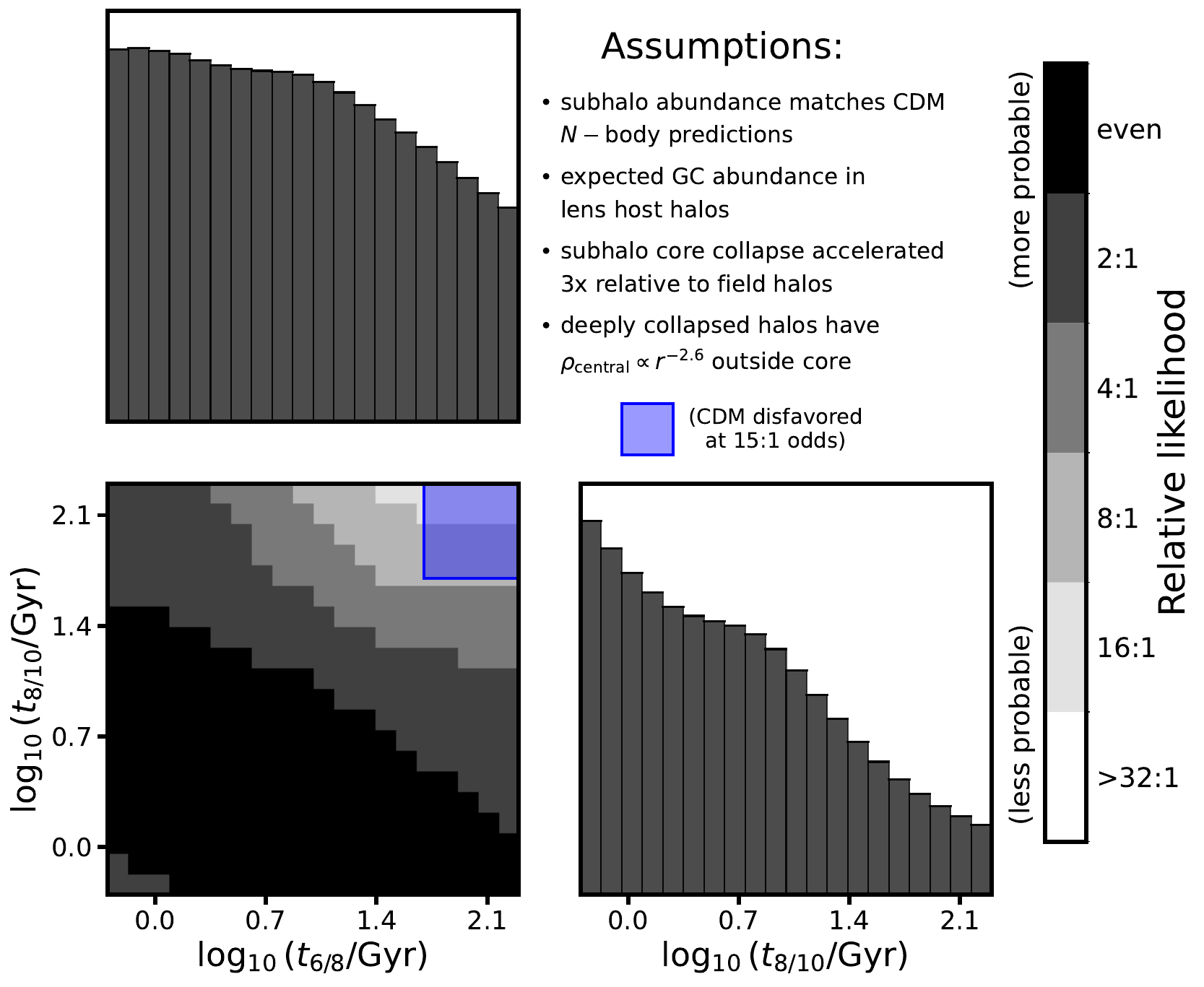}
		\includegraphics[width=0.48\textwidth]{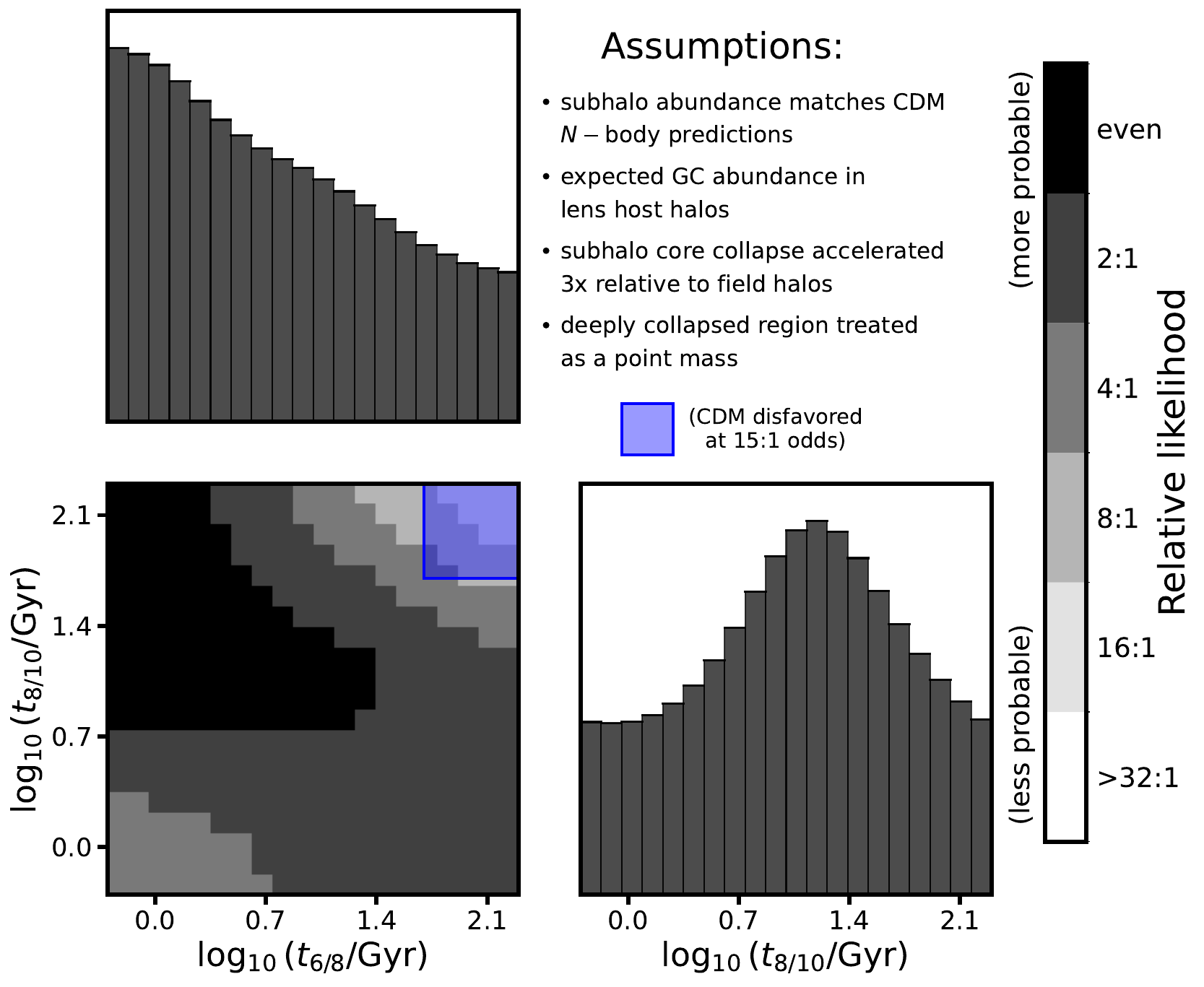}
		\caption{\label{fig:marginalpdfbestCD} The same as Figures \ref{fig:jointpdfbestA} and \ref{fig:jointpdfbestB}, but we project out parameters constrained by importance sampling weights and show only the $t_{6/8}$ and $t_{8/10}$ parameter space. As stated above the figure, we assume subhalo abundance in SIDM is the same as subhalo abundance predicted by CDM $N$-body simulations \citep{Du++25,Gannon++25}, corresponding to $\Sigma_{\rm{sub}}\sim 0.1 \ \rm{kpc^{-2}}$. The difference between the left and right marginal likelihoods is the assumed density profile of deeply collapsed halos controlled by $\xi_{\rm{core}}$. Table \ref{tab:priors} summarizes the importance sampling distributions noted above the figure.}.
	\end{figure*}
	\begin{figure*}
		\includegraphics[width=0.48\textwidth]{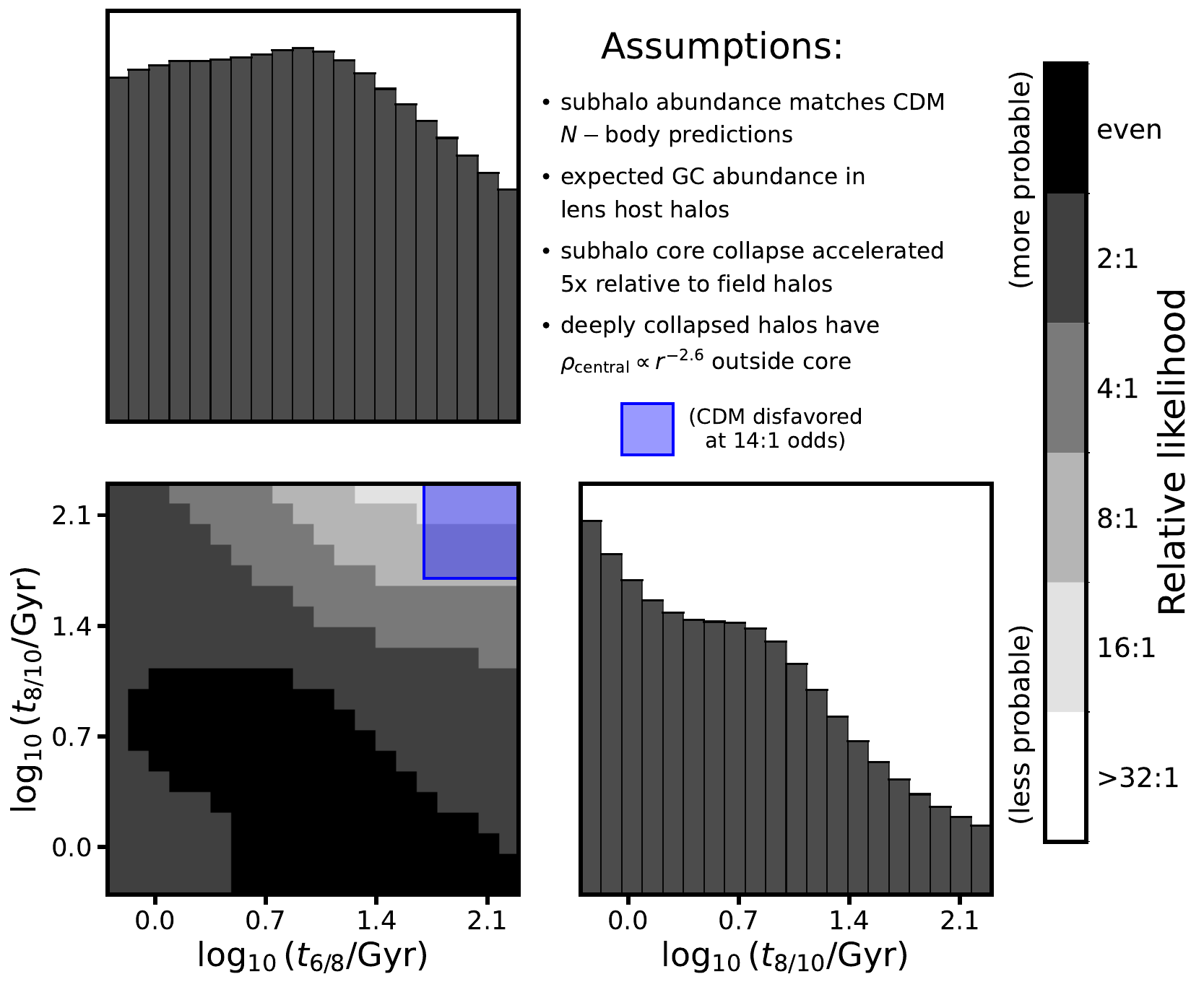}
		\includegraphics[width=0.48\textwidth]{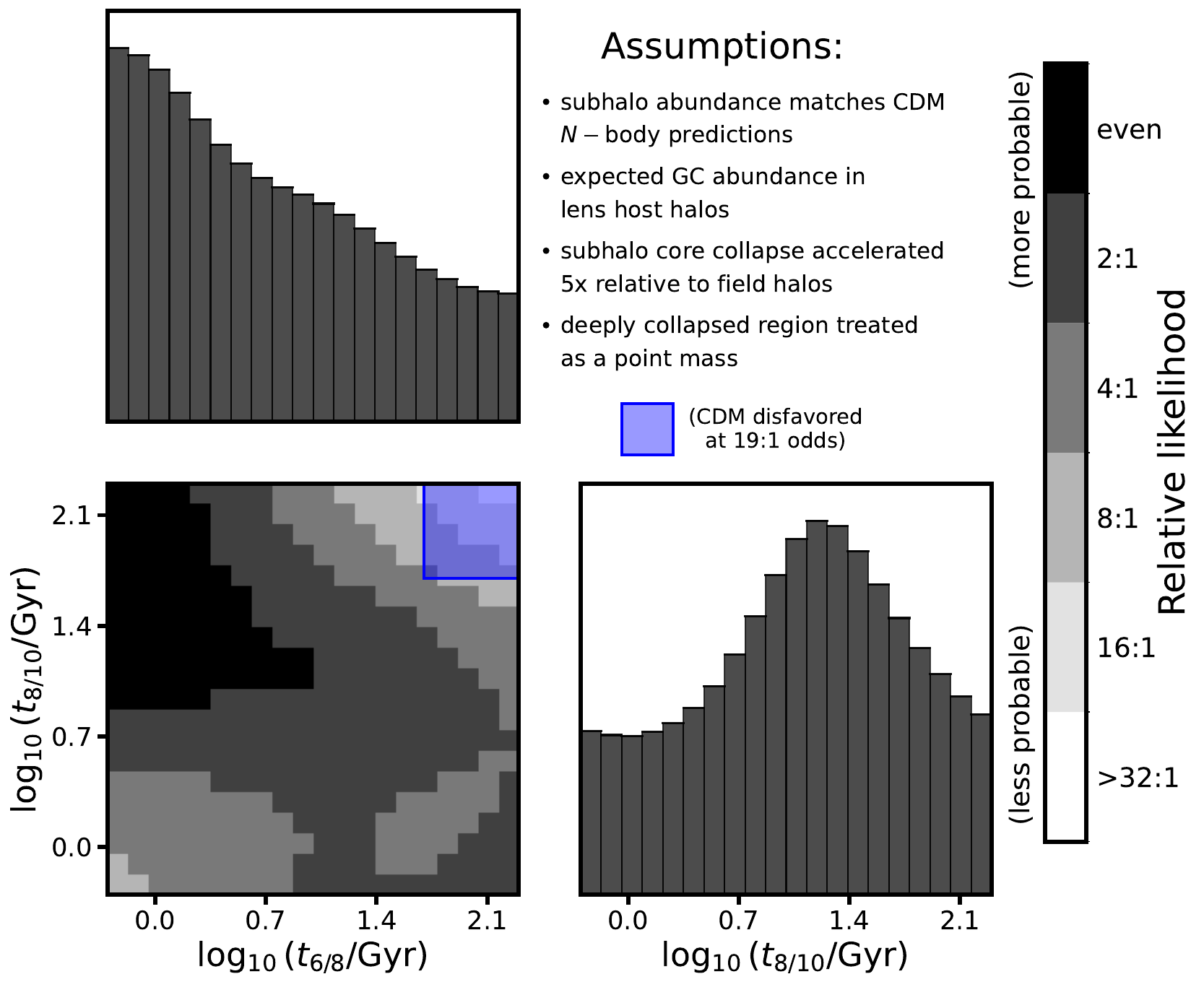}
		\caption{\label{fig:marginalpdfbestEF} The same as Figures \ref{fig:jointpdfbestA}-\ref{fig:marginalpdfbestCD}, but we assume subhalo abundance in SIDM is the same as subhalo abundance predicted by CDM $N$-body simulations, and that tidal stripping accelerates core collapse in SIDM subhalos by a factor of 5 relative to field halos with the same physical properties as the subhalo at infall. As in Figure \ref{fig:marginalpdfbestCD}, left and right marginal likelihoods make different assumptions for the internal structure of deeply collapsed halos. Table \ref{tab:priors} summarizes the importance sampling distributions noted above the figure.}.
	\end{figure*}
    \begin{figure*}
		\includegraphics[width=0.48\textwidth]{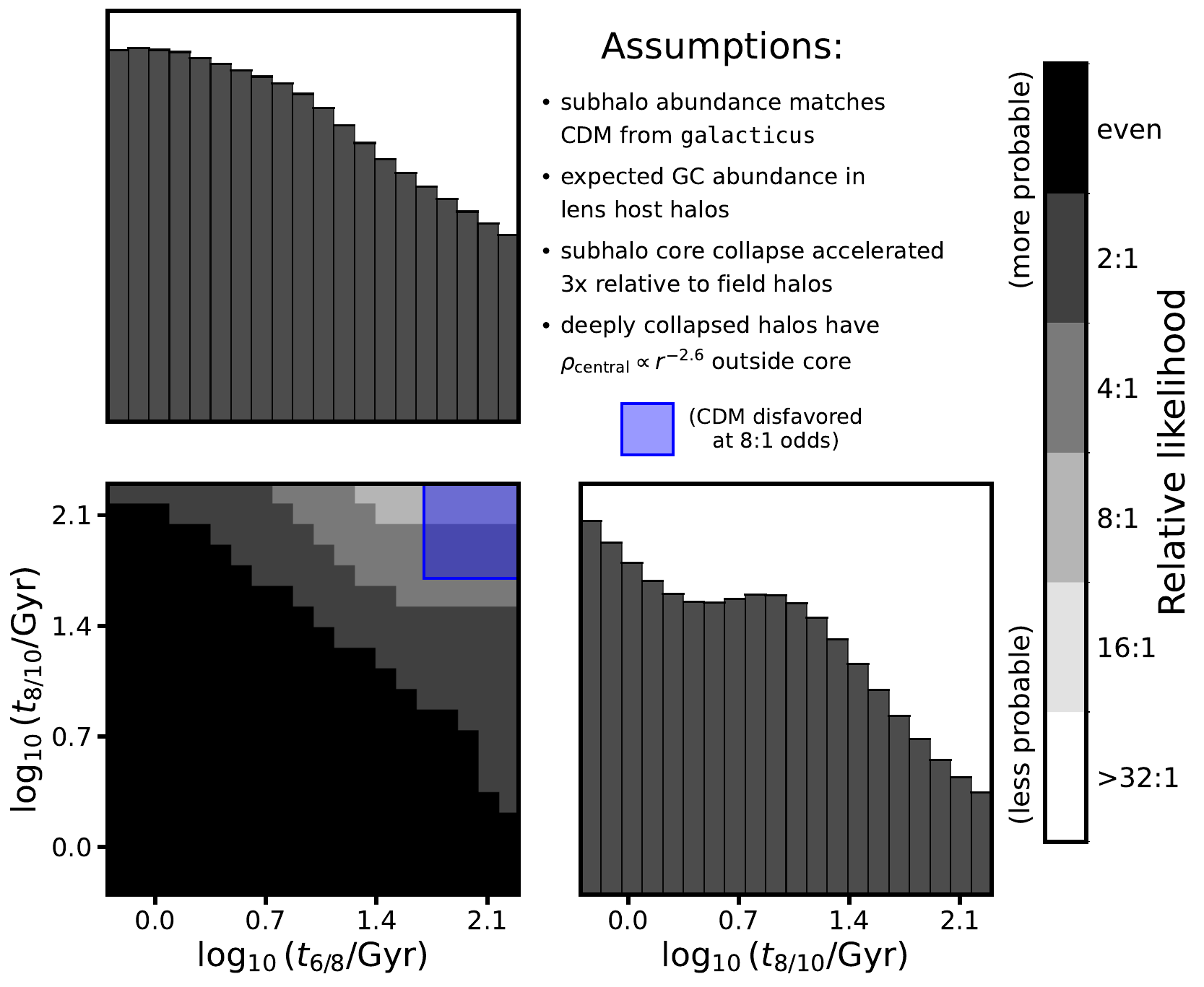}
		\includegraphics[width=0.48\textwidth]{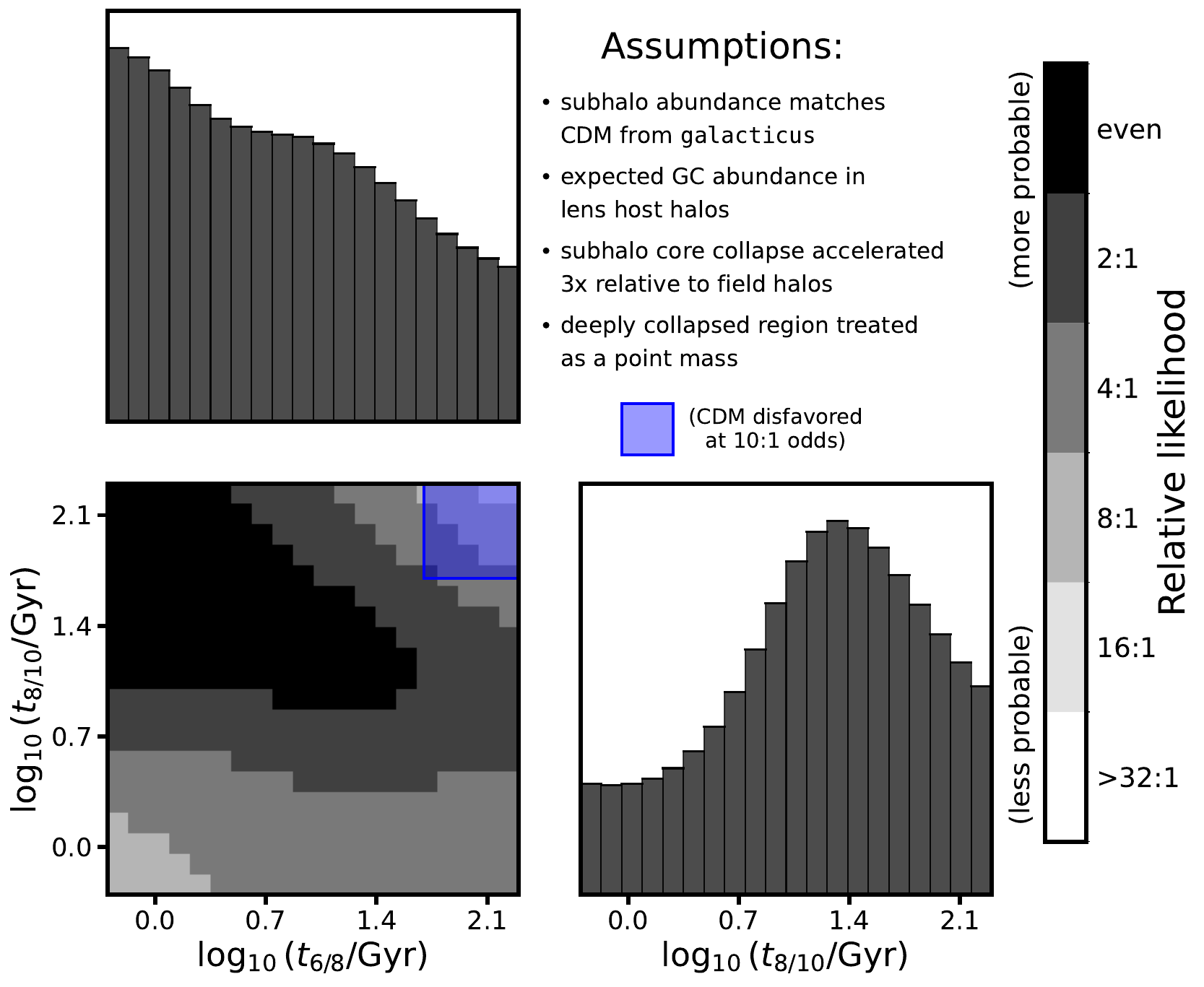}
		\caption{\label{fig:marginalpdfbestGH} The same as Figures \ref{fig:jointpdfbestA}-\ref{fig:marginalpdfbestEF}, but we assume subhalo abundance in SIDM is consistent with the predictions of the semi-analytic model {\tt{galacticus}}, which predicts $\sim 50 \%$ more subhalos than the Symphony $N$-body simulations \citep{Nadler++23b,Gannon++25}. We assume that tidal stripping accelerates core collapse in SIDM subhalos by a factor of 3 relative to field halos. Left and right marginal likelihoods make different assumptions for the internal structure of deeply collapsed halos. Table \ref{tab:priors} summarizes the importance sampling distributions noted above the figure.}.
	\end{figure*}
    
	We quote the results of our inference in terms of a Bayes factor computed between CDM and a model $M_{\rm{max}}\left(w\right)$. We define $M_{\rm{max}}\left(w\right)$ as the region of ${\boldsymbol{t}}$ parameter space within 0.25 dex of $\boldsymbol{t_{\rm{max}}}$, the collapse timescales that maximize the likelihood $\mathcal{L}_{w}\left({\data}|\boldsymbol{t},w\right)$. We choose a window of 0.2 dex around the maximum likelihood to match the volumes of the $M_{\rm{max}}$ and CDM parameter spaces, ensuring that the Bayes factor does not depend on prior volume and depends only on how well the region of parameter space explains the data. Averaging the marginal likelihood function over a region of parameter space, rather than simply quoting the relative likelihood between two points, makes our stated results more robust to shot noise in the projected likelihood function. 
	
	The Bayes factor, $BF\left(w\right)$, is given by
	\begin{equation}
			\label{eqn:bayesfactor}
				BF\left(w\right) = \frac{p\left({\data} | M_{\rm{max}}, w\right)}{p\left({\data} | {\rm{CDM}}, w\right)}.
	\end{equation}
	By construction, the Bayes factor defined in Equation \ref{eqn:bayesfactor} cannot be less than 1 when $\boldsymbol{t_{\rm{max}}}$ lies entirely outside the CDM region of parameter space. A Bayes factor close to $1$ means that the peak of $\mathcal{L}_{w}\left({\data}|\boldsymbol{t}\right)$ coincides with the region of parameter space we associate with CDM, meaning the data does not favor SIDM. On the other hand, a Bayes factor substantially greater than 1 indicates that the data disfavors CDM, and some other region of parameter space associated with SIDM is preferred. To aid the interpretation of our results, we convert this metric to the more familiar language of frequentist significance. Using the calibration of \citet{Sellke++01}, Bayes factors of $4.3$, $10$, $30$ correspond to $2.0\sigma$, $2.4\sigma$, and $2.9\sigma$, respectively. 
    
    In Appendix \ref{sec:appD} we calibrate the Bayes factor test under the null hypothesis (CDM). This calculation determines how often statistical fluctuations in the likelihood, or numerical effects associated with the use of a kernel density estimator, produce a spurious Bayesian preference for SIDM. Across 54 simulated surveys of the 29 lens systems under CDM ground truth we obtain values no larger than 3:1, well below the statistical significance of the Bayes factors inferred from the data. 
	
	We emphasize that our analysis does not assume a particular SIDM particle physics model. In fact, some areas of the $t_{6/8}$ and $t_{8/10}$ parameter space may be difficult to realize within the framework of an SIDM theory. For example, velocity-dependent cross sections generally predict $t_{6/8} > t_{8/10}$, so a model with $t_{8/10} >> t_{6/8}$ may be unusual from a particle physics standpoint, unless one invokes velocity-dependent inelastic processes, resonances \citep{Tulin++13,Gilman++23,Tran++25}, or environmental effects, such as baryons \citep[e.g.,][]{Zhong++23}, that affect only more-massive halos. However, as we will show, the constraints on $t_{6/8}$ and $t_{8/10}$ are strongly correlated, meaning a physically plausible combination of collapse timescales exists within the region of parameter space favored by the data. In a companion paper, we present an interpretation of our results in the context of velocity-dependent SIDM cross sections with light mediators, using a model that connects the particle physics parameters to the core collapse timescales in the long and short mean free path regimes. 
    
	\subsubsection{Inference on the collapse timescales}
	\label{ssec:results}
	
	Figure \ref{fig:jointpdfuniform} shows the likelihood function $\mathcal{L}\left({\data} | \qsub \right)$ computed without introducing importance weights. We note that this is equivalent to the Bayesian posterior distribution with a prior probability equal to the sampling distributions for each parameter listed in Table \ref{tab:tableqsub}. We have deliberately designed our inference with maximal flexibility in the abundance of subhalos ($\Sigma_{\rm{sub}}$), the rate of subhalo core collapse ($\lambda_{\rm{sub}}$), the surface mass density of globular clusters ($\Sigma_{\rm{gc}}$), and the internal structure of deeply collapsed halos ($\xi_{\rm{core}}$). As a result, the hyper-parameters shown in Figure \ref{fig:jointpdfuniform} have strong correlations throughout the 6D parameter space, and appear weakly constrained when projected into 1 or 2 dimensions. In the rest of this section, we will show that our data disfavors CDM relative to SIDM models with core collapse when we impose physically motivated assumptions for subhalo and globular cluster properties, which break covariances in the parameter space. While the different sets of assumptions for SIDM structure formation and globular cluster abundance affect the statistical significance of the result, they do not fully resolve the tension revealed by our analysis.
    
	Figure \ref{fig:jointpdfbestA} shows the marginal likelihood $\mathcal{L}_w\left(\data | \boldsymbol{t}, \boldsymbol{x} \right)$ with importance sampling that reflect our current best understanding of SIDM structure formation. We list assumptions enforced through the importance weights in the caption above the plot, and the importance sampling distributions are summarized in Table \ref{tab:priors}. Based on recent $N$-body simulations presented by \citet{Nadler++20b,Nadler++25a,Nadler++25b}, we assume a subhalo abundance is suppressed by $50 \%$ relative to CDM as a result of ram pressure stripping and the increased susceptibility of cored halos to complete tidal disruption. The CDM prediction for subhalo abundance corresponds to $\log_{10} \Sigma_{\rm{sub}} \sim -1.0$, so in Figure \ref{fig:jointpdfbestA} we use weights $\log_{10} \Sigma_{\rm{sub}} \sim \mathcal{N}(-1.3, 0.2)$ to add $50 \%$ disruption, relative to $N$-body simulations in CDM. We also assume that the surface mass density in globular clusters is consistent to within a factor of 2 with the expected abundance of globular clusters in a $6-12 \ \rm{kpc}$ projected annulus around massive elliptical galaxies, based on the globular cluster populations around elliptical galaxies in the local Universe (see Section \ref{ssec:globularclusters}). This assumption corresponds to importance weights $\log_{10} \Sigma_{\rm{gc}}\sim \mathcal{N}\left(5.6, 0.3\right)$. We further assume that tidal stripping causes subhalos to evolve towards core collapse three times faster than a field halo with the same physical properties as the subhalo at infall, based on numerical simulations that show tidal stripping accelerates subhalo core collapse by an order unity factor \citep{Nishikawa++20,Sameie++20,Turner++21,ShahAdhikari24}. This assumption corresponds to $\log_{10} \lambda_{\rm{sub}} \sim \mathcal{N}\left(0.5, 0.3\right)$. Finally, we assume that deeply collapsed halos have power-law profiles given by Equation \ref{eqn:rhosidm1} that match the internal structure of deeply collapsed halos simulated by \citet{Gurian++25}. This assumption corresponds to $\xi_{\rm{core}}\sim 0$, or importance weights $\xi_{\rm{core}}\sim \mathcal{N}\left(0, 0.3\right)$. With these sets of assumptions, CDM is disfavored with a Bayes factor of 24:1 relative to the most probable point in parameter space, which is an SIDM model with $\log_{10} \left(t_{6/8} / \rm{Gyr}\right) \sim 1$ and $\log_{10} \left(t_{8/10}/\rm{Gyr}\right) \sim 0.5$. 
    
    Figure \ref{fig:jointpdfbestB} shows the marginal likelihood obtained from the same set of assumptions as in Figure \ref{fig:jointpdfbestA} regarding subhalo and globular cluster abundance and the rate of subhalo core collapse, but we assume that deeply collapsed halos become denser than predicted by the simulations of \citet{Gurian++25}. This could occur if the density profile becomes steeper as the central region of the SIDM halo enters the short mean free path regime, if core collapse in SIDM halos results in the formation of a black hole at the center of the profile, or if scatter in $t_c$ tied to halo concentration preferentially drives core collapse in more-concentrated halos (see Appendix \ref{sec:appC}). For these assumptions, the data disfavors CDM with 24:1 odds. This is the same Bayes factor obtained for the inference shown in Figure \ref{fig:jointpdfbestA}, but we note that the most probable regions of parameter space shift towards models with shorter collapse timescales in low-mass halos. Our interpretation of this result is that certain features of the data, for example, the strong flux anomaly in J0924, require low-mass perturbers to be more compact than allowed by the $\rho\left(r\right) \propto r^{-2.6}$ model. If we allow SIDM halos to become denser ($\xi_{\rm{core}} \rightarrow 1$), then collapsed halos below $10^8 M_{\odot}$ can supply the required perturbation, and the likelihood shifts towards a higher abundance of such objects.
	
	Figures~\ref{fig:marginalpdfbestCD}, \ref{fig:marginalpdfbestEF} and~\ref{fig:marginalpdfbestGH} show the marginal likelihood of $t_{6/8}$ and $t_{8/10}$ under different permutations of the importance sampling weights and the physical assumptions for halo and globular cluster properties. Figure \ref{fig:marginalpdfbestCD} assumes that subhalo abundance in SIDM is consistent with the subhalo abundance in CDM predicted by $N$-body simulations \citep{Nadler++23b}, with the left and right panels changing the assumed density profile of deeply collapsed objects. In Figure~\ref{fig:marginalpdfbestEF}, we change the assumptions regarding the rate of subhalo core collapse, making them collapse five times faster that field halos with the same physical properties at infall. In Figure \ref{fig:marginalpdfbestGH}, we show the inferred collapse timescales under the same set of assumptions for subhalo collapse as in Figure~\ref{fig:marginalpdfbestCD}, but increase the normalization of the subhalo mass function to match subhalo abundance in CDM predicted by the semi-analytic model {\tt{galacticus}}, which corresponds to $\log_{10}\left(\Sigma_{\rm{sub}} / \rm{kpc^{-2}}\right) \sim -0.8$ \citep{Gannon++25,Du++25,Gilman++25a}. The joint distribution of collapse timescales has the same overall structure as the inferred collapse timescales in other cases, and making collapsed halos denser shifts the best-fitting region of parameter space towards models with shorter collapse timescales in halos less massive than $10^8 M_{\odot}$. Among the various sets of assumptions in Figures~\ref{fig:marginalpdfbestCD}-\ref{fig:marginalpdfbestGH}, the data consistently disfavors CDM with Bayes factors ranging from 8:1 to 19:1. 
    \begin{figure}
		\includegraphics[width=0.48\textwidth]{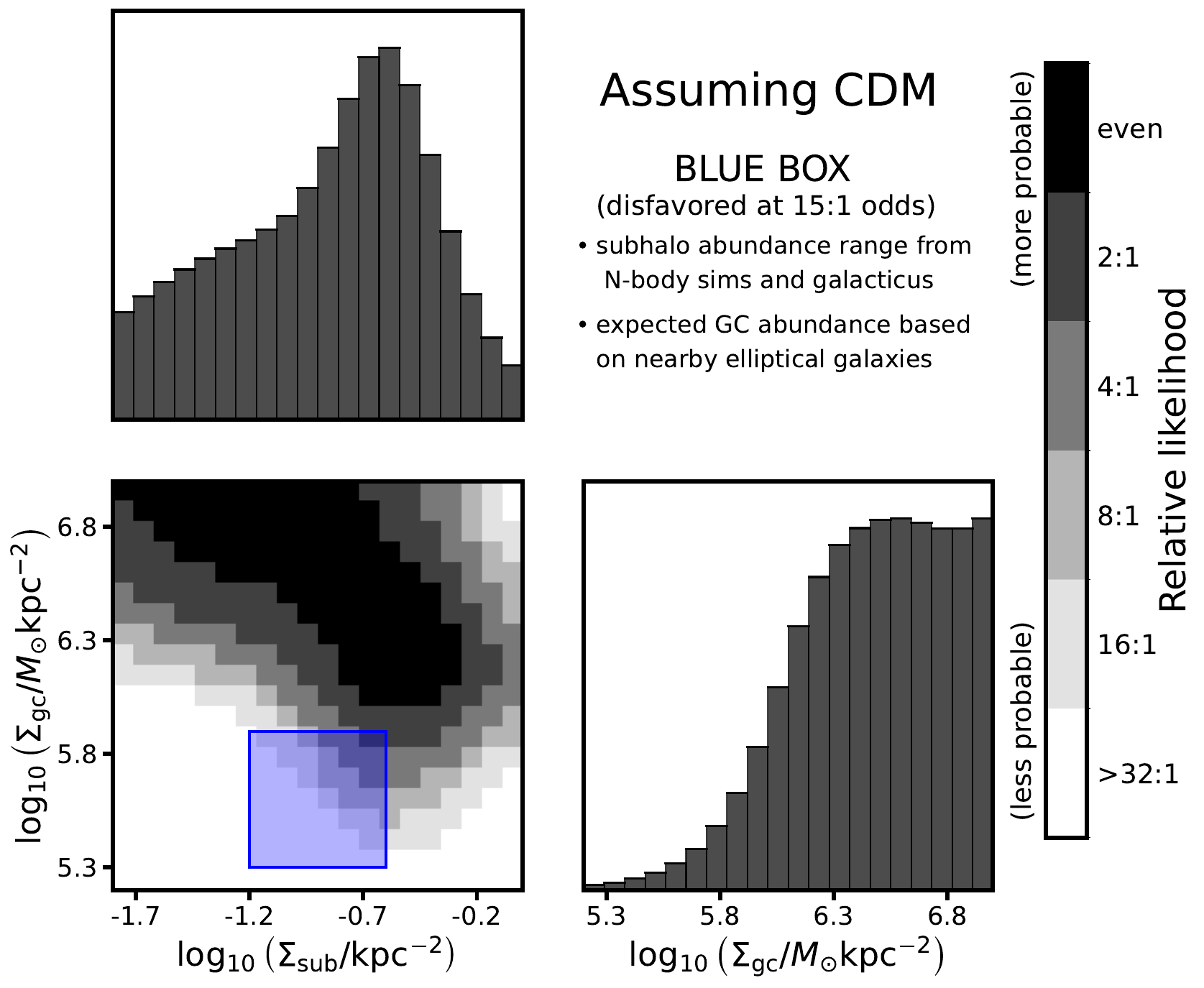}
		\caption{\label{fig:marginalpdfcdm} The inferred subhalo and globular cluster abundance around the 29 lenses in our sample with importance sampling weights that force both core collapse timescales to exceed $80 \ \rm{Gyr}$, resulting in structure formation outcomes indistinguishable from CDM, given our data. The blue shaded region encompasses the range of current uncertainties on the normalization of the subhalo mass function from $N$-body simulations and the semi-analytic model $\tt{galacticus}$ \citep{Gannon++25}, and spans a factor of 2 in either direction around the expected projected mass density of globular clusters in a $6-12 \ \rm{kpc}$ annulus based on observations of nearby ellipticals (see Section \ref{ssec:globularclusters}). This region of parameter space is disfavored at $15:1$ odds relative to the most probable point, which has $\log_{10} \Sigma_{\rm{sub}}\sim -0.5$ and $\log_{10} \Sigma_{\rm{gc}} \sim 6.3$.}.
	\end{figure}
    
    We note that our model also allows for delayed subhalo core collapse, corresponding to $\lambda_{\rm{sub}} < 1$. This could occur if scattering between subhalo and host halo dark matter particles transfers heat into infalling subhalos, which delays the onset of core collapse \citep[e.g.,][]{Zeng++22,Slone++23}. This process requires an appreciable cross section strength, likely greater than $10 \ \rm{cm^2} \ \rm{g^{-1}}$ \citep{Slone++23}, at a velocity scale corresponding to the relative velocity between an infalling subhalo and the halo ($v\sim 200 \ \rm{km} \ \rm{s^{-1}}$). Interpreting our inference with $\lambda_{\rm{sub}} <1$ results in a more statistically significant Bayesian penalty on CDM, because the collapse timescales must become shorter to produce a given number of collapsed subhalos in the lens model. Future analyses may use our publicly available likelihood functions with a subhalo evolution model that includes delayed core collapse, when such a model becomes available.   
	
	To conclude, we may turn the problem around and ask what properties of globular clusters and CDM halos are required to explain our dataset, without invoking SIDM. We use importance weights on $\log_{10} t_{6/8}$ and $\log_{10} t_{8/10}$ given by $\mathcal{N}\left(2.3, 0.25\right)$, and weights on the rate of subhalo core collapse $\log_{10} \lambda_{\rm{sub}} \sim \mathcal{N}\left(-1, 0.5\right)$, to isolate a region of parameter space in which no subhalos or field halos core collapse. The resulting marginal likelihood $\mathcal{L}_{w}\left({\data} | \Sigma_{\rm{sub}}, \Sigma_{\rm{gc}}\right)$ is shown in Figure \ref{fig:marginalpdfcdm}. The blue box highlights the range of theoretical predictions for the normalization of the subhalo mass function in CDM from $N$-body simulations and the semi-analytic model ${\tt{galacticus}}$ \citep{Gannon++25,Du++25}. The vertical extent of the blue region spans a factor of 4 around the expected abundance of globular clusters based on observations of nearby ellipticals in the same projected $6-12 \ \rm{kpc}$ annulus probed by our data. The most probable regions of parameter space with subhalo abundance consistent with theoretical predictions from $N$-body simulations and {\tt{galacticus}} have an abundance of globular clusters that exceeds the abundance measured in the local Universe by a factor of 5-10, depending on the assumed abundance of CDM subhalos. Models with globular cluster populations consistent with the local Universe are disfavored by the data, irrespective of subhalo abundance. A CDM explanation for our dataset therefore requires an implausibly high abundance of globular clusters, and more subhalos than predicted by $N$-body simulations and the semi-analytic model ${\tt{galacticus}}$. 
	
	\section{Discussion and conclusions}
		\label{sec:conclusions}
		We have developed a structure formation model that predicts the abundance of core-collapsed SIDM halos and subhalos given two core-collapse timescales, $t_{6/8}$ and $t_{8/10}$, affecting halos in the mass ranges $10^6-10^{8} M_{\odot}$ and $10^8 - 10^{10.7} M_{\odot}$, respectively. We constrain these parameters by forward modeling the flux ratios, image positions, and extended lensed arcs of a sample of 29 quadruply imaged quasars, while quantifying how uncertainties related to SIDM subhalo abundance, the effects of tidal stripping on core collapse, and the surface mass density in globular clusters around strong lenses affect our inference on the collapse timescales. Under different sets of modeling assumptions for SIDM structure formation, we compute a Bayes factor that quantifies to what degree the data favors SIDM over CDM, where we associate CDM with models in which the core collapse timescales significantly exceed a Hubble time. Our main results are summarized as follows: 
		
		\begin{itemize}
			\item The data prefer models in which some halos undergo core collapse, with Bayes factors disfavoring CDM ranging from 8:1 to 24:1. Our best current understanding of SIDM structure formation, in which subhalo abundance is suppressed by $50\%$ relative to CDM, tidal stripping accelerates subhalo core collapse, and globular cluster abundance around strong lenses is consistent with expectations based on elliptical galaxies in the Virgo cluster, results in Bayes factors that disfavor CDM with 24:1 odds. 
			\item The inferred core collapse timescales depend on the assumed internal structure of deeply collapsed halos. If core collapse results in denser structures than predicted by current simulations, the most probable regions of parameter space shift towards models with shorter collapse timescales in low-mass halos, $t_{6/8} < t_{8/10}$. 
			\item Assuming CDM, explaining our dataset requires a surface mass density in globular clusters around strong lenses that exceeds the expected abundance by a factor of 5--10, together with a higher subhalo abundance than predicted by $N$-body simulations or the semi-analytic model {\tt{galacticus}}.
		\end{itemize}

		Our analysis complements recent results from various probes of small-scale structure, including dwarf galaxies, stellar streams, and strong lensing, that point towards low-mass halos with higher central densities than predicted by CDM \citep[e.g.,][]{Nadler++23,Dutra++25,Natarajan++26,Nibauer++26,Vegetti++26,Yu26}. In the case of strong lensing on galaxy scales, the claimed tensions stem from the inferred properties of objects detected in lensed arcs through gravitational imaging, or by including individual perturbers in lens model reconstructions. Studies of individual halos provide easily interpretable outcomes that demonstrate the presence of compact dark objects, and partially motivate the population-level inference presented in this work. However, analyses of individual objects do not currently supply an interpretation of the detections in terms of fundamental physics. Dark matter theories predict the statistical properties of halo populations, not properties of any single object. A rigorous answer to whether a given detection disfavors CDM, or whether an observation favors SIDM, requires forward modeling the distribution of possible outcomes predicted by each theory. This enables statistical statements regarding which dark matter model provides a better explanation for the data. 
        
        Through the population-level inference presented in this work, we can quantify the relative likelihood between various competing explanations of the data, such as whether globular clusters or core-collapsed halos better account for the observed flux ratio anomalies. As discussed in Section~\ref{ssec:results}, we determine the abundance of CDM subhalos and globular clusters required to explain our data without invoking SIDM, and find that the required surface mass density in globular clusters exceeds the abundance measured in the local Universe by a factor of 5-10. We note that we also include scatter in the concentration-mass relation predicted by CDM $N$-body simulations, and therefore we account for the possibility that extreme flux ratio perturbations come from outliers in the concentration-mass relation. We also include angular structure in the main deflector mass profile to address concerns associated with overly-simplistic main deflector mass profiles as source of systematic uncertainty. Even when allowing for these possible alternative explanations, we recover a statistically significant preference for models in which halos undergo core collapse. 

        Our inference on the core collapse timescales implies an abundance of collapsed halos, which in principle predicts the number of dense perturbers expected to be detected in extended lensed arcs. Rigorously connecting single-halo detections to the properties of the population requires quantifying the selection function for individual objects, and accounting for non-detections, both within individual lenses and across the full sample of galaxy-galaxy lenses. This investigation could form the basis of future work, building on recent progress in characterizing the single-halo model selection functions \citep{VegettiKoopmans10,Ritondale++19,Despali++22,ORiordan++23,Nightingale++24} and developing population-level constraints from galaxy-galaxy lenses \citep{Birrer++17,CyrRacine++19,ORiordan26}.
        
        One of the challenges with carrying out a population-level inference on SIDM is that the structure formation model is less well calibrated than in CDM. While we understand some aspects of SIDM structure formation in broad strokes, such as the accelerated core collapse of subhalos due to tidal stripping, we do not currently have the tools to forward model these processes with the same level of fidelity as in CDM, or in warm dark matter \citep[e.g.][]{Gilman++25a}. To quantify how our inference depends on these sources of systematic uncertainty, we follow an empirical approach that assigns a considerable level of flexibility to subhalo abundance, the rate at which subhalos core collapse, and the internal structure of collapsed objects. As expected, the most probable combinations of the core collapse timescales depend on the assumptions for SIDM halo properties, and different sets of modeling assumptions affect the statistical significance with which CDM is disfavored by the data. In the future, tidal evolution models for SIDM subhalos that account for correlations between the density profile at infall and the subsequent tidal evolution, and how this tidal evolution affects the onset of core collapse, would obviate the need for an empirical treatment and lead to more precise inferences on core collapse timescales. We note that one assumptions that is built into our inference on $t_c$ is the assumed distribution of collapse times. As discussed in Section \ref{sec:structureformationmodel} we define $t_c$ as the median collapse time, and sample $\log_{10} t_c$ from a Gaussian distribution with a standard deviation 0.3 dex. The 0.3 dex standard deviation is tied to the scatter in the Knudsen number, which depends on $\Sigma_s = \rho_s r_s$, from the concentration-mass relation at a fixed halo mass. If the distribution of $t_c$ is highly non-Gaussian, our model may overestimate (underestimate) the abundance of collapsed halos for a given $t_c$, depending on whether the distribution has a long tail that extends to shorter (longer) collapse timescales. 
        
        Given the implications of our work for new physics, we highlight some caveats that could alleviate tension with CDM. First, as discussed in Section~\ref{ssec:results}, we can explain our data with only globular clusters, provided their abundance exceeds by a factor of 5--10 the expected abundance around lens-mass host halos based on scaling relations measured in the local Universe. With few recent exceptions \citep[e.g.,][]{Harris++24}, we lack observational constraints on globular cluster populations at cosmological distance, and in particular on the globular cluster populations around elliptical galaxies at cosmological distance. However, we are not aware of a mechanism that would lead to an increase in their projected number density by the amount required to explain our data. Second, an underestimation of CDM halo central density would lead us to unduly favor SIDM models. We regard deficiencies associated with the modeling of CDM structure as an unlikely cause for the tension, given that our structure formation and tidal evolution model for CDM (sub)halos accurately reproduces the internal structure of CDM halos predicted by state-of-the-art numerical simulations \citep{Gannon++25,Du++25}. We note that our data probe halo structure on mass scales where baryons are not expected to play a significant role in altering the density profile through supernova feedback \citep[e.g.,][]{DiCintio++14,Fitts++17}, which in any case tends to suppress central density rather than enhance it. Reproducing the lensing signal we associate with collapsed SIDM halos would require halos with stellar mass fractions of order $5\%$ of the total halo mass, or $M_{\star} \sim 5 \times 10^7 M_{\odot}$ in a $10^{9} M_{\odot}$ halo, contained within a few hundred parsecs. This exceeds the population-average stellar mass expected for halos on these scales by roughly two orders of magnitude \citep{Read++17,Nadler++20,Shao++26}, and would require a stellar distribution more compact than in observed galaxies of comparable stellar mass. Third, one of the main challenges with strong lensing inferences on dark matter properties pertains to controlling systematics associated with the lens modeling. Our lens modeling approach introduces azimuthal structure through a multipole expansion that matches azimuthal features in the baryonic mass profile of massive elliptical galaxies \citep{Paugnat++25b,Oh++26}. The next order of lens model complexity, which we have begun implementing in our analysis pipelines, involves radial variation and twists in the ellipticity of the mass distribution \citep{VandeVyvere++22,He++24}. Assuming light traces mass in the region constrained by strong lensing, an assumption supported by dynamical measurements of strong lenses \citep{Auger++10} and lens modeling studies \citep{Shajib++19}, we note that invoking these features to resolve flux anomalies \citep[e.g.,][]{Feng++26} requires mass ellipticity gradients that greatly exceed the measured gradients in nearby galaxies \citep{Hao++06,Goullaud++18} and in lens galaxies themselves \citep{He++24}. As emphasized in our analysis and in previous work \citep[see Appendix B][]{Gilman++24}, the degree to which complexity in the main deflector mass distribution affects substructure inferences requires population-level analyses that include both sources of perturbation in the lens model. Forthcoming papers that improve our lens modeling techniques \citep[e.g.][]{Paugnat++25}, and which analyze an expanded sample size of lensed quasars \citep{Shajib++25,Sheu++26,Paugnat++26}, will address concerns associated with these possible sources of systematic uncertainty. 
        
\begin{figure}
		\includegraphics[width=0.48\textwidth]{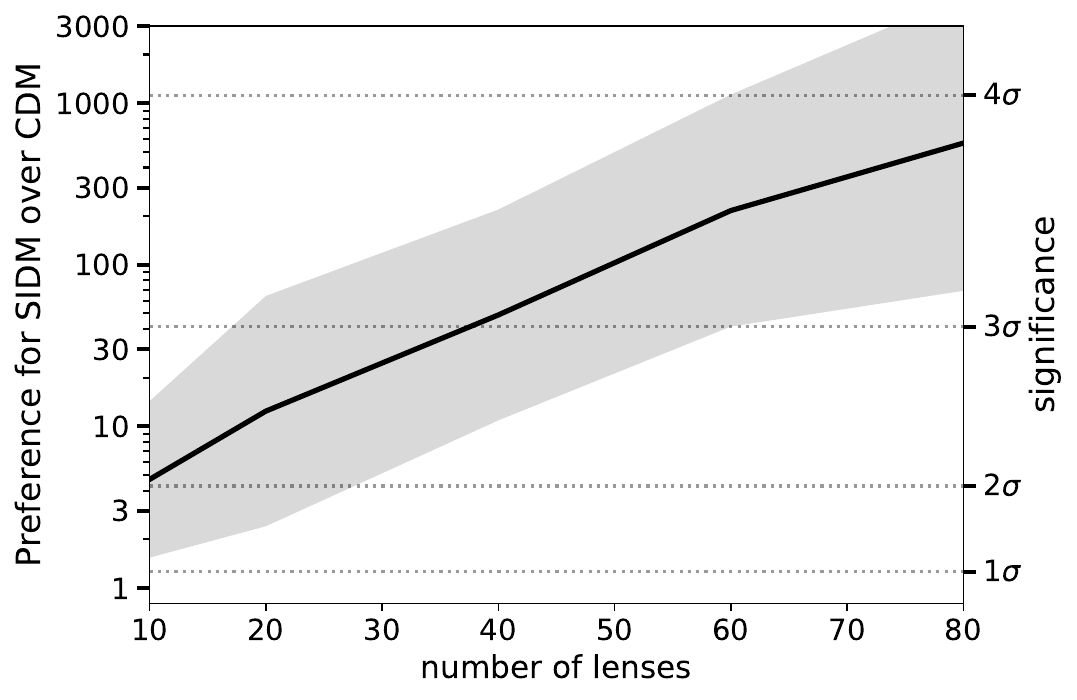}
		\caption{\label{fig:forecast} Forecast for the constraining power of lensed quasars obtained by bootstrap resampling of our sample. We assume SIDM structure formation parameters corresponding to the likelihood function shown in Figure \ref{fig:jointpdfbestA}, with the shaded band reflecting uncertainties in the lensing efficiency of collapsed halos. The $x$-axis shows the number of lenses, and the left and right $y$-axes show the Bayes factor disfavoring CDM and the corresponding frequentist significance using the calibration of \citet{Sellke++01}.}.
	\end{figure} 
    
        Cosmological surveys such as Euclid, Roman, and the Rubin Observatory will discover hundreds of new quadruply imaged quasars \citep{Oguri++10,Shajib++25}. A larger dataset will increase the statistical significance of our results. To forecast the constraining power from a larger sample, we bootstrap the sample of 29 systems analyzed in this work to estimate the information content per lens. Figure \ref{fig:forecast} shows the Bayes factor disfavoring CDM as a function of the number of lenses. The solid curve shows the median from 200 bootstrap iterations at a given sample size, and the shaded band is the $1 \sigma$ uncertainty in $\log_{10} \rm{BF}$ as a function of $N$. Assuming the 29 lenses in our sample are representative of the broader population of known and soon-to-be-discovered quadruply imaged quasars, we expect that doubling our sample size would result in a Bayes factor $\sim 100$, exceeding the $3 \sigma$ threshold obtained by translating the Bayes factor to a frequentist significance \citep{Sellke++01}. 

        In closing, our analysis provides the first population-level constraint on SIDM core collapse timescales from a sample of 29 lensed quasars, and favors (Bayes factors 8:1--24:1) SIDM with core collapse over CDM. However, just as our confidence in the existence of particle dark matter stems from multiple independent lines of evidence, our inferences on its particle properties should derive from multiple independent experiments with distinct sources of systematic uncertainty. In this context, inferences on halo properties from strong lensing should be interpreted as predictions for other small-scale probes of cosmic structure. In particular, the collapse timescales favored by our analysis predict a population of core collapsed halos around the Milky Way. Their elevated central densities should impart detectable perturbations to stellar streams, and increase the inner densities of low-mass galaxies, relative to CDM expectations. Further improvements in the statistical significance of our results, and confirmation of the signal from independent probes of dark matter substructure, are needed to claim definitive evidence for a self-interacting dark sector. 
        
		\section*{Acknowledgments}
        We thank Kim Boddy, Francis-Yan Cyr-Racine, Alex Drlica-Wagner, Vera Gluscevic, Matt Malkan, Manoj Kaplinghat, Kallia Petraki, Thomas Pignard, and Ethan Nadler for helpful discussions during the course of this project. 
	
    DG acknowledges support provided by the Brinson Foundation through a Brinson Prize Fellowship grant, and gratefully acknowledges their continued support of his research. DG acknowledges support from the Kavli Foundation under Grant No. PS-2025-GR-249-3072. AMN and CG acknowledge support from the National Science Foundation through the grant ``CAREER: An order of magnitude improvement in measurements of the physical properties of dark matter'' NSF-AST-2442975.
    Research at the Perimeter Institute is supported in part by the Government of Canada through the Department of Innovation, Science and Economic Development Canada and by the Province of Ontario through the Ministry of Colleges and Universities.
    TT acknowledges support from NSF through grant AST-2205100, from NASA through grant JWST-GO-2046, JWST-GO-7184, HST-GO-17916, and from the Gordon and Betty Moore Foundation.
    KNA is partially supported by the U.S. National Science Foundation (NSF) Theoretical Physics Program Grant No.\ PHY-2609954.
    TA acknowledges support from ANID-FONDECYT Regular Project 1240105 and the ANID BASAL project FB210003.
    SB acknowledges support by the Stony Brook Department of Physics \& Astronomy.
    SGD acknowledges support from the Ajax Foundation.
    SFH acknowledges support through UK Research and Innovation (UKRI) under the UK government's Horizon Europe Funding Guarantee (EP/Z533920/1, selected in the 2023 ERC Advanced Grant round) and an STFC Small Award (ST/Y001656/1).
    AK was supported by the U.S. Department of Energy (DOE) Grant No.\ DE-SC0009937, by World Premier International Research Center Initiative (WPI), MEXT, Japan, and by Japan Society for the Promotion of Science (JSPS) KAKENHI Grant No.\ JP20H05853.
    MM acknowledges support by the SNSF (Swiss National Science Foundation) through Ambizione grant PZ00P2\_223738.
    TM acknowledges the support by JSPS KAKENHI Grant Number 25K24918.
    Part of this work was carried out at the Jet Propulsion Laboratory, California Institute of Technology, under a contract with the National Aeronautics and Space Administration (80NM0018D0004).
    DS acknowledges the support of the Fonds de la Recherche Scientifique-FNRS, Belgium, under grant No.\ 4.4503.1.
    KCW is supported by JSPS KAKENHI Grant Numbers JP24H00221, JP24K07089.
    
    This work is based on observations made with the James Webb Space Telescope through the Cycle 1 program JWST GO-2046 (PI: Nierenberg), and the Hubble Space Telescope through HST-GO-15320, HST-GO-15652, HST-GO-17916 (PI: Treu) and HST-GO-13732 (PI: Nierenberg). Funding from NASA through these programs is gratefully acknowledged.\\
    
		\section*{Software} 
        This work made use of {\tt{astropy}}:\footnote{\url{http://www.astropy.org}} a community-developed core Python package and an ecosystem of tools and resources for astronomy \citep{astropy};  {\tt{cobyqa}} \citep{rago_thesis,cobyqa}; {\tt{colossus}} \citep{Diemer18}; {\tt{lenstronomy}}\footnote{\url{https://github.com/lenstronomy/lenstronomy}} \citep{Birrer++18,Birrer++21}; {\tt{numpy}} \citep{numpy}; {\tt{pyHalo}}\footnote{\url{https://github.com/dangilman/pyHalo}} \citep{Gilman++20}; {\tt{trikde}}\footnote{\url{https://github.com/dangilman/trikde}}; {\tt{samana}}\footnote{\url{https://github.com/dangilman/samana}}; and {\tt{scipy}} \citep{scipy}. 

        The calculations for this paper used {\tt{pyHalo}} version 1.4.7 commit 7fc0362, {\tt{lenstronomy}} commit 38906b9 from fork \url{https://github.com/dangilman/lenstronomy.git}, and {\tt{samana}} commit 398ede3.

        \section*{Computing} 
        The total computing time invested to perform the calculations presented in this work exceeded 4M CPU hours. This work was completed in part with resources provided by the University of Chicago Research Computing Center. This work also used computational and storage services associated with the Hoffman2 Shared Cluster provided by the UCLA Institute for Digital Research and Education's Research Technology Group. Finally, this work used the Pinnacles cluster at the Cyberinfrastructure and Research Technologies unit at the University of California, Merced, supported by NSF award OAC-2019144. 
        
		\section*{Data availability}
		The data used in this article come from HST-GO-15320, HST-GO-15652, HST-GO-17917, HST GO-17916, HST-GO-13732, JWST GTO-1198, JWST GO-2046, JWST GO-7184. The raw data are publicly available online. Astrometry and flux ratio measurements are presented by \citet{Nierenberg++14,Nierenberg++20,Keeley++24,Keeley++25}. The data and analysis scripts used in this work are available through the open-source software package {\tt{samana}}. 
		
		\bibliography{main}
		\bibliographystyle{apsrev4-2}
		
		\appendix 
		\section{Modeling imaging data of J0607-2152, J0659+1629, 2M1134-2103, B2045+265, and J2205-3727}
        \label{sec:appA}
\begin{figure*}
			\centering
			\includegraphics[trim=0cm 0cm 0cm
			0cm,width=0.195\textwidth]{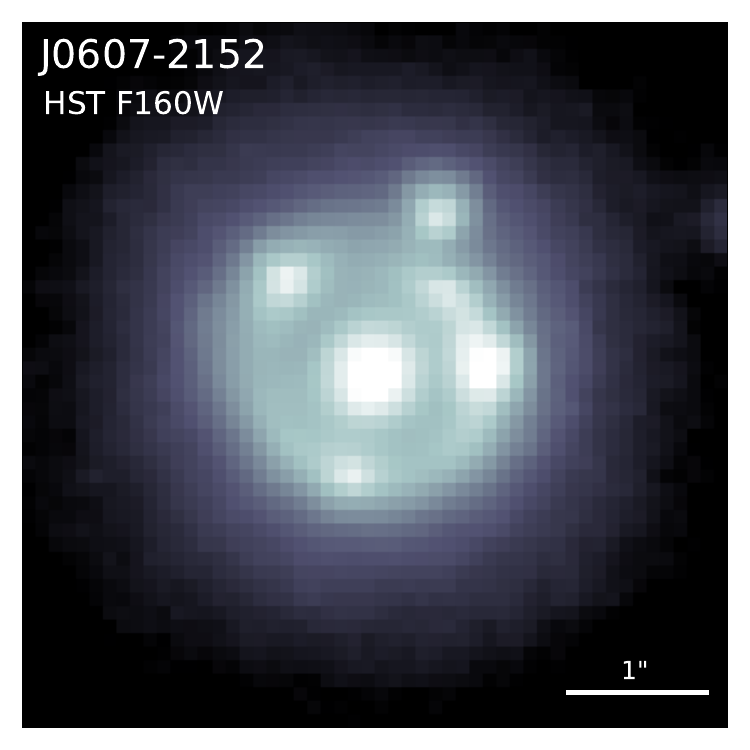}
			\includegraphics[trim=0cm 0cm 0cm
			0cm,width=0.195\textwidth]{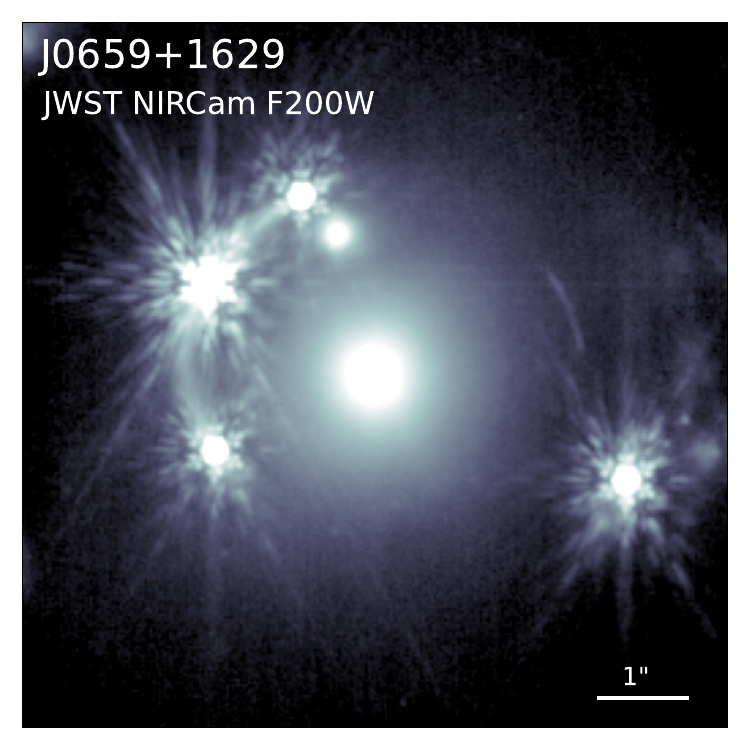}
			\includegraphics[trim=0cm 0cm 0cm
			0cm,width=0.195\textwidth]{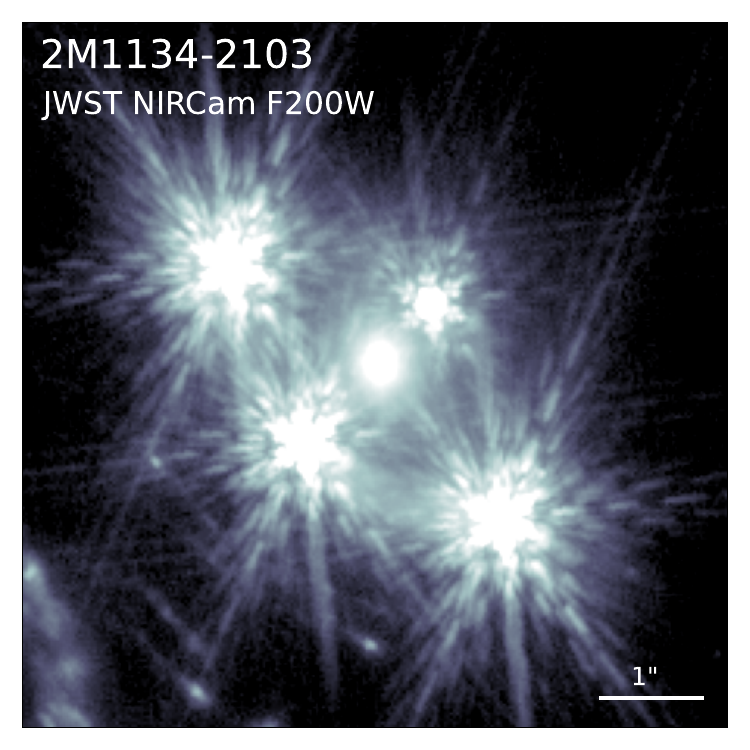}
			\includegraphics[trim=0cm 0cm 0cm
			0cm,width=0.195\textwidth]{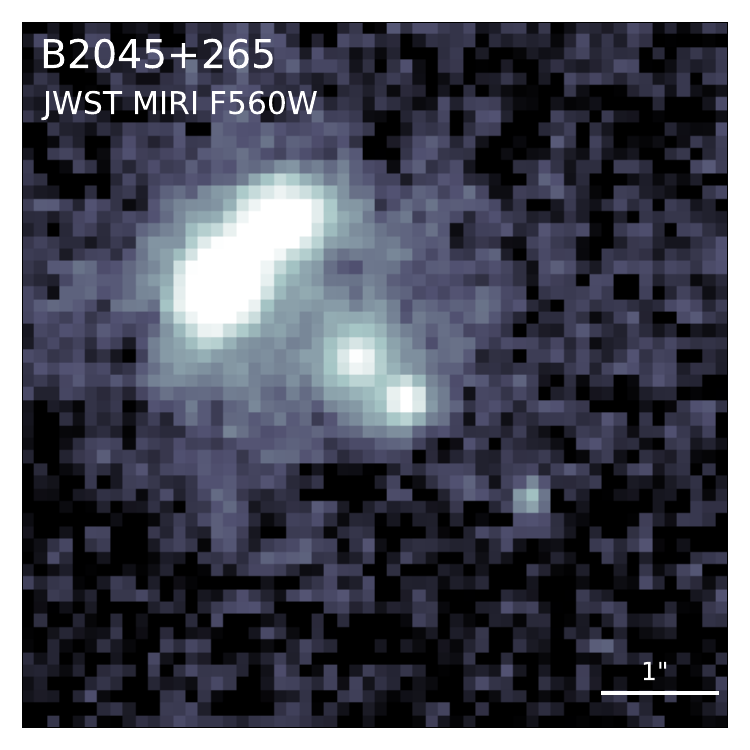}
			\includegraphics[trim=0cm 0cm 0cm
			0cm,width=0.195\textwidth]{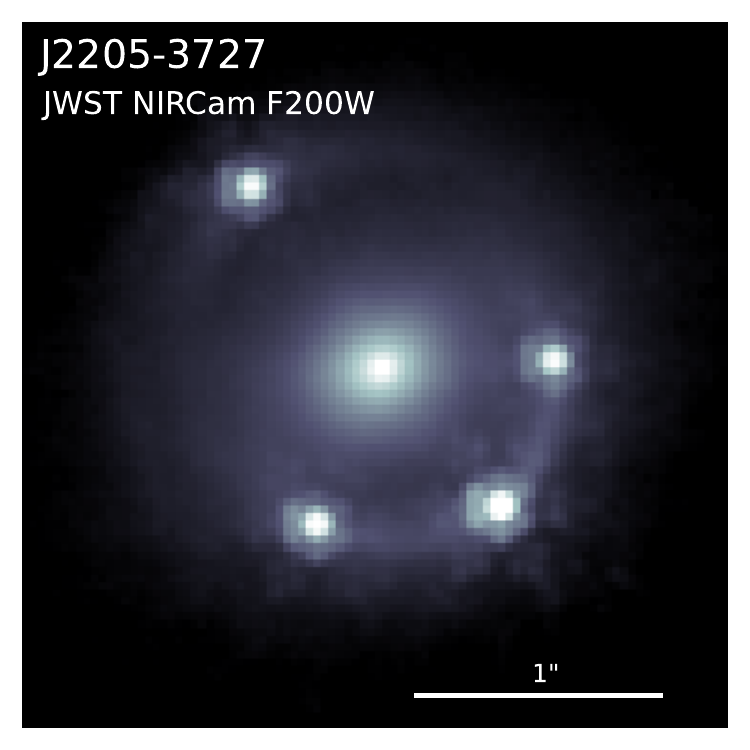}
			\caption{\label{fig:nircamcutouts} Imaging data cutouts of J0607 (HST F160W), J0659 (JWST NIRCam F200W), 2M1134 (JWST NIRCam F200W), B2045 (MIRI F560W), and J2205 (JWST NIRCam F200W). The color scale is logarithmic and spans 2 dex. We note that the luminous feature in B2045, second from right, was later identified as a foreground star.}
		\end{figure*}

\begin{figure*}
			\centering
			\includegraphics[trim=0cm 0cm 0cm
			0cm,width=0.95\textwidth]{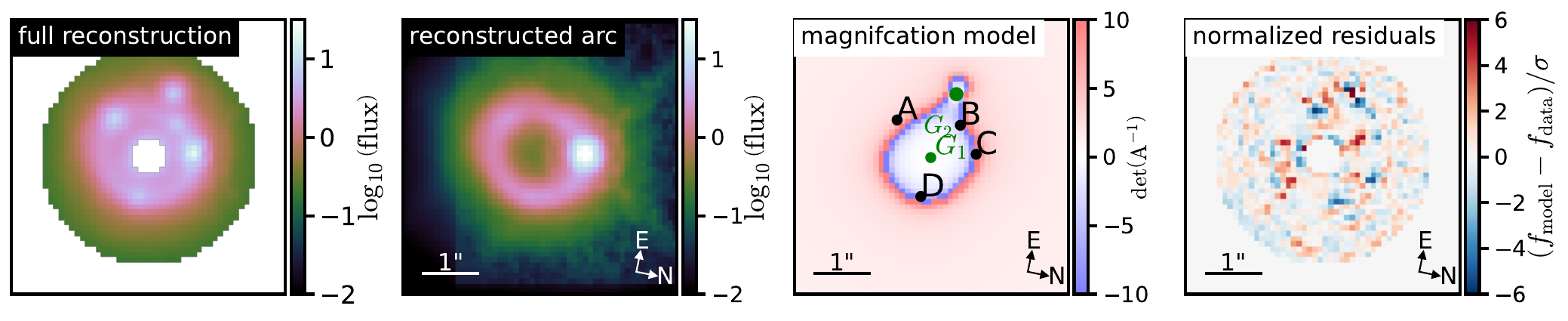}
			\includegraphics[trim=0cm 0cm 0cm
			0cm,width=0.95\textwidth]{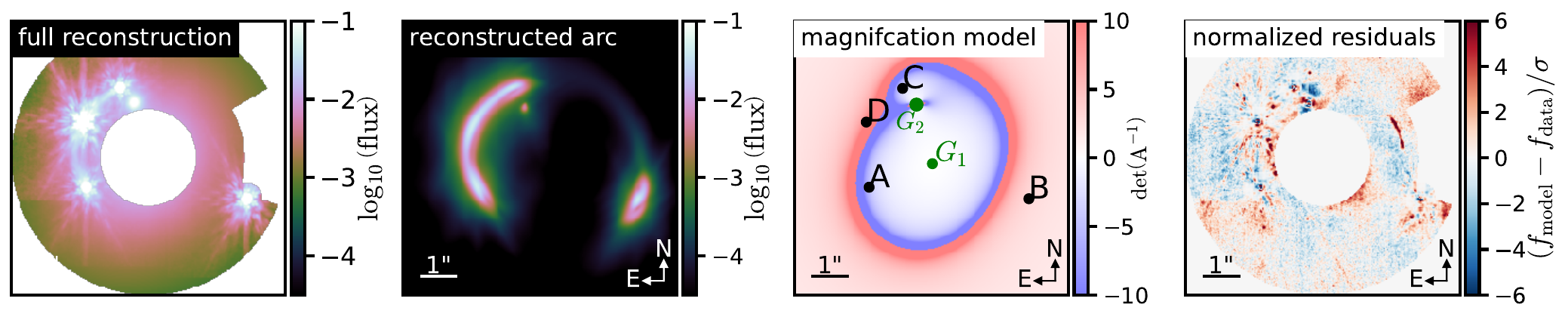}
			\includegraphics[trim=0cm 0cm 0cm
			0cm,width=0.95\textwidth]{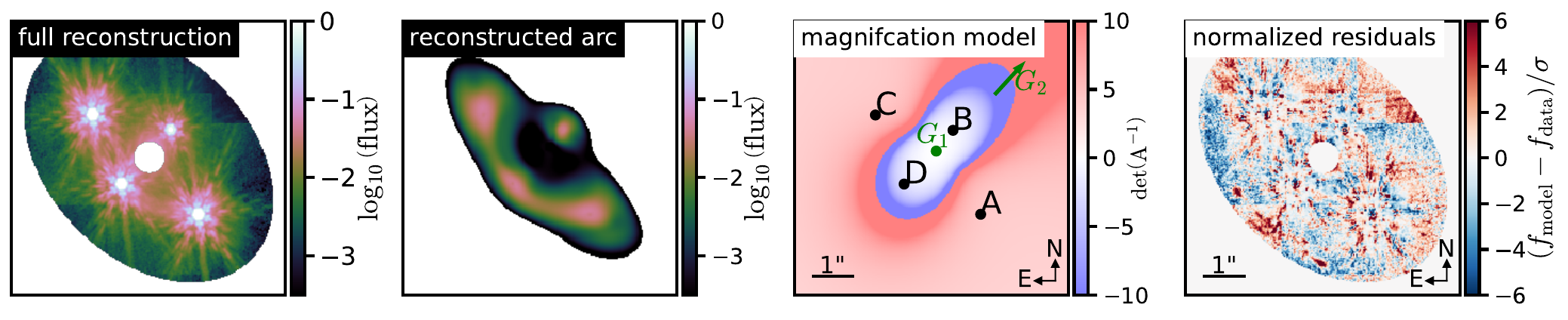}
			\includegraphics[trim=0cm 0cm 0cm
			0cm,width=0.95\textwidth]{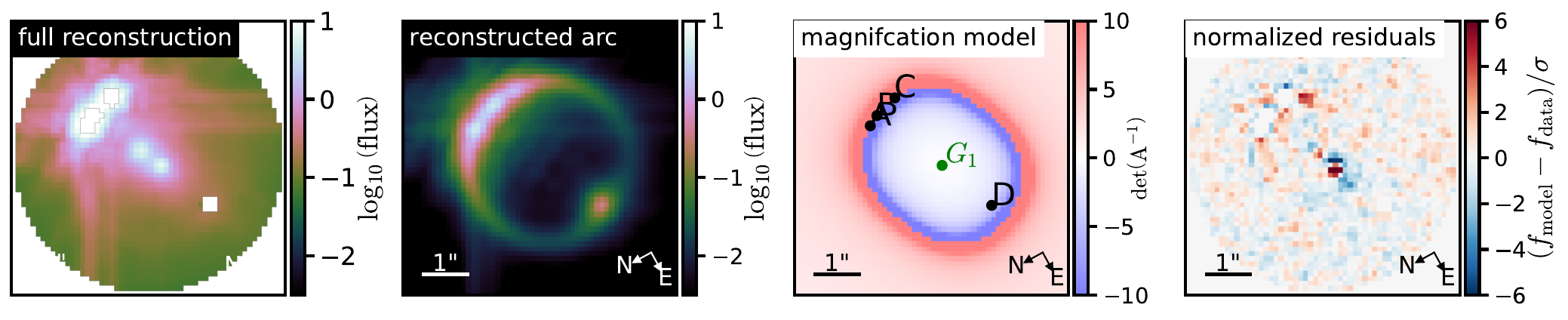}
			\includegraphics[trim=0cm 0cm 0cm
			0cm,width=0.95\textwidth]{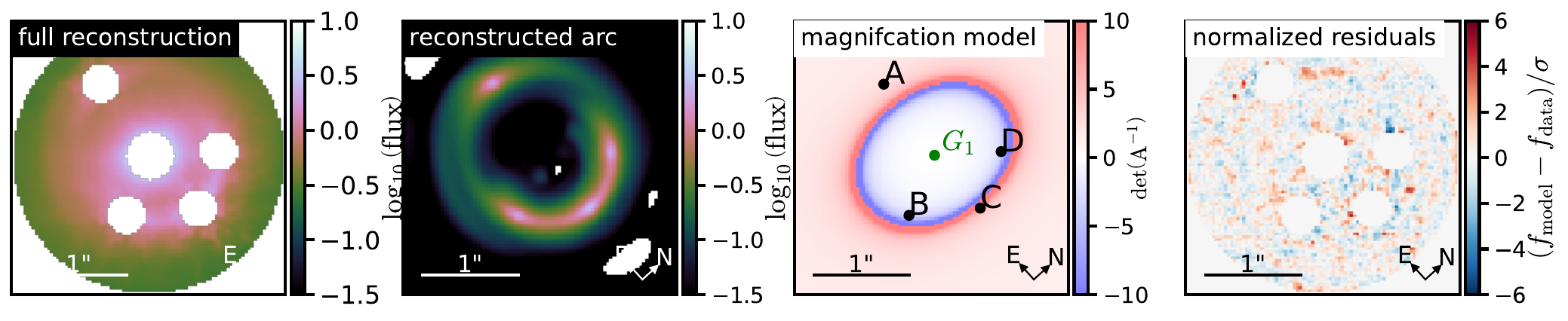}
			\caption{\label{fig:bmodels} From top to bottom, each row shows lens and light models for J0607 (HST F160W), J0659 (JWST NIRCam F200W), 2M1134 (JWST NIRCam F200W), B2045 (MIRI F560W), and J2205 (JWST NIRCam F200W). From left, we show the reconstructed lensed image, the reconstructed lensed quasar host galaxy, the magnification model, and the normalized residuals of all light components. These lens models are the starting point for the forward modeling pipeline, and the lens models used for the dark matter inference reproduced the morphology of the lensed arcs to a comparable level of fidelity. For additional details we refer to Ref. \citep{Gilman++25a}.}
		\end{figure*}

		We update previous imaging data of J0659+1629, 2M1134-2103, and J2205-3727 with NIRCam observations obtained through JWST GO-7184 (PI: Millon). We also replace the MIRI imaging data for J0607 with imaging data in F160W obtained through HST GO 17916 (PI: Treu) \citep{Sheu++26}. Cutouts of the five lens systems are shown in Figure \ref{fig:nircamcutouts}. As in Paper IV, we set up a \textit{baseline model} for each lens system, which determines the starting point for each re-optimization of the flux ratios and imaging data for different substructure realizations. Figure \ref{fig:bmodels} shows the baseline lens models for the five systems. With the exception of 2M1134 (see discussion below), the lens models are the same as those used in \citet{Gilman++25a}. We generate the PSFs used to model the NIRCam imaging data using the {\tt{starred}} software package \citep{Michalewicz++23,Millon++24}. 
        \begin{itemize}
        \item J0607: Our lens model for J0607 is consistent with the one constructed using MIRI data by \citet{Gilman++25a}, and from the HST image data modeling presented by \citet{Sheu++26}. We use the reconstructed PSF from \citet{Sheu++26} in our lens model. The HST data has a higher signal to noise ratio, which aids in constraining the mass of the nearby satellite galaxy. We use shapelets \citep{Birrer++15} with $n_{\rm{max}}=10$ concentric with an elliptical S{\'e}rsic profile to model the lensed quasar host galaxy. 
        \item J0659: The NIRCam imaging of J0659 exhibits the same morphological features as the MIRI imaging data used by \citet{Gilman++25a}. However, the higher angular resolution and signal to noise per pixel enable more precise determination of the mass of the satellite galaxy near image C. Our lens model predicts an Einstein radius for the satellite $\theta_{\rm{E}}\sim 0.32 \pm 0.05$. We use an elliptical S{\'e}rsic profile plus shapelets with $n_{\rm{max}}=5$ to model the lensed quasar host galaxy surface brightness. 
        \item 2M1134: A massive galaxy is located $\sim 20$ arcsec away from the main deflector, in the same direction as the large external shear $\gamma_{\rm{ext}} \sim 0.4$ required to model this system. When modeling the imaging data for this lens, we include an SIS profile at the position of the observed galaxy with an Einstein radius of $16 \pm 4$ arcseconds, corresponding to a velocity dispersion of $\sim 800 \ \rm{km} \ \rm{s^{-1}}$ at the lens redshift. Including this cluster-mass object reduces the external shear from $\sim 0.4$ to $\sim \mathcal{O}\left(10^{-2}\right)$, which suggests 2M1134 lies on the outskirts of a cluster halo responsible for the large external shear. We note that the flux ratio predictions from a lens model that includes the cluster are not significantly different from the flux ratios predicted by lens models with a large external shear. We use an elliptical S{\'e}rsic profile plus shapelets with with $n_{\rm{max}}=5$ to model the lensed quasar host galaxy surface brightness. 
        \item B2045: We included this system in our sample for this work because we developed techniques to efficiently and accurately compute the magnifications of the merging image triplet (see next section). We model the lensed arc visible in the MIRI F560W band, and use an elliptical S{\'e}rsic profile to model the lensed quasar host galaxy.  
        \item J2205: The NIRCam imaging data has a higher signal to noise per pixel than the MIRI imaging data used by \citet{Gilman++25a}. We use an elliptical S{\'e}rsic profile with $n_{\rm{max}}=10$ to model the lensed quasar host galaxy surface brightness. 
		\end{itemize}
		
\section{Computing image magnifications and lensing formalism}
\label{sec:appB}
This section discusses efficient algorithms for multi-plane ray tracing. These are intended to give accurate population-level predictions for the probability distribution of the flux ratios, which depend on the image magnifications of a finite size background source. If the surface brightness of the warm dust region or the nuclear narrow-line region surrounding the background quasar is $\mathcal{I}\left(\boldsymbol{\beta}\right)$, the magnification is
\begin{equation}
	\label{eqn:magnification}
	\mu = \frac{\int \mathcal{I}\left(\beta\left(\boldsymbol{\theta}\right)\right) d^2 \theta }{\int \mathcal{I}\left(\boldsymbol{\beta}\right) d^2 \beta}
\end{equation}
which depends on $\boldsymbol{\beta}\left(\boldsymbol{\theta}\right)$, the lensed coordinates on the source plane, and where $\boldsymbol{\theta}$ represents an angle on the sky as seen by the observer. 

We begin in Section \ref{ssec:decoupledmultiplane} by reviewing the decoupled multi-plane formalism, an approximation for exact ray tracing that we use to solve the lens equation. This formalism was developed by \citet{Gilman++24} and used in previous analyses \citep{Keeley++24,Keeley++25,Gilman++25a}. In Section \ref{ssec:fastmags} we describe a further approximation for multi-plane ray tracing designed to efficiently calculate the image magnifications. Section \ref{ssec:adaptivetiling} describes an adaptive tiling algorithm that we use to accelerate magnification calculations, given a lens model. 

The improvements to our modeling pipeline discussed in this section reduce the time required to forward model a single realization by a factor of 5--10, depending on the number of halos in the model and the typical magnification of a lensed image. The average time to generate a lens model, which involves rendering a (sub)halo population, solving for a set of macromodel parameters that satisfies the lens equation, and ray-tracing through the lens model to compute $\mu$, is reduced to $\sim 5$ seconds, on average. 

\subsection{Decoupled multi-plane ray tracing}
\label{ssec:decoupledmultiplane}
For a multi-plane system with $N$ planes between observer and source, the exact form of the multi-plane lens equation is given by \citep{BlandfordNarayan86}
\begin{eqnarray}
	\label{eqn:raytracingexact}
	\boldsymbol{\beta} &=& \boldsymbol{\theta} - \frac{1}{D_s} \sum_{i=1}^{N} D_{i,s} \boldsymbol{\alpha_i}\left(D_i \boldsymbol{\theta_{i}}\right) \\ 
	&=& \boldsymbol{\theta} - \boldsymbol{\alpha_{\rm{eff}}}\left(\boldsymbol{\theta}, \boldsymbol{\alpha_{\rm{m}}}, \boldsymbol{\alpha_{\rm{r}}}\right).
\end{eqnarray}
In the second line, we have defined a multi-plane deflection angle $\boldsymbol{\alpha_{\rm{eff}}}$, which depends on $\boldsymbol{\theta}$, the macromodel deflection field $\boldsymbol{{\alpha_{\rm{m}}}}$, and the deflection angles $\boldsymbol{\alpha_{\rm{r}}}$ produced by the substructure realization $\rsub$ (using notation from Section \ref{sec:inference}).  Here $D_{i,s}$, $D_s$, and $D_i$ represent angular diameter distance from the $i$th lens plane to the source plane, from the observer to the source plane, and from an observer to the $i$th lens plane, respectively. 

For each realization of a given lens, we solve for $\boldsymbol{\alpha_{\rm{m}}}$ by requiring that the resulting lens model (main deflector plus halos) satisfy the lens equation for the observed image positions. Using Equation \ref{eqn:raytracingexact}, this optimization problem becomes computationally expensive because each new proposed set of macromodel parameters changes the path of a light ray through the background lens planes, requiring the continuous recalculation of all deflection angles produced by halos between the main deflector and the source. The computational expense stems from the recursive nature of Equation \ref{eqn:raytracingexact}. 

To speed up the optimization problem for $\boldsymbol{\alpha_{\rm{m}}}$ we use the decoupled multi-plane formalism, an approximation for multi-plane ray tracing that breaks the recursive nature of Equation \ref{eqn:raytracingexact} while preserving non-linear properties of multi-plane ray tracing. We begin by constructing a macromodel that satisfies the lens equation for the image positions through an optimization algorithm {\tt{cobyqa}} \citep{cobyqa}. Because we sample a subset of macromodel parameters and astrometric uncertainties for each realization, each starting point for the macromodel will be unique. The deflection field associated with this initial guess for the macromodel is $\boldsymbol{\hat{\alpha}_{\rm{m}}}$. Using $\boldsymbol{\hat{\alpha}_{\rm{m}}}$, the deflection field from the static halo population $\boldsymbol{\alpha_{\rm{r}}}$, and Equation \ref{eqn:raytracingexact}, we ray trace through background lens planes until reaching a set of coordinates on the source plane $\boldsymbol{\hat{\beta}}$:
\begin{equation}
	\label{eqn:betahat}
	\boldsymbol{\hat{\beta}} = \boldsymbol{\theta} - \boldsymbol{\alpha_{\rm{eff}}}\left(\boldsymbol{\theta}, \boldsymbol{\hat{\alpha}_{\rm{m}}}, \boldsymbol{\alpha_{\rm{r}}}\right).
\end{equation}
We then write the lens equation as
\begin{equation}
	\label{eqn:raytracingdecoupled}
	\boldsymbol{\beta} = \boldsymbol{\hat{\beta}} + \frac{T_{d,s}}{T_s}\left(\boldsymbol{\hat{\alpha}_{\rm{m}}} - \boldsymbol{{\alpha_{\rm{m}}}}\right)
\end{equation}
Here, $T_{d,s}$ and $T_s$ represent comoving distances from the main deflector to the source plane, and from the observer to the source. $\boldsymbol{\hat{\beta}}$ encodes perturbations by halos along the line of sight computed from the exact multi-plane lens equation. However, the perturbations by halos couple to $\boldsymbol{\hat{\alpha}}_{\rm{m}}$, rather than to each new proposal of $\boldsymbol{\alpha_{\rm{m}}}$. In this sense halos ``decouple'' from the main deflector. For a given realization, this approximation predicts different flux ratios than exact ray tracing, but it predicts the same population-level statistics when used to compute flux ratios \citep{Gilman++24}. 

In this work, we use the decoupled multi-plane formalism and Equation \ref{eqn:raytracingdecoupled} to optimize the macromodel parameters such that they satisfy the lens equation for the image positions for each realization. This involves one multi-plane ray tracing calculation per image to compute $\boldsymbol{\hat{\beta}}$, and then an optimization problem in which we solve for $\boldsymbol{\alpha}_{\rm{m}}$ such that the four images map to the same source position. Because Equation \ref{eqn:raytracingdecoupled} is linear in $\boldsymbol{\alpha_{\rm{m}}}$, this optimization problem is computationally inexpensive and is accomplished in $\mathcal{O}\left(1 \ \rm{sec}\right)$. 
\begin{figure*}
	\centering
	\includegraphics[trim=0cm 0cm 0cm
	0cm,width=0.46\textwidth]{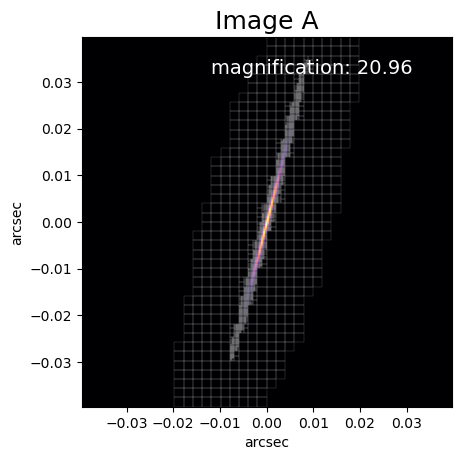}
	\includegraphics[trim=0cm 0cm 0cm
	0cm,width=0.46\textwidth]{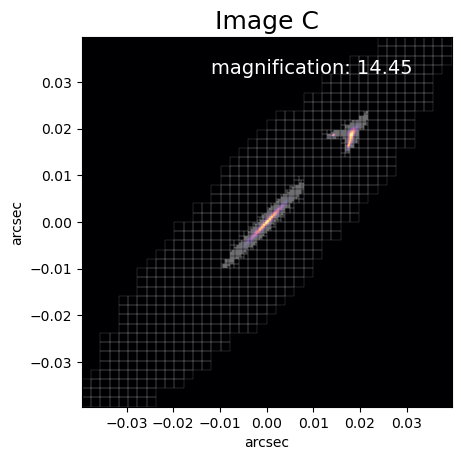}
	\caption{\label{fig:tiling} Lensed quasar images A and C for a particular realization of the lens system RXJ1131. Panels illustrate the adaptive tiling of the image plane used to efficiently compute image magnifications, as discussed in Section \ref{ssec:adaptivetiling}. White lines in each panel show the pixel sizes, which sub-divide into smaller pixels to produce a high resolution image of the source. A core collapsed SIDM halo on top of Image C splits the image in two. }
\end{figure*}

\subsection{An efficient algorithm for computing image magnifications}
\label{ssec:fastmags}
Solving the lens equation requires calculations using four central rays, one ray per image, with the location of the central ray determined by ray tracing backwards with Equation \ref{eqn:raytracingexact} through each observed image position. To evaluate Equation \ref{eqn:magnification} and compute a magnification for a finite source, however, we must propagate a bundle of light rays through the lens system. The magnification depends on the distortion of the bundle around the central ray, meaning the magnification depends on derivatives of the deflection angles. To evaluate Equation \ref{eqn:magnification}, we tile a finely sampled grid with $N_{\rm{pix}}$ pixels around the central ray. Each pixel requires $N_{\rm{halo}}$ deflection angle evaluations, where $N_{\rm{halo}}$ is the number of halos in the lens system. The total number of deflection angle evaluations therefore scales as $N_{\rm{pix}} \times N_{\rm{halos}}$. For a typical lens system $N_{\rm{pix}} \sim \mathcal{O}\left(10^4\right)$ and $N_{\rm{halo}}\sim \mathcal{O}\left(10^2\right)$ at each lens plane.  This calculation requires $\mathcal{O}\left(1 \ \rm{min}\right)$ per realization, and is the main computational bottleneck when generating millions of realizations per lens.  

Image magnifications depend on derivatives of the deflection angles, so we only expect perturbers close to a central ray to strongly affect the magnification. Suppose that at each of the $i$th lens planes, $ N_{\rm{near}}+N_{\rm{far}}$ total halos contribute to the deflection field, where $N_{\rm{near}}$ and $N_{\rm{far}}$ represent the number of halos that are ``near'' and ``far'' from the central ray, respectively. We will delineate near and far later in this section, but it turns out that for a typical system $N_{\rm{far}}>> N_{\rm{near}}$, meaning that most of the computation time is spent evaluating deflection angles for halos that do not significantly affect the magnification. This motivates our development of a new perturbative technique to compute image magnifications. 

Let the set of coordinates $\left(\boldsymbol{\theta_{0,1}}, \boldsymbol{\theta_{0,2}}, ..., \boldsymbol{\theta_{0,N}}\right)$ represent the angular position of the central ray along the line of sight, and $\left(\boldsymbol{\alpha_{\rm{ref,1}}, \alpha_{\rm{ref,2}}, ..., \alpha_{\rm{ref,N}}}\right)$ represent the deflection angle produced by all halos at the $i$th lens plane at the center of the ray bundle, $\boldsymbol{\theta_{i,0}}$. We approximate the lensing effect of each ``far'' halo at the $i$th lens plane through its tidal tensor, $\boldsymbol{\Gamma_h}$, evaluated at the position of the central ray, such that $\boldsymbol{\Gamma_h}$ has components $\Gamma_{ab} = \partial \alpha_{a} / \partial \theta_b \vert_{\boldsymbol{\theta_{i,0}}} $. We model the deflection field from halos close to the central ray exactly, without the tidal approximation. We can then write the deflection field across the $i$th lens plane, $\alpha_{i}\left(\boldsymbol{\theta_i}\right)$, as $\boldsymbol{\alpha_i} = \boldsymbol{\alpha_{\rm{ref,i}}} + \boldsymbol{\tilde{\alpha}_i}$, where 
\begin{equation}
	\label{eqn:tidalapprox}
	\boldsymbol{\tilde{\alpha}_{i}}\left(\boldsymbol{\theta_i}\right) = \sum_{h=1}^{N_{\rm{near}}} \left[\boldsymbol{\alpha_h} \left( \boldsymbol{\theta_i} \right) - \boldsymbol{\alpha_h} \left( \boldsymbol{\theta_{0,i}} \right)\right] + \sum_{h=1}^{N_{\rm{far}}} \boldsymbol{\Gamma_h}  \cdot \left(\boldsymbol{\theta_i} - \boldsymbol{\theta_{0,i}} \right). 
\end{equation}
Note that $\boldsymbol{\tilde{\alpha}_i}$ has zero deflection at each $\boldsymbol{\theta_{0,i}}$, and it therefore represents a distortion field around the central ray caused by nearby perturbers (the first term in Equation \ref{eqn:tidalapprox}), and the collective effects of many distant ones (the second term). At the main lens plane, we include an additional term $\boldsymbol{\tilde{\alpha}_{\rm{m}}} = \boldsymbol{\alpha_{\rm{m}}} \left(\boldsymbol{\theta_i}\right) - \boldsymbol{\alpha_{\rm{m}}} \left(\boldsymbol{\theta_{0,i}}\right)$ to account for the distortion produced by the macromodel around the central ray. We define the macromodel deflection of the central ray in terms of $\boldsymbol{\hat{\alpha}_{\rm{m}}}$, such that the ray bundle propagates down the same path through background lens planes as the light ray used to compute $\boldsymbol{\hat{\beta}}$. 

We use Equation \ref{eqn:tidalapprox} to propagate ray bundles down the path specified by the $\boldsymbol{\theta_{i,0}}$, accumulating distortions from halos along the line of sight until we reach the source plane. From the structure of Equation \ref{eqn:tidalapprox} we can also define thresholds for the ``near'' and ``far'' contributions. The second term represents the first order expansion of the deflection field around $\boldsymbol{\theta_{i,0}}$, which will be accurate provided the components of $\Gamma_h$ do not vary strongly across the extent of the lensed image. In practice, we connect the notion of ``near'' and ``far'' to the mass of a halo, with a distance threshold defining the far away halos: $R = R_0 \left( m /10^8 M_{\odot}\right)^{1/3}$ with $R_0 = 0.4 \ \rm{arcsec}$. The $1/3$ exponent stems from the next-order term in the expansion of the far halos around $\boldsymbol{\theta_{i,0}}$, the flexion, which decays as $r^{-3}$. Our choice of $R_0 = 0.4 \ \rm{arcsec}$ is already very conservative, and includes many more perturbers in the ``near'' group than necessary to accurately predict the flux ratios. 

Our approach is similar to the methodology considered by \citet{McCully++14}, specifically their Section 3.3 titled ``Multiple main planes'', in which there are several dominant lens planes and several lens planes where the deflection field is computed from the tidal tensor. Our approach differs from the methodology discussed by \citet{McCully++14} because a given lens plane in our simulations typically includes both a tidal component (the ``far'' halos) and a subset of perturbers (the ``near'' halos) for which we compute the exact deflection field $\boldsymbol{\alpha_{h}}\left(\boldsymbol{\theta_{i}}\right)$.  

The algorithm discussed in this section turns the calculation of image magnifications, previously the largest computational bottleneck in our analysis pipeline, into one of the fastest parts of the code. The first term in Equation \ref{eqn:tidalapprox} involves a sum over each halo's individual deflection angle at each coordinate, and involves $N_{\rm{near}} \times N_{\rm{pix}}$ deflection angle evaluations. The second term requires us to compute $\boldsymbol{\Gamma_h}$ for each of the ``far'' halos, but the calculation of $\boldsymbol{\Gamma_h}$ involves only a few deflection angle calculations to compute derivatives at $\boldsymbol{\theta_{i,0}}$. Relative to the computation time required for exact ray tracing, this approach is faster by a factor $N_{\rm{far}} \sim \mathcal{O}\left(10^2\right)$, meaning the calculation of an image magnification is reduced from minutes to seconds or a fraction of a second. 
\begin{figure}
	\centering
	\includegraphics[trim=0cm 0cm 0cm
	0cm,width=0.46\textwidth]{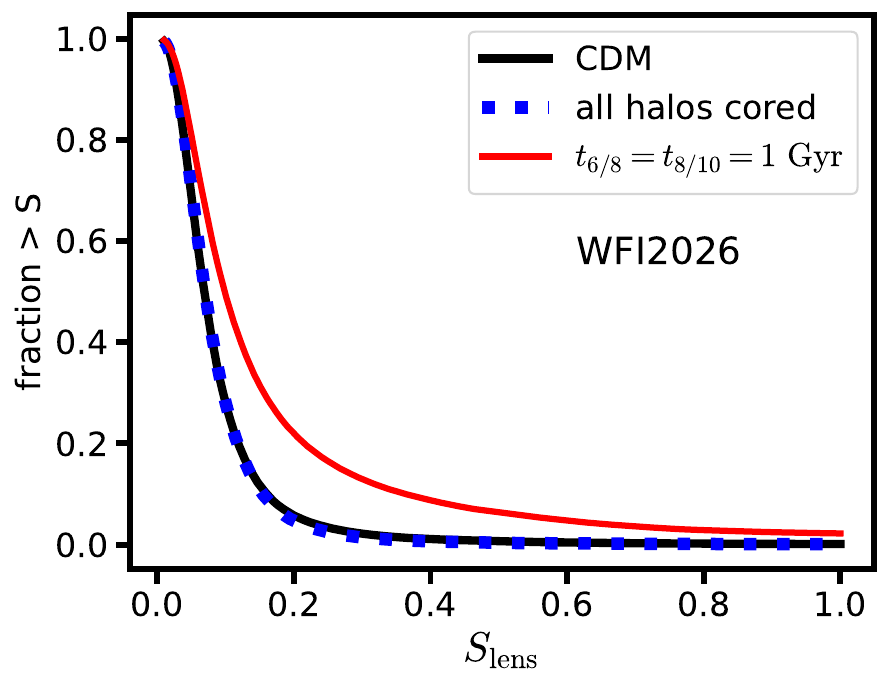}
	\includegraphics[trim=0cm 0cm 0cm
	0cm,width=0.46\textwidth]{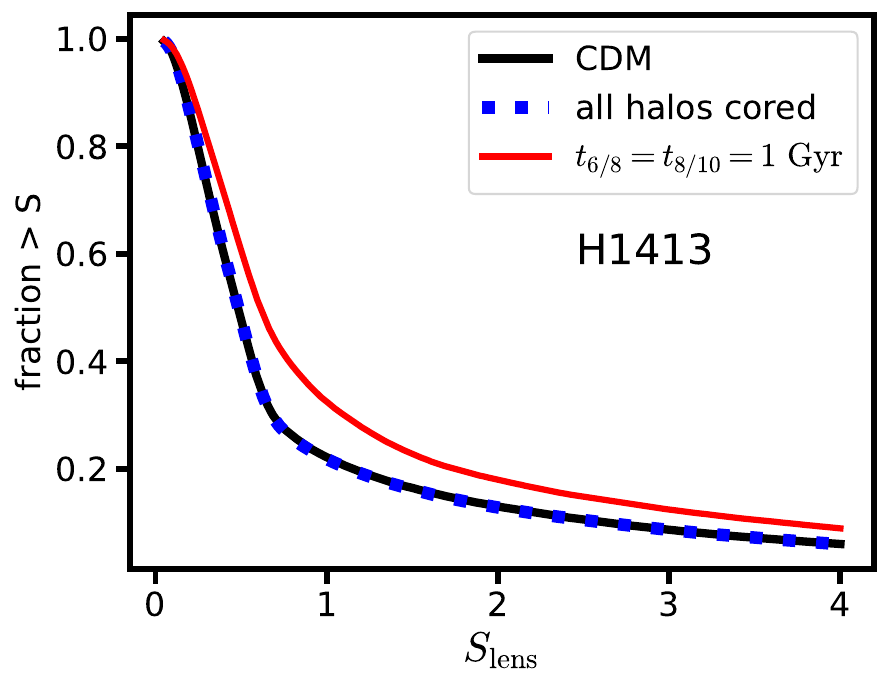}
	\caption{\label{fig:cores} The cumulative distribution of the flux ratio anomaly summary statistic defined in Equation \ref{eqn:slens}. The $y$-axis shows the fraction of lens models with a flux ratio anomaly statistic greater than the values along the $x$-axis, such that models with more small-scale perturbation produce broader distributions.}
\end{figure}

\subsection{Propagating ray bundles with an adaptive tiling}
\label{ssec:adaptivetiling}
As discussed in the previous section, evaluating Equation \ref{eqn:magnification} involves $N_{\rm{pix}} \times N_{\rm{halo}}$ deflection angle calculations. The previous subsection described an approximation for multi-plane ray tracing that effectively reduces $N_{\rm{halo}}$. This section describes a numerical procedure applied to the ray bundles that reduces $N_{\rm{pix}}$. 

To compute an image magnification, we use a high angular resolution tiling of pixels around each image position, and map each pixel back to the source plane. Previous work used a brute-force approach in which we ray trace through every pixel near the image position that could plausibly carry flux. Our code uses a resolution of 0.01-0.1 milli-arcseconds per pixel, depending on the source size, and a window size of order 10 - 100 milli-arcseconds, meaning $N_{\rm{pix}} \sim \mathcal{O}\left(10^4\right)$. 

A brute-force approach results in a large amount of wasted CPU time, because most pixels in the image plane carry zero flux. However, we do not know ahead of time which pixels will carry zero flux. We implement a new algorithm to compute image magnifications that uses an adaptive tiling intended to identify areas in the image plane where a high resolution tiling is required. The algorithm proceeds as follows: 
\begin{itemize}
	\item Step 1: Build a low resolution grid centered on the center of each ray bundle; the default setting of this grid is 20 times coarser than the high-resolution grid, and therefore is faster to compute by a factor of $20^2$.
	\item Step 2: Compute the flux in each pixel, $F_0$, and identify pixels where $F_0$ exceed some threshold $F_{\rm{tol}}$. 
	\item Step 3: Subdivide each pixel where $F_0 > F_{\rm{tol}}$ into four equal parts. Recast each of these four sub-divided pixels back to the source plane; compute the new flux per pixel, $F_i$, where $i$ runs from 1 to 4. 
	\item Step 4: For each of the sub-divided pixels compute the new flux $F_{\rm{i}}$, and the sum $F_{\rm{tot}} = \sum_{i=1}^4 F_{\rm{i}}$. 
	\item Step 5: If $|F_{\rm{tot}} / F_0 - 1|$ exceeds some threshold $\Delta F_{\rm{tol}}$, sub-divide each of the four previously divided pixels and repeat the procedure from Step 3. If $|F_{\rm{tot}} / F_0 - 1| < \Delta F_{\rm{tol}}$, stop sub-dividing these pixels. 
\end{itemize} 
The algorithm described above generates a high resolution tiling in the image plane around pixels where flux appears, and uses a coarse grid in areas where no flux appears. This method therefore significantly reduces $N_{\rm{pix}}$, as the high-resolution tiling is only used in areas where it is needed. 

Figure \ref{fig:tiling} shows an example of lensed quasar images for the lens system RXJ1131 computed using the adaptive tiling algorithm. The coarse grid becomes finely sampled only around the lensed image, where flux appears in the image plane. For this particular realization Image C experiences a ``direct hit'' by a core collapsed halo and is split in two; this is the same phenomenon that causes the jagged features in the magnification curve shown in Figure \ref{fig:sidmmagcross}. Our adaptive tiling algorithm correctly identifies and evaluates the total integrate flux of both components. 

\section{SIDM structure formation model}
\label{sec:appC}

In this Appendix we provide additional details on aspects of the SIDM structure formation model related to halo density profiles. In Section \ref{ssec:cores} we quantify the extent to which cored halos impact flux ratio statistics. Section \ref{ssec:directionalscatter} comments on the role of directional scatter in the core collapse timescale tied to the concentration-mass relation. 
\begin{figure*}
	\centering
	\includegraphics[trim=0cm 0cm 0cm
	0cm,width=0.95\textwidth]{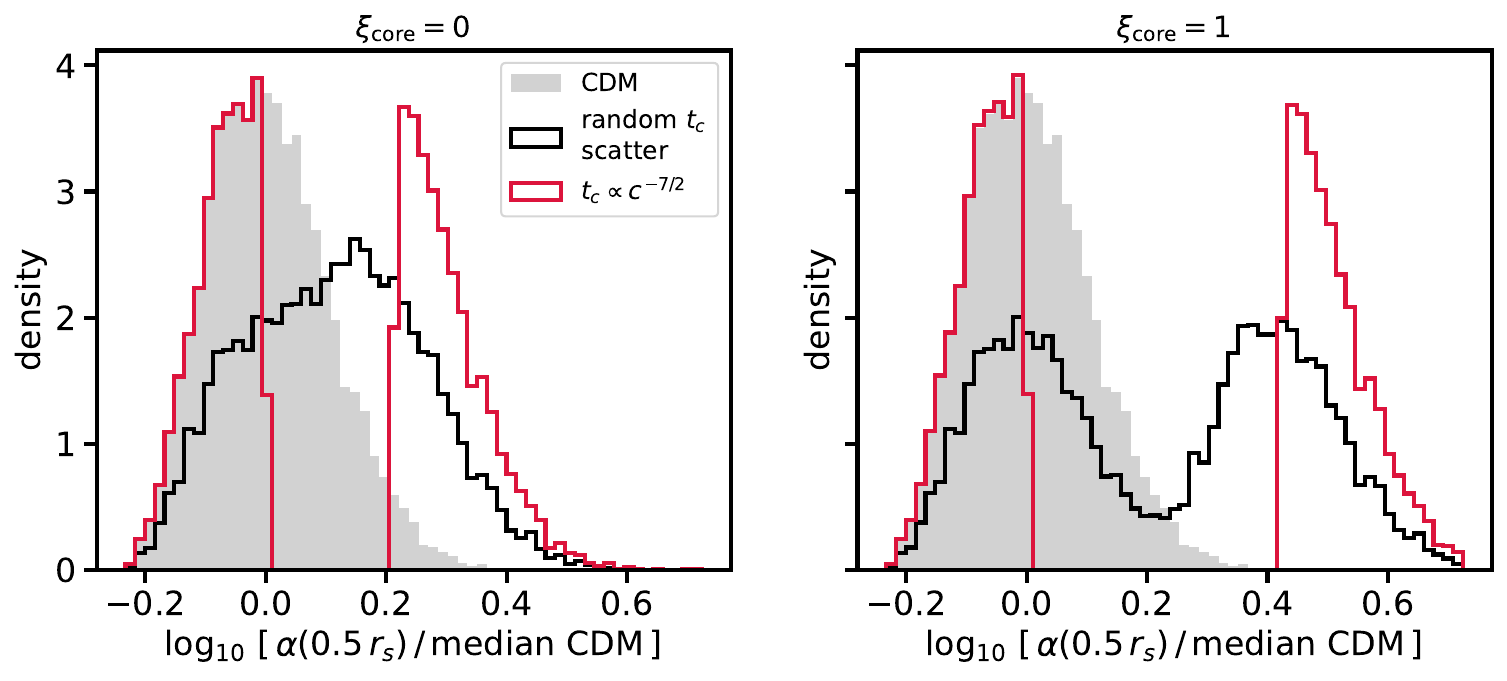}
	\caption{\label{fig:directionalscatter} The effects of directional scatter in the core collapse timescale on the lensing efficiency of core collapsed halos. The gray distribution shows the deflection angle at $0.5 r_s$ for a population of CDM halos with masses $10^8 M_{\odot}$ and a scatter of 0.2 dex in the concentration-mass relation. The black distribution shows the deflection angle at $0.5 r_s$ the results from randomly selecting which halos core collapse. The red distribution includes directional scatter based on the $t_c \propto c^{-7/2}$ scaling of the core collapse timescale in the long mean free path limit (Equation \ref{eqn:tclmfp}). Left and right panels show the effect of changing the internal structure of collapsed halos through $\xi_{\rm{core}}$. }
\end{figure*}

\subsection{Cored halo effects}
\label{ssec:cores}
Our analysis models SIDM halo populations as collapsed and un-collapsed halos, modeled as truncated NFW profiles and with the model discussed in Section \ref{ssec:collapsedsidm}, respectively. We do not model the core formation stage of SIDM halos because the flux ratio perturbations from a population of cored halos are nearly indistinguishable on the population level from CDM halos. To demonstrate this, we simulate the flux ratios in two lenses WFI2026 and H1413, with three different halo populations. First, we consider a CDM population, where field halos and subhalos are modeled as truncated NFW profiles. Second, we consider a population of halos with a density profile given by 
\begin{equation}
\label{eqn:sidmcored}
\rho(r) = \frac{\rho_s r_s^3 r_t^2}
{\left(r_c^2 + r^2\right)^{1/2} \left(r_s^2 + r^2\right)\left(r_t^2 + r^2\right)}
\end{equation}
where $r_s$, $r_c$, and $r_t$ are the scale, core, and truncation radii. The profile has a core inside $r_c$, drops as $r^{-3}$ between $r_s$ and $r_t$, and is truncated as $r^{-5}$ beyond $r_t$. We evaluate this model for $r_c \sim 0.2 r_s$, which reproduces the parametric model for SIDM halo evolution presented by \citet{YangD++24} at $t/t_c = 0.4$ to within $\sim 30\%$ in density at fixed enclosed mass. We use Equation \ref{eqn:sidmcored} instead of the exact parametric model proposed by \citet{YangD++24} because Equation \ref{eqn:sidmcored} admits analytic solutions for the projected mass and deflection angles. We have implemented the profile in Equation \ref{eqn:sidmcored} in $\tt{pyHalo}$ and $\tt{lenstronomy}$. Finally, as a point of comparison, we compute the flux ratio statistics for an SIDM model in which most halos core collapse $t_{6/8} = t_{8/10} = 1 \ \rm{Gyr}$. 

To quantify differences between these halo populations, we compute the cumulative distribution of a summary statistic
\begin{equation}
\label{eqn:slens}
S_{\rm{lens}} = \sqrt{\sum_{i=1}^3 \left(f_{i} - f_{\rm{ref,i}}\right)^2}
\end{equation}
where $f_{\rm{ref,i}}$ represents some reference flux ratio, and $f_i$ represents a flux ratio computed by forward modeling the data. For this experiment, we choose $f_{\rm{ref,i}}$ as the median of the flux ratio distribution in CDM. The cumulative distribution of $S_{\rm{lens}}$ for the lens systems WFI2026 and H1413 are shown in Figure \ref{fig:cores}. While core collapse produces more flux ratio perturbations and therefore a more extended distribution of $S_{\rm{lens}}$, core halos and truncated NFW profiles produce nearly identical flux ratio statistics on the population level. 

\subsection{Directional scatter in core collapse timescales}
\label{ssec:directionalscatter}
In the long mean free path limit the core collapse timescale has an explicit dependence on halo structure and the thermally-averaged cross section strength, $\sigma$, given by 
\begin{equation}
\label{eqn:tclmfp}
t_c = \frac{150}{C}\frac{1}{\sigma \rho_s r_s \sqrt{4 \pi G \rho_s}},
\end{equation}
where $C \sim 0.75$ is a factor calibrated to fluid simulations\citep{Balberg++02,Essig++19}. This expression scales as $t_c \propto \sigma^{-1} c^{-7/2}$ at fixed halo mass. In the short mean free path limit the scaling changes to $t_c \propto \sigma \ c^{1/2}$, meaning scatter in the concentration mass relation propagates differently onto the core collapse timescale in a way that also depends on $\sigma$.  

In this work we assign halos a collapse time drawn from a distribution $\sim \mathcal{N}\left(\log_{10}t_c,0.3\right)$, where $t_c$ represents either $t_{6/8}$ or $t_{8/10}$, depending on the halo mass. As discussed in Section \ref{ssec:timescale}, we adopt this methodology so that we can interpret our inference on $t_c$ without making explicit assumptions for how the collapse timescale depends on $\sigma$, and other factors such as inelastic scattering or dissipation. Our approach also avoids making explicit assumptions for how the median collapse time in different halo mass ranges depends on environmental effects. 
\begin{figure*}
	\centering
	\includegraphics[trim=0cm 0cm 0cm
	0cm,width=0.48\textwidth]{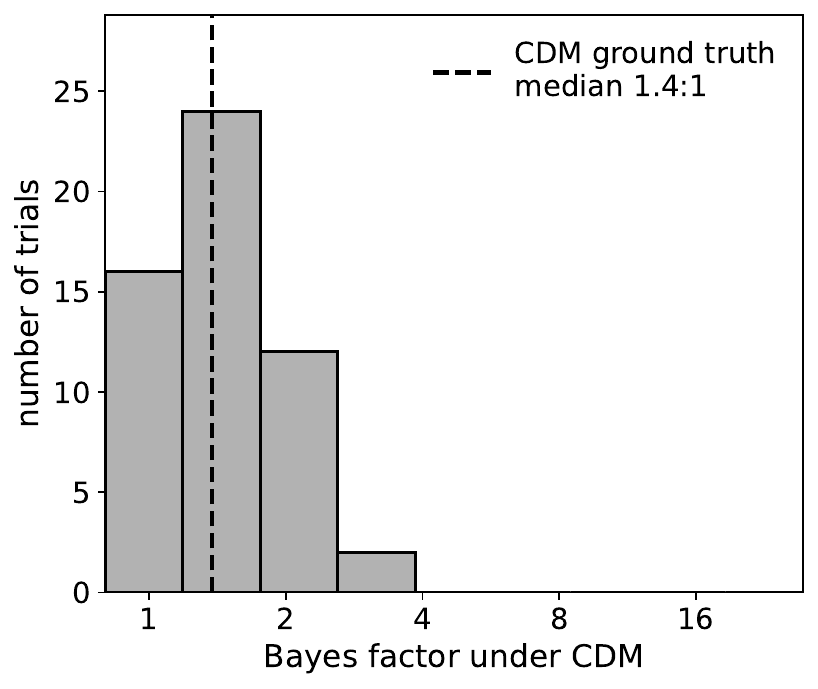}
    \includegraphics[trim=0cm 0cm 0cm
	0cm,width=0.48\textwidth]{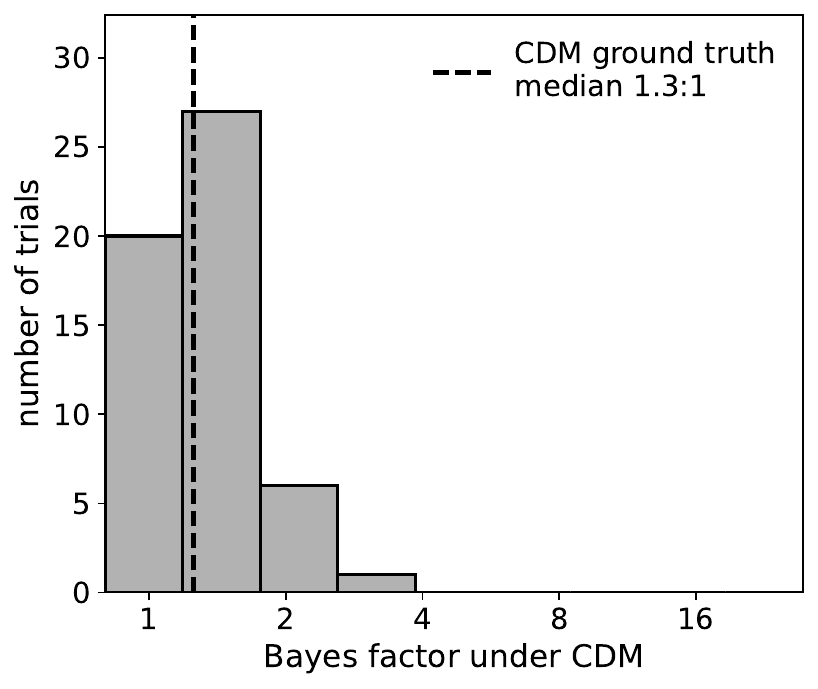}
	\caption{\label{fig:null} Bayes factors computed for $54$ surveys under a CDM ground truth for importance sampling weights corresponding to those in Figures~\ref{fig:jointpdfbestA} (left panel) and the $\xi_{\rm{core}}\sim1$ panel of Figure~\ref{fig:marginalpdfbestEF} (right panel). The largest value obtained was 3:1, which occurred twice for the importance weight assumptions in the left panel and once for the assumptions corresponding to the experiments in the right panel. The Bayes factors obtained from analyzing the real dataset for the same set of importance sampling weights ranged from 14:1 to 24:1.}
\end{figure*}

While we can choose $t_{6/8}$ and $t_{8/10}$ to match a given abundance of collapsed halos, by not including directional scatter tied to $c$ our approach differs from the outcomes one would obtain from Equation \ref{eqn:tclmfp} regarding \textit{which} halos collapse. Figure \ref{fig:directionalscatter} illustrates how the deflection angles from collapsed halos depends on concentration, as predicted by our model for collapsed halo density profiles presented in Section \ref{ssec:collapsedsidm}. We first generate a population of CDM halos with masses of $10^8 M_{\odot}$ with halo concentrations sampled according to the model of \citet{DiemerJoyce19} with a scatter of 0.2 dex, and compute the deflection angle produced by each halo at $0.5 r_s$. The distribution of deflection angles in CDM, relative to the median of the distribution, is shown as the gray distribution in Figure \ref{fig:directionalscatter}. Next, we generate a population of core collapsed halos. We begin by assigning each halo a core collapse timescale $t_c = t_{\rm{halo}} + \delta t$, such that $t_{\rm{halo}} / t_c = 1$ if $\delta t=0$. To create the population of halos shown as the black distribution we compute $\delta t$ as a random draw per halo, as done in this work. The red distribution computes $\delta t / t_{\rm{halo}} = \left(c /c_{\rm{med}}\right)^{-7/2}$, which depends on the concentration, $c$, of each halo sampled the concentration-mass relation. The total number of collapsed halos is the same between the red and black distributions, but the red distribution selects more-concentrated halos as the ones that undergo core collapse. The collapsed population with $t_c$ tied to concentration (red) therefore becomes systematically denser than a population of halos in which core collapsed is assigned at random. This effect is clearly visible in Figure \ref{fig:directionalscatter} as a the split between deflection angles at $0.5 r_s$ between the collapsed and un-collapsed populations. 

Part of the motivation for introducing a flexible model for the internal structure of collapsed halos is to account for factors such as directional scatter, which may alter the typical lensing efficiency of core collapsed halos. As shown by the left and right panel of Figure~\ref{fig:directionalscatter}, the effect of introducing directional scatter has a similar effect to changing the internal structure of collapsed halos. Our model with $\xi_{\rm{core}}=1$, shown on the right, better captures the systematic shift in lensing efficiency associated with directional scatter. Using importance sampling weights $\xi_{\rm{core}}\sim\mathcal{N}\left(1,0.3\right)$ may therefore give a better approximation of the lensing signal of SIDM halo populations in the long mean free path limit, where $t_c \propto c^{-7/2}$. A promising avenue for future work would involve modeling this effect alongside the correlated processes of subhalo tidal evolution and core collapse. 

\section{Null hypothesis calibration of the Bayes factor}
\label{sec:appD}
In this Appendix, we compute the distribution of Bayes factors, as defined in Equation \ref{eqn:bayesfactor}, under a CDM ground truth. These tests address the question of how frequently we would infer a statistically significant Bayesian preference for SIDM over CDM purely by chance, or as a result of possible numerical effects in the calculation of the likelihood functions, such as the use of a kernel density estimator (see Appendix B of \citet{Gilman++25a}).  

Using the simulations for each of the 29 lenses in our sample, we select a realization consistent with CDM, and use the corresponding flux ratios as the ground truth for the inference. We perturb each set of flux ratios according to the measurement uncertainties in each lens system, and omit the flux ratios and model parameters selected as ground truth from the resulting inference. We perform $54$ trials on the full sample of 29 lenses, and calculate the Bayes factors using Equation \ref{eqn:bayesfactor}. 

Figure \ref{fig:null} shows the distribution of Bayes factors computed under a CDM ground truth for the importance sampling weights used in Figure \ref{fig:jointpdfbestA} (left panel) and in Figure \ref{fig:marginalpdfbestEF} (right panel). The median of the distributions are 1.4 and 1.3, and the largest value obtained in 54 trials was 3:1, twice for the left panel, and once for the right panel. These values are well below the Bayes factors we extract from the real data. This demonstrates that our results are robust to statistical fluctuations in the likelihood, or numerical effects that enter the calculation.

	\end{document}